\documentclass[a4paper,fleqn]{cas-sc}
\usepackage[authoryear,longnamesfirst]{natbib}
\usepackage{amsmath,amssymb}
\usepackage{graphicx}
\usepackage{fontspec}
\usepackage{booktabs,array,multirow,longtable}
\usepackage{hanging}
\usepackage{float}
\usepackage{placeins}
\usepackage{xcolor}
\definecolor{myvioletblue}{RGB}{73,67,168}
\definecolor{bestred}{RGB}{180,32,42}
\definecolor{secondblue}{RGB}{34,86,164}
\newcommand{\best}[1]{\textcolor{bestred}{\textbf{#1}}}
\newcommand{\secondbest}[1]{\textcolor{secondblue}{\textbf{#1}}}
\hypersetup{colorlinks=true,linkcolor=myvioletblue,citecolor=myvioletblue,urlcolor=myvioletblue}
\DeclareSymbolFont{correctsymbols}{OMS}{cmsy}{m}{n}
\DeclareMathSymbol{\in}{\mathrel}{correctsymbols}{"32}
\DeclareMathSymbol{\times}{\mathbin}{correctsymbols}{"02}
\DeclareMathSymbol{-}{\mathbin}{correctsymbols}{"00}
\DeclareMathSymbol{\ast}{\mathbin}{correctsymbols}{"03}
\DeclareMathSymbol{\vee}{\mathbin}{correctsymbols}{"5F}
\DeclareMathSymbol{\wedge}{\mathbin}{correctsymbols}{"5E}
\newcommand{\setopen}{\mathopen{\text{\textbraceleft}}}
\newcommand{\setclose}{\mathclose{\text{\textbraceright}}}
\makeatletter\@removefromreset{equation}{section}\makeatother

\newcommand{\bluefigref}[1]{\hyperref[fig:#1]{\textcolor{myvioletblue}{\textbf{Figure~\ref*{fig:#1}}}}}
\newcommand{\bluetabref}[1]{\hyperref[tab:#1]{\textcolor{myvioletblue}{\textbf{Table~\ref*{tab:#1}}}}}
\begin{document}
\shortauthors{Yu et al.}
\shorttitle{Action-Conditioned Multi-Ramp Metering}
\title[mode=title]{Uncertainty-Aware Coordinated Multi-Ramp Metering Based on an Action-Conditioned Traffic Prediction Operator Under Sparse Observations}
\author[1,2]{Hongxin Yu}[orcid=0000-0002-1009-2995]
\author[1,2]{Fengyue Jin}
\author[1]{Tao Jiang}
\author[1]{Yibing Wang}[orcid=0000-0001-9937-2055]\cormark[1]
\ead{wangyibing@zju.edu.cn}
\author[2]{Klaus Bogenberger}
\author[3]{Jingqiu Guo}
\author[1]{Lihui Zhang}
\cortext[1]{Corresponding author.}
\affiliation[1]{organization={Institute of Intelligent Transportation Systems, Zhejiang University},city={Hangzhou},postcode={310058},country={China}}
\affiliation[2]{organization={Department of Traffic Engineering and Control, Technical University of Munich},city={Munich},postcode={80333},country={Germany}}
\affiliation[3]{organization={School of Transportation Engineering, Tongji University},city={Shanghai},postcode={201804},country={China}}
\begin{abstract}Coordinated multi-ramp metering (CRM) is an important approach to mitigating freeway congestion, but sparse observations in practice complicate the state estimation and model calibration required by predictive control. This study develops an Action-Conditioned Traffic Prediction Operator (ATPO) driven by sparse, heterogeneous observations and candidate multi-ramp actions, and embeds it in an uncertainty-aware CRM framework (ATPOUCRM). ATPO forms a four-channel history from the density and speed measurements of fixed detectors and probe vehicles and their separate observation masks. Queries at future spatiotemporal points attend to all valid historical observations before incorporating upstream mainline demand and on-ramp queue information. Candidate metering actions condition the latent field through feature-wise linear modulation, a spatial Fourier operator, and a causal temporal kernel. The operator produces action-conditioned predictive distributions of future density and speed throughout the corridor; a bootstrap ensemble additionally characterizes disagreement among models. Using these predictions, ATPOUCRM primarily minimizes expected total time spent, including mainline travel and on-ramp waiting, while penalizing predictive uncertainty near merge areas. It uses the cross-entropy method to search candidate actions and applies a safety filter before execution. For five-minute global traffic state prediction on a simulated 7-km freeway corridor, ATPO achieves root mean square errors of 2.54 veh/km/lane, 4.56 km/h, and 0.76 veh·h for density, speed, and total time spent over the forecast horizon, respectively, outperforming the comparison methods. Across 100 control scenarios, ATPOUCRM reduces mean total time spent by 26.17\% relative to no control. ATPO also generates predictions on finer spatiotemporal grids without retraining.\end{abstract}
\maketitle
\setlength{\abovedisplayskip}{7pt plus 2pt minus 2pt}
\setlength{\belowdisplayskip}{7pt plus 2pt minus 2pt}
\setlength{\abovedisplayshortskip}{4pt plus 2pt minus 1pt}
\setlength{\belowdisplayshortskip}{5pt plus 2pt minus 1pt}
\section{Introduction}\label{sec:1}

Freeways are essential corridors for urban and intercity travel. Increasing vehicle ownership, however, has contributed to recurrent congestion that reduces network efficiency and service reliability. Ramp metering regulates the flow entering the mainline, reduces density near merge areas, and thereby limits the formation and propagation of bottleneck congestion. Its practical value was demonstrated by the local-ramp feedback controller ALINEA (\textcolor{myvioletblue}{\textbf{\hyperlink{bib:32}{Papageorgiou et al., 1991}}}). Yet local benefits from metering one ramp may come at the expense of downstream traffic conditions or longer on-ramp queues. This has motivated Coordinated multi-Ramp Metering (CRM), which allocates metering and queue storage across ramps to improve corridor-wide performance. The field deployment of HERO (\textcolor{myvioletblue}{\textbf{\hyperlink{bib:34}{Papamichail et al., 2010b}}}) further showed that coordination among adjacent ramps can improve traffic operations along an entire freeway corridor.

CRM can be organized into three methods according to the information and predictive model used for decision-making. The first coordinates local feedback controllers through predefined rules. HERO uses ALINEA locally and links upstream and downstream ramps according to mainline occupancy and on-ramp queues, drawing on upstream queue storage to relieve a downstream bottleneck. Such controllers require only real-time measurements at ramps and nearby mainline locations, rather than online traffic-model optimization or demand forecasts. Their heuristic rules, however, do not directly maximize network-wide throughput (\textcolor{myvioletblue}{\textbf{\hyperlink{bib:34}{Papamichail et al., 2010b}}}). The second approach optimizes predicted traffic evolution with a traffic-flow model. AMOC and subsequent model predictive control (MPC) methods simulate candidate metering sequences and optimize multiple ramps subject to demand, queue, and control constraints (Kotsialos et al., 2002, \textcolor{myvioletblue}{\textbf{\hyperlink{bib:14}{2004}}}; \textcolor{myvioletblue}{\textbf{\hyperlink{bib:12}{Hegyi et al., 2005}}}; \textcolor{myvioletblue}{\textbf{\hyperlink{bib:33}{Papamichail et al., 2010a}}}). This provides a systematic framework for anticipatory coordination, but the ranking of candidate actions depends on accurate estimates of the current network state, model parameters, and boundary demand. Performance can deteriorate when these conditions depart from the model assumptions. The third approach learns action-driven traffic dynamics from data and uses the learned model for predictive optimization. \textcolor{myvioletblue}{\textbf{\hyperlink{bib:8}{Gu et al. (2023)}}} learned ramp-action-dependent transitions of link-aggregated states with a deep Koopman model and embedded it in MPC for coordinated ramp metering. \textcolor{myvioletblue}{\textbf{\hyperlink{bib:66}{Zhou et al. (2026)}}} additionally learned latent dynamics, reward, and value functions for offline planning. These methods reduce reliance on explicit traffic-flow equations, but their training and control use state vectors covering all modeled links, which are difficult to obtain under sparse real-world observations.

This leads to a central question: how can the strengths of these approaches be combined to achieve effective coordinated ramp metering with sparse detectors? Traffic MPC links two tasks. A prediction model estimates the future state under each candidate control sequence, conditioned on available measurements and external demand; an optimizer compares the resulting system costs and executes the selected action in a receding horizon. A model that predicts the future global traffic state under alternative metering actions from sparse observations would therefore provide the foundation for predictive multi-ramp control under realistic sensing conditions.

Global Traffic State Prediction (GTSP) from sparse observations is one route to such an action-conditioned model. It requires both inference of states at unobserved locations and prediction of their subsequent evolution. Existing methods have mainly followed data-driven and traffic-flow-model-based approaches.

Among data-driven methods, IGNNK uses an inductive graph neural network for spatiotemporal kriging to reconstruct states at unobserved locations (\textcolor{myvioletblue}{\textbf{\hyperlink{bib:56}{Wu et al., 2021}}}). FUNS  (Roth and Liebig, 2022) and MoGERNN extend the target to future states at sensor-free locations (\textcolor{myvioletblue}{\textbf{\hyperlink{bib:65}{Zhou et al., 2025}}}). These methods improve spatial coverage with sparse observations. Their original tasks, however, are primarily conditioned on historical traffic measurements and do not jointly represent the effects of candidate ramp-metering sequences, mainline demand, and ramp queues on the future global traffic state. They therefore cannot directly compare the traffic consequences of alternative actions.

Model-based studies instead rely on traffic-flow dynamics. The cell transmission model (Daganzo, 1994) and METANET-type models (\textcolor{myvioletblue}{\textbf{\hyperlink{bib:49}{Wang and Papageorgiou, 2005}}}) describe traffic evolution through conservation and wave propagation. Coupled with filtering, they can estimate the current global state from limited observations and then propagate it forward. Forecast accuracy nevertheless depends on how well the assumed fundamental diagram, merging rules, and boundary demand match actual traffic. Parameter estimation and online correction can reduce errors under changing conditions but cannot fully eliminate model mismatch (\textcolor{myvioletblue}{\textbf{\hyperlink{bib:50}{Wang et al., 2008}}}, \textcolor{myvioletblue}{\textbf{\hyperlink{bib:52}{2022}}}).

Other studies combine data learning and traffic-flow models for GTSP. In one staged design, a learning model first predicts boundary conditions or detector states, after which a traffic-flow model propagates states into unobserved locations (\textcolor{myvioletblue}{\textbf{\hyperlink{bib:58}{Yang et al., 2023}}}, \textcolor{myvioletblue}{\textbf{\hyperlink{bib:59}{2024}}}). The traffic flow model provides spatial propagation, but errors in the first-stage forecast can be carried into later prediction steps and degrade the future global state.

Neural operators (NO) offer another way to learn mappings from sparse traffic observations to global traffic state fields. DeepONet encodes an input function and queries output coordinates (\textcolor{myvioletblue}{\textbf{\hyperlink{bib:23}{Lu et al., 2021}}}), whereas the Fourier neural operator (FNO) parameterizes nonlocal interactions in the frequency domain (\textcolor{myvioletblue}{\textbf{\hyperlink{bib:19}{Li et al., 2021}}}). For traffic applications, \textcolor{myvioletblue}{\textbf{\hyperlink{bib:43}{Thodi et al. (2024)}}} used a physics-informed FNO for forward and inverse problems of a macroscopic traffic model and recovered density fields from sparse inputs. In traffic control, \textcolor{myvioletblue}{\textbf{\hyperlink{bib:63}{Zhang et al. (2025)}}} approximated control-gain calculations for PDE-based boundary control with a NO, reducing the computational burden of backstepping design. \textcolor{myvioletblue}{\textbf{\hyperlink{bib:24}{Lv et al. (2025)}}} used DeepONet to learn the mapping from model parameters to coupled control kernels and accelerate online gain updates. These studies demonstrate the potential of NOs for traffic modeling, estimation and control design. The NO-based traffic control, however, start from prescribed macroscopic equations and mainly learn boundary-control gains or backstepping kernels. Predicting the future global traffic state under different candidate ramp actions directly from heterogeneous sparse observations from fixed detectors and probe vehicles remains an open task.

To address this task, we propose an Action-Conditioned Traffic Prediction Operator (ATPO) and an ATPO-based Uncertainty-Aware Coordinated Multi-Ramp Metering framework (ATPOUCRM). The operator encodes density and speed observations from fixed detectors and probe vehicles together with their separate masks as a four-channel historical input. Future spatiotemporal coordinates query valid historical observations to form a corridor-wide latent field. Two independent cross-attention branches then incorporate mainline demand and ramp-queue information. Candidate metering sequences modulate latent features layer by layer through feature-wise linear modulation (FiLM). Each causal FNO block combines a Fourier layer that captures spatial dependence along the road with a causal temporal kernel that aggregates only the current and preceding forecast steps. The resulting operator predicts conditional distributions of future density and mean speed throughout the corridor, while a bootstrap deep ensemble characterizes model disagreement. The controller then searches feasible actions with the cross-entropy method, primarily minimizing expected total time spent on the mainline and at ramp queues while accounting for uncertainty near merge areas and congestion risk. A pre-execution safety filter checks the selected action. The controller predicts and optimizes again after receiving new observations.

The principal contributions are as follows:

(1) We develop ATPO, an action-conditioned traffic prediction operator driven jointly by heterogeneous sparse observations and candidate multi-ramp actions. Channel-specific observation masks and future spatiotemporal queries integrate fixed detector and probe vehicle data to predict density and speed at unobserved locations. Separate encodings of mainline demand and ramp queues enable comparison of traffic evolution under alternative actions from the same initial state. ATPO achieves lower action-conditioned GTSP errors than the comparison methods.

(2) We combine FiLM with causal FNO blocks to preserve the temporal ordering of control effects while allowing changes in grid resolution. Candidate control fields modulate each point before the spatial Fourier layer; a temporal kernel defined by physical time differences then restricts each forecast step to current and earlier actions. Continuous-coordinate queries and shared spectral parameters permit the same learned weights to operate on different uniform temporal and spatial grids over the same physical domain. Cross-resolution experiments assess accuracy on finer grids.

(3) We introduce ATPOUCRM, a CRM framework that uses predictive uncertainty in decision-making. Heteroscedastic outputs and a bootstrap deep ensemble represent data noise and model disagreement, respectively. The controller uses the cross-entropy method to search candidate actions, primarily minimizing expected total time spent, including mainline travel and on-ramp waiting. An uncertainty penalty and a pre-execution safety filter reduce the selection of high-risk actions. Control experiments show reduced mainline congestion and better CRM performance than the baselines.

\section{Literature Review}\label{sec:2}

\subsection{Coordinated Multi-Ramp Metering}\label{sec:2.1}

Feedback ramp metering adjusts on-ramp discharge using real-time measurements and is computationally efficient for online deployment. \textcolor{myvioletblue}{\textbf{\hyperlink{bib:32}{Papageorgiou et al. (1991)}}} proposed ALINEA, which uses downstream merging area occupancy/density as feedback to maintain a target occupancy. \textcolor{myvioletblue}{\textbf{\hyperlink{bib:51}{Wang et al. (2010)}}} studied proportional-integral feedback in PI-ALINEA to respond to bottlenecks at different downstream locations. Local feedback alone cannot coordinate several ramps that affect the same bottleneck. \textcolor{myvioletblue}{\textbf{\hyperlink{bib:34}{Papamichail et al. (2010b)}}} therefore proposed HERO, adding queue-coordination rules across neighboring ramps to local ALINEA controllers. Upstream queue storage shares the metering burden of a downstream bottleneck, and the method was deployed on the Monash Freeway. These methods select actions from real-time measurements and predefined rules without online evaluation of multiple control sequences.

Model-based control instead incorporates the predicted consequences of candidate actions. \textcolor{myvioletblue}{\textbf{\hyperlink{bib:15}{Kotsialos et al. (2002)}}} established a nonlinear optimal-control framework for freeway networks and developed the AMOC tool for coordinated control. \textcolor{myvioletblue}{\textbf{\hyperlink{bib:33}{Papamichail et al. (2010a)}}} placed AMOC in the optimization layer of a hierarchical predictive controller, with ALINEA providing local feedback; \textcolor{myvioletblue}{\textbf{\hyperlink{bib:12}{Hegyi et al. (2005)}}} used MPC to optimize ramp metering and variable speed limits jointly. \textcolor{myvioletblue}{\textbf{\hyperlink{bib:7}{Gomes and Horowitz (2006)}}} incorporated queue lengths and metering constraints into optimization based on an asymmetric cell transmission model. \textcolor{myvioletblue}{\textbf{\hyperlink{bib:29}{Muralidharan and Horowitz (2015)}}} used a model of capacity drop and weaving effects and improved computational efficiency through trajectory constraints and optimization relaxations. \textcolor{myvioletblue}{\textbf{\hyperlink{bib:37}{Roncoli et al. (2016)}}} studied hierarchical coordination of ramp metering, mainline flow control, and lane changing in a connected-vehicle setting. \textcolor{myvioletblue}{\textbf{\hyperlink{bib:9}{Han et al. (2020)}}} used an aggregated traffic model to reduce the prediction and optimization complexity of a multi-bottleneck corridor. Together, these studies show that predictive control performance depends on both efficient action optimization and accurate prediction of action-dependent traffic evolution.

Machine learning has entered ramp metering through policy learning and dynamics modeling. \textcolor{myvioletblue}{\textbf{\hyperlink{bib:10}{Han et al. (2022)}}} combined historical observations with samples generated by a traffic flow model to train local and coordinated metering policies using physics-informed reinforcement learning. \textcolor{myvioletblue}{\textbf{\hyperlink{bib:13}{Hu and Ma (2024)}}} used a conventional controller as a teacher to guide learning of CRM and boundary control in a large network. \textcolor{myvioletblue}{\textbf{\hyperlink{bib:-4}{Airaldi et al. (2025)}}} used an MPC optimization problem as a function approximator for reinforcement learning, addressing model mismatch and changing demand in single-ramp control. A separate route learns action-driven traffic dynamics for control planning. \textcolor{myvioletblue}{\textbf{\hyperlink{bib:8}{Gu et al. (2023)}}} used a deep Koopman model to learn transitions of link-aggregated states and embedded it in MPC based CRM. \textcolor{myvioletblue}{\textbf{\hyperlink{bib:66}{Zhou et al. (2026)}}} combined offline reinforcement learning with latent dynamics, reward, and value models for planning. These two studies can predict consequences of candidate actions, but both take state vectors covering the modeled links as inputs. They do not resolve how such states can be constructed from heterogeneous sparse fixed detector and probe vehicle observations or directly predict future global density and speed fields under candidate actions.

\subsection{Global Traffic State Prediction}\label{sec:2.2}

GTSP must produce future states beyond observed locations; it therefore involves both reconstruction of the current state from limited measurements and prediction of subsequent evolution. Model-based approaches propagate detector information into unobserved links through vehicle conservation, traffic wave propagation, and merging and diverging equations. \textcolor{myvioletblue}{\textbf{\hyperlink{bib:4}{Daganzo's (1994)}}} cell transmission model represents state propagation through conservation and supply-demand relations. METANET, introduced by \textcolor{myvioletblue}{\textbf{\hyperlink{bib:27}{Messmer and Papageorgiou (1990)}}}, simulates freeway networks with macroscopic density and speed equations; \textcolor{myvioletblue}{\textbf{\hyperlink{bib:15}{Kotsialos et al. (2002)}}} demonstrated its use in large freeway networks. \textcolor{myvioletblue}{\textbf{\hyperlink{bib:49}{Wang and Papageorgiou (2005)}}} applied an extended Kalman filter (EKF) to estimate the complete current freeway state from sparse detectors. Subsequent online parameter-estimation studies updated the model with new observations. Accurate current state estimation, however, does not imply accurate open-loop multistep prediction: errors in future boundary demand, fundamental diagram parameters, capacity, and merging behavior may shift the position and strength of predicted congestion waves as the model is propagated.

Data-driven research first established spatiotemporal forecasting on fixed sensor networks. \textcolor{myvioletblue}{\textbf{\hyperlink{bib:18}{Li et al. (2018)}}} proposed DCRNN, which represents inter-node information transfer as bidirectional diffusion on a road graph. \textcolor{myvioletblue}{\textbf{\hyperlink{bib:57}{Wu et al. (2019)}}} proposed Graph WaveNet, which learns node dependencies beyond predefined road connections through an adaptive graph. Both primarily forecast states at observed locations. Subsequent work extended prediction to unobserved locations. \textcolor{myvioletblue}{\textbf{\hyperlink{bib:56}{Wu et al. (2021)}}} proposed IGNNK for spatiotemporal kriging with inductive graph learning. \textcolor{myvioletblue}{\textbf{\hyperlink{bib:38}{Roth and Liebig (2022)}}} proposed FUNS to forecast nodes without local observations throughout the historical window. \textcolor{myvioletblue}{\textbf{\hyperlink{bib:65}{Zhou et al. (2025)}}} proposed MoGERNN, combining a mixture of graph experts with spatiotemporal encoding to forecast unobserved nodes under changing sensor layouts. This sequence of work expanded spatial coverage from monitored nodes to unobserved nodes in road networks.

Combining traffic flow models with data learning offers another route. \textcolor{myvioletblue}{\textbf{\hyperlink{bib:35}{Pereira et al. (2022)}}} embedded a discretized traffic reaction model in a recurrent network and learned parameters that vary in space and time, allowing prediction to draw on both traffic dynamics and observations. \textcolor{myvioletblue}{\textbf{\hyperlink{bib:0}{Briani et al. (2024)}}} used an LSTM to predict congestion near detectors and future vehicle counts, which support fundamental diagram inversion and LWR boundary conditions, respectively. \textcolor{myvioletblue}{\textbf{\hyperlink{bib:58}{Yang et al. (2023)}}} combined boundary prediction, a variable-speed cell transmission model, and an EKF to predict network speed. \textcolor{myvioletblue}{\textbf{\hyperlink{bib:59}{Yang et al. (2024)}}} further combined a spatiotemporal dependency network, the cell transmission model, and EKF for network-state prediction. \textcolor{myvioletblue}{\textbf{\hyperlink{bib:48}{Wang et al. (2025)}}} first predicted detector states with an LSTM and then propagated them to unobserved links using METANET-EKF. Traffic flow equations give these methods an explicit spatial propagation mechanism, but errors in predicted boundaries or detector states can propagate through subsequent global predictions.

The observation source also determines the input structure of GTSP. Fixed detectors repeatedly measure flow, speed, or occupancy at the same cross section, whereas probe vehicles provide position and speed observations at irregular points along their trajectories. \textcolor{myvioletblue}{\textbf{\hyperlink{bib:54}{Wright and Horowitz (2016)}}} fused loop-detector and GPS probe vehicle data to estimate freeway density; \textcolor{myvioletblue}{\textbf{\hyperlink{bib:30}{Nantes et al. (2016)}}} used heterogeneous detector data for urban traffic state estimation; \textcolor{myvioletblue}{\textbf{\hyperlink{bib:52}{Wang et al. (2022)}}} studied joint state and parameter estimation from fixed sensors and connected vehicle information. These studies indicate that fixed and mobile observations are complementary in temporal continuity and spatial coverage and may improve GTSP prediction.

Predictive uncertainty also affects whether predictions are suitable for control. Local errors in congestion onset, wave propagation speed, or queue discharge time can change the predicted errors and ranking of candidate actions. \textcolor{myvioletblue}{\textbf{\hyperlink{bib:40}{Sengupta et al. (2024)}}} quantified uncertainty in traffic forecasts with a Bayesian approach and examined generalization. \textcolor{myvioletblue}{\textbf{\hyperlink{bib:25}{Mallick et al. (2024)}}} used an ensemble of spatiotemporal graph models to characterize data and model uncertainty separately. In general model based control, \textcolor{myvioletblue}{\textbf{\hyperlink{bib:3}{Chua et al. (2018)}}} propagated disagreement across candidate trajectories with an ensemble of probabilistic dynamics models and used it for sampling-based planning. These studies provide methodological support for using predictive distributions in traffic control.

\subsection{Neural Operators in Traffic Applications}\label{sec:2.3}

NOs learn mappings between input and output functions, rather than only between fixed length vectors. DeepONet, proposed by \textcolor{myvioletblue}{\textbf{\hyperlink{bib:23}{Lu et al. (2021)}}}, uses a branch network to encode samples of an input function and a trunk network to represent output locations. The FNO proposed by \textcolor{myvioletblue}{\textbf{\hyperlink{bib:19}{Li et al. (2021)}}} parameterizes nonlocal interactions in Fourier space and returns to the physical domain through an inverse transform. \textcolor{myvioletblue}{\textbf{\hyperlink{bib:16}{Kovachki et al. (2023)}}} formalized NO parameterization and its relation to numerical discretization from a function space perspective, providing a framework for reusing learned parameters across grid resolutions.

In traffic applications, NOs have been used for traffic state estimation, traffic flow equation solution, and traffic control. \textcolor{myvioletblue}{\textbf{\hyperlink{bib:6}{Gao and Di (2024)}}} compared different NO architectures for traffic state estimation. \textcolor{myvioletblue}{\textbf{\hyperlink{bib:20}{Li et al. (2025)}}} introduced vehicle-conservation and fundamental diagram constraints into a physics-informed DeepONet for traffic state estimation. \textcolor{myvioletblue}{\textbf{\hyperlink{bib:1}{Chen et al. (2026)}}} combined a mixed traffic model with DeepONet to estimate traffic states in a connected-vehicle environment. \textcolor{myvioletblue}{\textbf{\hyperlink{bib:43}{Thodi et al. (2024)}}} used a physics-guided FNO to learn the forward and inverse problems of macroscopic traffic flow equations and reconstruct density fields from sparse inputs. \textcolor{myvioletblue}{\textbf{\hyperlink{bib:11}{Harting et al. (2025)}}} combined an open-loop FNO predictor with sparse sensor feedback to form a closed-loop density observer. \textcolor{myvioletblue}{\textbf{\hyperlink{bib:47}{Wang et al. (2026b)}}} further examined reconstruction on finer spatiotemporal grids after training on low-resolution data. \textcolor{myvioletblue}{\textbf{\hyperlink{bib:21}{Li et al. (2026)}}} combined graph convolution and FNO to extract spatiotemporal and spectral features for traffic flow forecasting.

NO based control has followed two main approaches. The first approximates mappings needed in control design to reduce online computation. \textcolor{myvioletblue}{\textbf{\hyperlink{bib:-2}{Bhan et al. (2024)}}} used a neural operator to approximate feedback gains and control calculations in backstepping. For freeway traffic control, \textcolor{myvioletblue}{\textbf{\hyperlink{bib:63}{Zhang et al. (2025)}}} learned the feedback mapping for backstepping boundary control of the ARZ traffic flow equations, and \textcolor{myvioletblue}{\textbf{\hyperlink{bib:24}{Lv et al. (2025)}}} studied adaptive boundary control kernels with unknown model parameters. The second approach approximates future system evolution for action evaluation. \textcolor{myvioletblue}{\textbf{\hyperlink{bib:-1}{Bhan et al. (2025)}}} learned a predictor needed for predictive feedback in nonlinear delay systems. \textcolor{myvioletblue}{\textbf{\hyperlink{bib:5}{De Jong et al. (2025)}}} proposed a multistep DeepONet predictor embedded in MPC so that an optimizer can compare candidate input sequences. The existing traffic control operators in this set mainly accelerate computation of a control law or kernel for a prescribed macroscopic traffic equation. An action-conditioned NO that predicts a global traffic field from sparse heterogeneous observations for multi-ramp optimization remains to be studied.

Action-conditioned prediction also requires the correct temporal ordering of control effects. FNO can represent nonlocal dependence in traffic fields. However, if an entire future control sequence is injected into a latent field and the network subsequently mixes all forecast steps without a causal constraint, a later action may influence an earlier prediction through the network's computation path. Causal sequence modeling restricts this path by aggregating only the current and preceding information at each forecast time (\textcolor{myvioletblue}{\textbf{\hyperlink{bib:-3}{Bai et al., 2018}}}).

\section{Methodology}\label{sec:3}

\subsection{Problem Formulation}\label{sec:3.1}

Consider a freeway corridor partitioned into $N_{x}$ cells. At discrete times $t=0,1,\text{...} ,T-1$, let the global traffic state be $S_{t}=({\rho}_{t},v_{t})$, where ${\rho}_{t}=({\rho}_{t,1},\text{...} ,{\rho}_{t,N_{x}})$ and $v_{t}=(v_{t,1},\text{...} ,v_{t,N_{x}})$; ${\rho}_{t,i}$ and $v_{t,i}$ denote the density and mean speed, respectively, of cell $i$ at time $t$.

Suppose the corridor contains $N_{r}$ controlled on-ramps. The vector $U_{t}=(r_{1,t},\text{...} ,r_{N_{r},t})$ contains their actual metering rates at time$t$, and each rate is limited by ramp capacity:

\begin{equation}\label{eq:1}\begin{aligned}r_{\operatorname{min}}\leq r_{b,t}\leq r_{\operatorname{max}},b=1,\text{...} ,N_{r}\end{aligned}\end{equation}

In Eq. (1), $b$ indexes the on-ramps, while $r_{\operatorname{min}}$ and $r_{\operatorname{max}}$ are the lower and upper bounds on the metering rate.

Let $Z_{t}$ denote the exogenous conditions known at time $t$, including upstream mainline demand and the queues at each on-ramp. Ramp metering changes merging flows, whereas external demand changes the number of vehicles entering the system. Together they determine subsequent traffic conditions. The evolution of the global state under control and exogenous inputs can therefore be written abstractly as:

\begin{equation}\label{eq:2}\begin{aligned}S_{t+1}=\mathcal{F}(S_{t},U_{t};Z_{t})\end{aligned}\end{equation}

In Eq. (2), $\mathcal{F}$ is the traffic dynamics operator that maps the current state $S_{t}$, metering action $U_{t}$, and exogenous conditions $Z_{t}$ to the next state $S_{t+1}$. Thus, given the current global state, actual metering rates, and external conditions, $\mathcal{F}$ produces the density and mean speed one time step later.

Let the discrete simulation interval be ${\Delta}t$ hours. Over $[0,T)$, the network total time spent (TTS) by vehicles under dynamics $\mathcal{F}$ is:

\begin{equation}\label{eq:3}\begin{aligned}TTS=\sum _{t=0}^{T-1}[\sum _{i=1}^{N_{x}}n_{i}L_{i}{\rho}_{t,i}+\sum _{b=1}^{N_{r}}w_{b,t}]{\Delta}t\end{aligned}\end{equation}

In Eq. (3), $n_{i}$ and $L_{i}$ are the number of lanes and length of cell $i$, respectively; $w_{b,t}$ is the number of vehicles waiting at on-ramp $b$ at time $t$. The term $n_{i}L_{i}{\rho}_{t,i}$ gives the vehicle count in mainline cell $i$, and $w_{b,t}$ gives the waiting vehicles at on-ramp $b$. Summing both components, multiplying by the time interval ${\Delta}t$, and accumulating over time yields network TTS.

As we know, minimizing TTS improves network operating efficiency. CRM therefore seeks an action sequence satisfying the metering constraints in Eq. (1) that minimizes Eq. (3), written as $\operatorname{min}\mathrm{TTS}$.

However, detector installation costs limit measurement coverage. Suppose detectors are installed in only $N_s$ cells (the set $\mathcal D_d\subseteq\setopen 1,\text{...},N_x\setclose$, with $|\mathcal D_d|=N_s$), while probe vehicles provide intermittent speed samples at moving locations. Let $O_{t}$ denote the sparse observations available from both sources at time $t$. Because the current states of unobserved cells are unknown, the complete state cannot be propagated directly through Eq. (2) to evaluate the global TTS under candidate actions.

We therefore construct a short-horizon predictor $P_{{\theta}}$-ATPO conditioned on future metering actions. It uses $H_{p}$ historical steps of sparse observations $O_{t-H_{p},t}$ through time $t$ and a candidate action sequence $U_{f}$ over the next $H_{f}$ steps to output density and speed distributions at every future spatiotemporal point. Eq. (4) expresses this mapping compactly as a state trajectory.

\begin{equation}\label{eq:4}\begin{aligned}\hat{S}_{t+1:t+H_{f}}=P(O_{t-H_{p}+1,t},U_{f},Z_{t};{\theta})\end{aligned}\end{equation}

where ${\theta}$ comprises the learnable predictor parameters. Changing the future control sequence while keeping historical observations fixed allows the same model to compare predicted TTS under alternative actions.

\subsection{Overall Framework}\label{sec:3.2}

\begin{figure}[htbp]\centering
\includegraphics[width=0.95\textwidth,height=0.73\textheight,keepaspectratio]{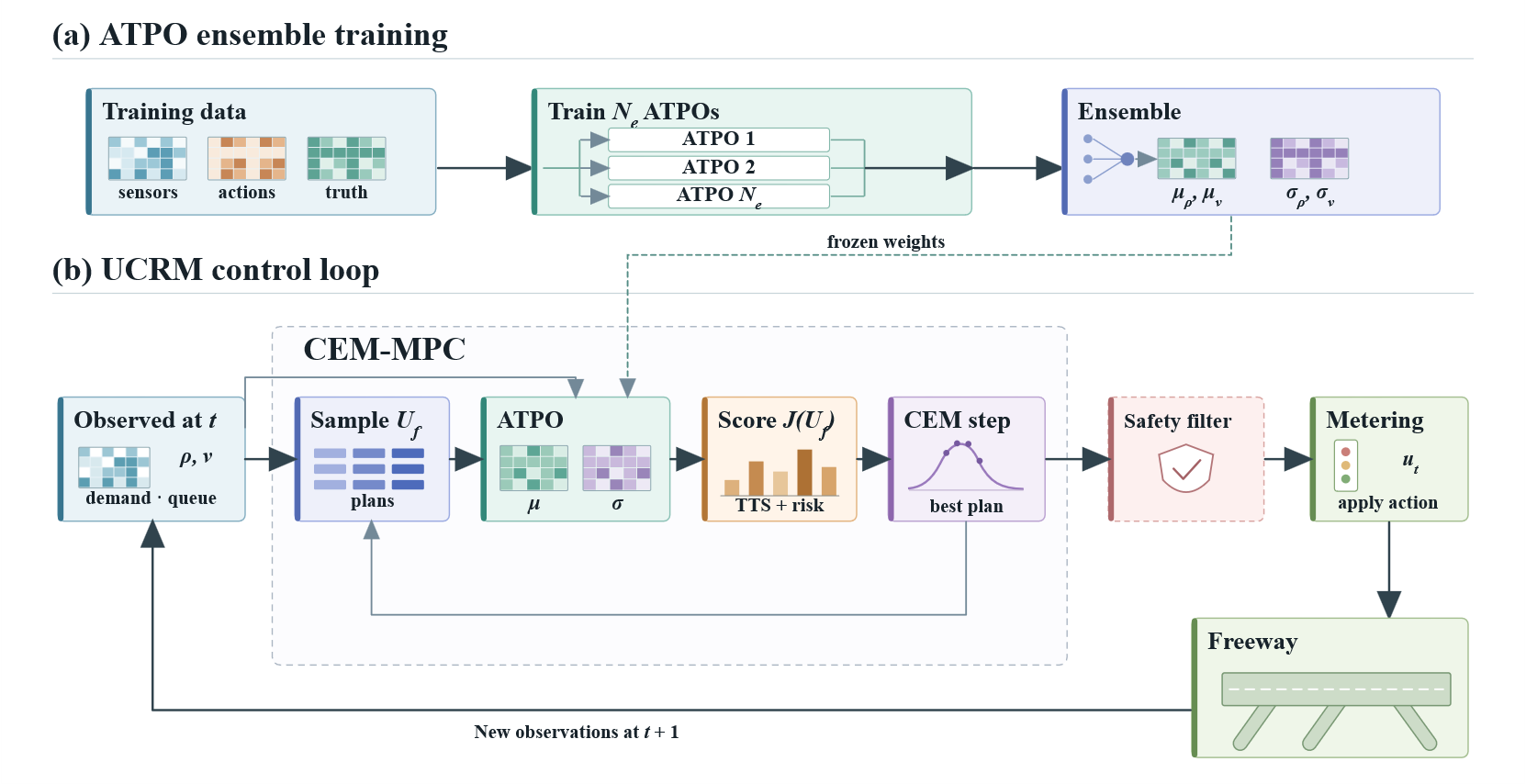}
\caption{\textbf{Overall architecture of ATPOUCRM.}}\label{fig:1}\end{figure}

As shown in \hyperref[fig:1]{\textcolor{myvioletblue}{\textbf{Figure 1}}}, the architecture of ATPOUCRM comprises the Action-Conditioned Traffic Prediction Operator (ATPO), $P_{{\theta}}$, and an Uncertainty-Aware Coordinated Multi-Ramp Metering controller (UCRM).

\hyperref[fig:1]{\textcolor{myvioletblue}{\textbf{Figure 1}}} connects two modules. ATPO provides future traffic distributions conditional on candidate actions, while UCRM uses those distributions to select and execute metering rates. At each control update, they form a closed loop of prediction, evaluation, and replanning.

ATPO integrates historical observations from fixed detectors and probe vehicles with upstream mainline demand, on-ramp queues, and candidate metering sequences. It outputs the mean and standard deviation of future density and speed throughout the corridor. The mean of density estimates mainline vehicle accumulation; a queue recurrence adds waiting vehicles at the ramps, yielding a TTS surrogate for each candidate action. Section 3.3 describes the network and training loss.

UCRM jointly optimizes actions at all controlled on-ramps. It measures efficiency by predicted TTS and also considers predicted uncertainty at merge areas, excessive density, and action smoothness when scoring candidates. After CEM selects a plan, a pre-execution safety filter checks the action to be implemented. Section 3.4 gives the optimization procedure.

The framework has offline training and online receding-horizon phases.

In the offline phase, simulation trajectories under different demands and metering plans provide training samples. Inputs retain sparse observations and candidate actions, and supervision uses the future global traffic states produced under those actions. Multiple ATPO members are trained independently, and their ensemble represents the predictive mean and disagreement at inference.

At each online control update, new observations refresh the historical input and ramp queues. CEM then generates and scores feasible metering plans. The selected plan passes through the safety filter, after which only its current control interval is executed. The procedure repeats when new measurements arrive.

\subsection{Action-Conditioned Traffic Prediction Operator Under Sparse Observations}\label{sec:3.3}

This section describes how ATPO maps sparse historical observations and candidate actions to a future global traffic state field. It first creates latent features on the future grid from sparse history, incorporates upstream mainline demand and on-ramp queues, conditions a causal FNO backbone on the candidate metering rates, and finally outputs conditional density and speed distributions.

\begin{figure}[htbp]\centering
\includegraphics[width=0.95\textwidth,height=0.73\textheight,keepaspectratio]{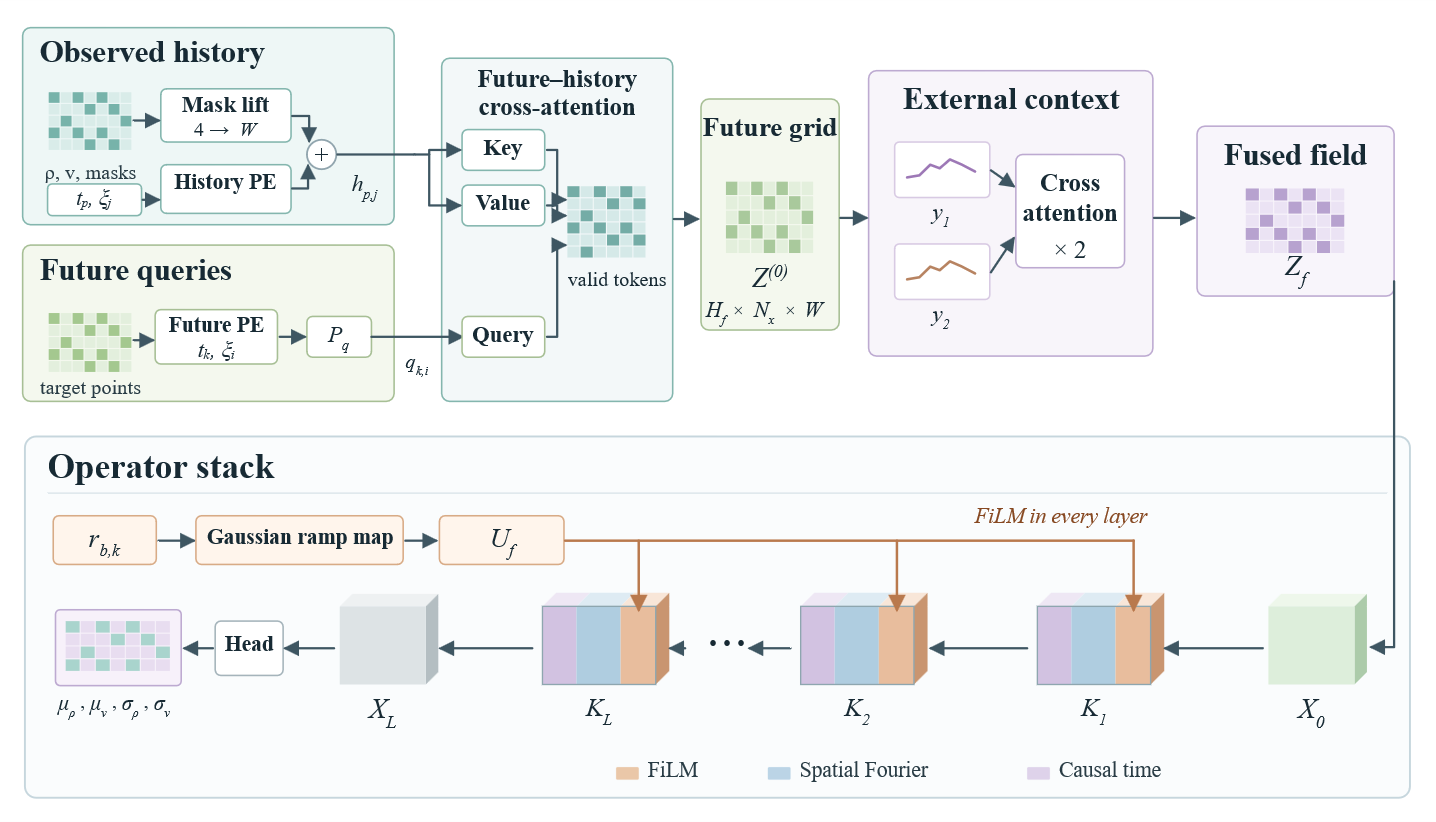}
\caption{\textbf{Architecture of the Action-Conditioned Traffic Prediction Operator.}}\label{fig:2}\end{figure}

The upper part of \hyperref[fig:2]{\textcolor{myvioletblue}{\textbf{Figure 2}}} constructs the fused latent field $Z_{f}$ from historical observations and future spatiotemporal coordinates, before candidate actions are introduced. The lower part takes $Z_{f}$ as the operator input $X_{0}$ and feeds the same candidate control field $U_{f}$ to the FiLM branch of every causal FNO block. Each block applies pointwise control modulation, spatial propagation along the road, and causal temporal propagation. A shared pointwise head then produces four predictive parameters.

\subsubsection{Sparse Historical Observations and Future Query Features}\label{sec:3.3.1}

\noindent\textbf{Historical observations and valid points}\par

Let the historical and forecast windows contain $H_{p}$ and $H_{f}$ time steps, respectively, with grids $H_{p}\times N_{x}$ and $H_{f}\times N_{x}$. Their physical durations are $T_{p}$ and $T_{f}$, and the corridor length is $L_{x}$. A future point $(k,i)$ lies at forecast step $k$ and cell $i$, with time $t_{k}\in [0,T_{f}]$ and cell-center position ${\xi}_{i}\in [0,L_{x}]$. A historical point $(p,j)$ lies at historical step $p$ and cell $j$, with time $t_{p}\in [-T_{p},0]$ and cell-center position ${\xi}_{j}\in [0,L_{x}]$. The forecast origin is the last row of the historical window; all times are defined relative to it. The quantities $H_{p}$, $H_{f}$, and $N_{x}$ specify grid resolution, whereas $T_{p}$, $T_{f}$, and $L_{x}$ specify the physical domain. Cross-resolution use therefore resamples the two grids without changing the physical domain.

At each historical grid point $(p,j)$, density ${\rho}_{p,j}$ and mean speed $v_{p,j}$ have separate binary observation masks $m_{p,j}^{{\rho}},m_{p,j}^{v}$. A fixed detector is assumed to provide both variables at its location, so both masks are one. A probe vehicle is assumed to report speed only at its position and reporting time; the speed mask is one and the density mask is zero. At a historical point observed by neither source, both values and both masks are zero. Thus, the four-channel network input and the set of valid historical points are defined as:

\begin{equation}\label{eq:5}\begin{aligned}s_{p,j}=[m_{p,j}^{{\rho}}{\rho}_{p,j},m_{p,j}^{v}v_{p,j},m_{p,j}^{{\rho}},m_{p,j}^{v}]\end{aligned}\end{equation}

\begin{equation}\label{eq:6}\begin{aligned}\mathcal H_{\mathrm{obs}}=\setopen(p,j):\;m_{p,j}^{\rho}+m_{p,j}^{v}>0\setclose\end{aligned}\end{equation}

In Eq. (6), $\mathcal H_{\mathrm{obs}}$ denotes the set of historical spatiotemporal points at which at least one of density or speed is observed. Because each mask is binary, $m_{p,j}^{\rho}+m_{p,j}^{v}>0$ means that at least one observation is available. Keeping both the zero-filled values and their observation indicators prevents the network from interpreting the missing density as a genuine zero-density state.

\noindent\textbf{Position encoding for historical and future points}\par

To distinguish when and where observations were made and forecasts are requested, fixed-frequency sinusoidal and cosine Fourier encodings are constructed for the continuous spatiotemporal coordinates of historical and future points (\textcolor{myvioletblue}{\textbf{\hyperlink{bib:45}{Vaswani et al., 2017}}}; \textcolor{myvioletblue}{\textbf{\hyperlink{bib:42}{Tancik et al., 2020}}}). Time and position are first normalized by $T_{f}$ and $L_{x}$ and then encoded at predetermined frequencies. Separate learnable projections map the historical and future grids to width $W$. Here, $W$ is the number of latent feature channels. Both sides use the same fixed frequencies, but separate projections distinguish observed positions from positions to be predicted.

\begin{equation}\label{eq:7}\begin{aligned}PE(t,{\xi})=[PE_{t}(\frac{t}{T_{f}});PE_{{\xi}}(\frac{{\xi}}{L_{x}})]\end{aligned}\end{equation}

\begin{equation}\label{eq:8}\begin{aligned}PE_{t}(a)=[sin(2{\pi}{\omega}_{r}a),cos(2{\pi}{\omega}_{r}a)]\end{aligned}\end{equation}

\begin{equation}\label{eq:9}\begin{aligned}PE_{{\xi}}(b)=[sin(2{\pi}{\nu}_{r}b),cos(2{\pi}{\nu}_{r}b)]\end{aligned}\end{equation}

In Eqs. (8) and (9), $R$ is the preset number of frequencies per coordinate axis, with $r=0,\text{...} ,R-1$. The quantities ${\omega}_{r}$ and ${\nu}_{r}$ are the fixed frequencies for time and space at index $r$. Concatenating both axes gives $PE(t,{\xi})\in \mathbb{R} ^{4R}$. As shown in \hyperref[fig:2]{\textcolor{myvioletblue}{\textbf{Figure 2}}}, the Fourier positional encodings of historical grid point $(p,j)$ and future grid point $(k,i)$ are referred to as History PE and Future PE, respectively.

\noindent\textbf{Historical feature lifting and future query projection}\par

For historical observations, the Mask lift layer in \hyperref[fig:2]{\textcolor{myvioletblue}{\textbf{Figure 2}}} applies the same pointwise linear map to every historical grid point. It converts the four-channel traffic values and validity indicators in Eq. (5) to a feature of width $W$. A projection of the historical time and road position is then added to form the history feature $h_{p,j}\in \mathbb{R} ^{W}$:

\begin{equation}\label{eq:10}\begin{aligned}h_{p,j}=W_{h}s_{p,j}+b_{h}+P_{h}PE(t_{p},{\xi}_{j})\end{aligned}\end{equation}

In Eq. (10), $W_{h}\in \mathbb{R} ^{W\times 4}$ and $b_{h}\in \mathbb{R} ^{W}$ are the weight matrix and bias of the Mask lift layer, and $P_{h}\in \mathbb{R} ^{W\times 4R}$ projects the historical coordinate encoding.

For each future point $(k,i)$, Future PE is projected by $P_{q}$ to a future positional feature $q_{k,i}\in \mathbb{R} ^{W}$ of width $W$:

\begin{equation}\label{eq:11}\begin{aligned}q_{k,i}=P_{q}PE(t_{k},{\xi}_{i})\end{aligned}\end{equation}

In Eq. (11), $P_{q}\in \mathbb{R} ^{W\times 4R}$ is the projection matrix for future coordinate encodings.

\subsubsection{Space--Time Cross-Attention from Future Points to Historical Observations}\label{sec:3.3.2}

\noindent\textbf{Historical Key/Value and future Query projections}\par

The history feature $h_{p,j}$ at each valid historical point is then projected separately into a Key and a Value:

\begin{equation}\label{eq:12}\begin{aligned}K_{p,j}=W_{K}h_{p,j}\end{aligned}\end{equation}

\begin{equation}\label{eq:13}\begin{aligned}V_{p,j}=W_{V}h_{p,j}\end{aligned}\end{equation}

In Eqs. (12) and (13), $W_{K}\in \mathbb{R} ^{d\times W}$ and $W_{V}\in \mathbb{R} ^{W\times W}$ are independently learned projection matrices, and $d$ is the channel dimension used for query-key dot products. The Key $K_{p,j}\in \mathbb{R} ^{d}$ identifies the relevance of a historical point to a future query, while the Value $V_{p,j}\in \mathbb{R} ^{W}$ carries the traffic information aggregated into that future point.

A historical point $(p,j)$ enters the valid-token set $\mathcal H_{\mathrm{obs}}$ only when $m_{p,j}^{\rho}+m_{p,j}^{v}>0$; points missing both channels are masked out of attention. Even if a cell has no local measurement throughout the historical window, its future query can still attend to valid historical points at other cells.

The future positional feature $q_{k,i}$ at each target point $(k,i)$ is mapped by $W_{Q}$ to a Query vector $Q_{k,i}$ of dimension $d$:

\begin{equation}\label{eq:14}\begin{aligned}Q_{k,i}=W_{Q}q_{k,i}\end{aligned}\end{equation}

In Eq. (14), $W_{Q}\in \mathbb{R} ^{d\times W}$ is an independently learned projection matrix. The future Query $Q_{k,i}\in \mathbb{R} ^{d}$ specifies when and where a prediction is required; it does not contain the true future traffic state at that point.

\noindent\textbf{Attention weights and historical context}\par

Each future Query computes a scaled dot product with every valid historical point in the same sample. A distance bias measured along the corridor favors nearby observations while retaining access to distant historical information:

\begin{equation}\label{eq:15}\begin{aligned}a_{k,i,p,j}=\frac{Q_{k,i}^{T}K_{p,j}}{\sqrt{d}}-(\frac{{\xi}_{i}-{\xi}_{j}}{\ell _{i}})^{2}\end{aligned}\end{equation}

In Eq. (15), $\ell _{i}>0$ is a tunable distance scale in units of road length. The quantity $a_{k,i,p,j}$ is the unnormalized attention score from future point $(k,i)$ to historical point $(p,j)$; the spatial-bias term $(\frac{{\xi}_{i}-{\xi}_{j}}{\ell _{i}})^{2}$ lowers scores for more distant points. The future point $(k,i)$ queries all valid historical points to form its historical context $C_{k,i}$:

\begin{equation}\label{eq:16}\begin{aligned}{\alpha}_{k,i,p,j}=\frac{\exp(a_{k,i,p,j})}{\sum _{(p,j)\in \mathcal H_{\mathrm{obs}}}\exp(a_{k,i,p,j})}\end{aligned}\end{equation}

\begin{equation}\label{eq:17}\begin{aligned}C_{k,i}=\sum _{(p,j)\in \mathcal H_{\mathrm{obs}}}{\alpha}_{k,i,p,j}V_{p,j}\end{aligned}\end{equation}

In Eqs. (16) and (17), ${\alpha}_{k,i,p,j}$ is a scalar weight normalized over the valid set $\mathcal H_{\mathrm{obs}}$, and $C\in \mathbb{R} ^{H_{f}\times N_{x}\times W}$ is the context tensor over the complete future spatiotemporal domain. The global historical query can retrieve observations from other road cells but never uses true states after the forecast origin.

\subsubsection{Future Latent-Field Initialization and Fusion of Two Exogenous Sources}\label{sec:3.3.3}

The future latent field is initialized by combining the historical context $C_{k,i}$ and positional query $q_{k,i}$ at each target point and applying a shared linear map to obtain $Z^{(0)}_{k,i}$:

\begin{equation}\label{eq:18}\begin{aligned}Z^{(0)}_{k,i}=W_{Z}(C_{k,i}+q_{k,i})+b_{Z}\end{aligned}\end{equation}

In Eq. (18), $W_{Z}\in \mathbb{R} ^{W\times W}$ and $b_{Z}\in \mathbb{R} ^{W}$ are the learnable weight matrix and bias of the initialization layer, and $Z^{(0)}_{k,i}\in \mathbb{R} ^{W}$ is the initial latent feature at future point $(k,i)$. Computing $Z^{(0)}_{k,i}$ at each point produces the initial field $Z^{(0)}\in \mathbb{R} ^{H_{f}\times N_{x}\times W}$ over all $H_{f}\times N_{x}$ future spatiotemporal points.

GTSP also depends on boundary traffic and vehicles waiting at on-ramps. The predictor therefore receives two exogenous sequences known at the forecast origin: $y_{1}$ is the history of upstream mainline demand, representing traffic entering the corridor, and $y_{2}$ is the history of on-ramp queues, representing vehicles awaiting release. Separate Key-Value sequences encode these sources. Features at each future grid point query and fuse them so the network can represent their distinct influence on future traffic (\textcolor{myvioletblue}{\textbf{\hyperlink{bib:45}{Vaswani et al., 2017}}}).

At a future point $(k,i)$, $Z^{(0)}_{k,i}$ serves as the Query input to both exogenous cross-attention branches. The sequences $y_{1}$ and $y_{2}$ are encoded as $Y_{1}\in \mathbb{R} ^{n_{1}\times W}$ and $Y_{2}\in \mathbb{R} ^{n_{2}\times W}$, with $n_{1}$ and $n_{2}$ valid demand and queue tokens, respectively. Each branch forms Keys and Values from its own sequence and applies standard multi-head cross-attention (\textcolor{myvioletblue}{\textbf{\hyperlink{bib:45}{Vaswani et al., 2017}}}), producing feature vectors $E_{1,k,i},E_{2,k,i}\in \mathbb{R} ^{W}$. Two learned scalars, $a_{1}$ and $a_{2}$, determine global fusion weights ${\pi}_{1}$ and ${\pi}_{2}$. Pointwise weighted fusion followed by $LayerNorm$ gives the fused latent feature $Z_{f,k,i}$ at $(k,i)$:

\begin{equation}\label{eq:19}\begin{aligned}{\pi}_{s}=\frac{\exp(a_{s})}{\exp(a_{1})+\exp(a_{2})}\end{aligned}\end{equation}

\begin{equation}\label{eq:20}\begin{aligned}Z_{f,k,i}=LayerNorm({\pi}_{1}E_{1,k,i}+{\pi}_{2}E_{2,k,i})\end{aligned}\end{equation}

Here and in Eq. (19), $s\in\setopen 1,2\setclose$ identifies the demand and queue branches. The pointwise fused features $Z_{f,k,i}\in \mathbb{R} ^{W}$ in Eq. (20) are arranged on the future grid, with channels moved to the second axis, yielding the fused latent field $Z_{f}\in \mathbb{R} ^{W\times H_{f}\times N_{x}}$.

\subsubsection{Control-Conditioned Spatial Fourier and Causal Temporal Operator}\label{sec:3.3.4}

\noindent\textbf{Spatial mapping of ramp metering actions}\par

The future latent field from the preceding section does not yet contain a candidate action. To obtain distinct forecasts under different metering plans from the same history, the metering rate at each future time is mapped to a control field whose temporal and spatial coordinates align with the fused latent field.

A ramp metering rate primarily affects its merge area. The spatial weight assigned to future cell $i$ for on-ramp $b$ is therefore:

\begin{equation}\label{eq:21}\begin{aligned}g_{b,i}=\exp\!\left(-\frac{({\xi}_{i}-{\xi}_{b})^{2}}{2{\sigma}_{b}^{2}}\right)\end{aligned}\end{equation}

In Eq. (21), ${\xi}_{b}$ is the position of the merge area center for ramp $b$, ${\sigma}_{b}>0$ sets the spatial width, and $g_{b,i}$ is a dimensionless weight. The Gaussian kernel reaches one at the merge center and decreases with road distance. It encodes the spatial location of the ramp action; its weights are not normalized to sum to one across road cells. The actions at all ramps are superposed to form the control feature $u_{k,i}$ at future grid point $(k,i)$:

\begin{equation}\label{eq:22}\begin{aligned}u_{k,i}=\sum _{b=1}^{N_{r}}r_{b,k}g_{b,i}\end{aligned}\end{equation}

In Eq. (22), $r_{b,k}$ is the metering rate (veh/h) at ramp $b$ in forecast step $k$. Thus, a higher metering rate and a position closer to the merge area both increase that ramp’s contribution to the local control feature. The field encodes control influence rather than allocating the ramp flow among road cells. Arranging all $u_{k,i}$ on the future grid gives the control field $U_{f}$, with $U_{f}\in \mathbb{R} ^{H_{f}\times N_{x}}$. To change grid resolution, the Gaussian weights $g_{b,i}$ are recalculated at the new cell centers.

\noindent\textbf{Stacked control-conditioned causal FNO blocks}\par

The fused field $Z_{f}$ enters the first control-conditioned causal FNO block. Set $X_{0}=Z_{f}$. Block $\mathcal{K} _{l}$ at layer $l$ receives $X_{l-1}$ and the future control field $U_{f}$ and outputs $X_{l}$. There are $L$ blocks with independent parameters, and the last produces $X_{L}\in \mathbb{R} ^{W\times H_{f}\times N_{x}}$:

\begin{equation}\label{eq:23}\begin{aligned}X_{l}=\mathcal{K} _{l}(X_{l-1};U_{f}),l=1,\text{...} ,L\end{aligned}\end{equation}

\begin{figure}[htbp]\centering
\includegraphics[width=0.95\textwidth,height=0.73\textheight,keepaspectratio]{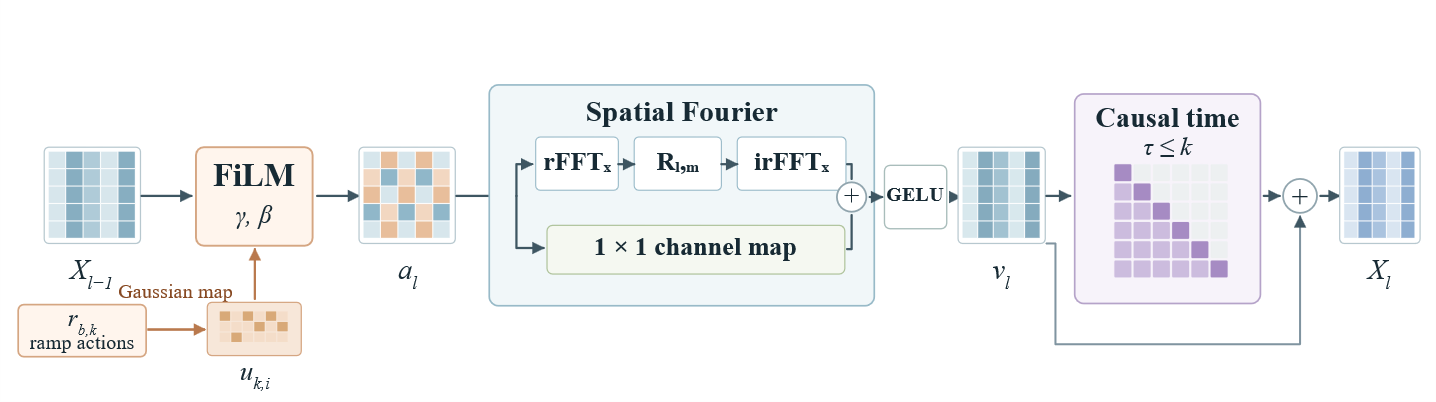}
\caption{\textbf{Control-conditioned causal FNO block.}}\label{fig:3}\end{figure}

\hyperref[fig:3]{\textcolor{myvioletblue}{\textbf{Figure 3}}} details a block $\mathcal{K} _{l}$. FiLM (\textcolor{myvioletblue}{\textbf{\hyperlink{bib:36}{Perez et al., 2018}}}) first inserts the candidate metering rate into the latent feature at the corresponding future spatiotemporal point. The spatial Fourier operator then propagates information along the road at that forecast time. Finally, a causal temporal kernel carries effects of earlier actions into subsequent forecast steps. This order represents spatial dependence while preserving the temporal ordering of control and response.

\noindent\textbf{Pointwise FiLM modulation}\par

The value $u_{k,i}$ of the future control field $U_{f}$ at point $(k,i)$ is passed through a parameter generator $G_{{\theta}}$ to produce ${\gamma}_{k,i}$ and ${\beta}_{k,i}$:

\begin{equation}\label{eq:24}\begin{aligned}({\gamma}_{k,i},{\beta}_{k,i})=G_{{\theta}}(u_{k,i})=W_{u,2}GELU(W_{u,1}u_{k,i}+b_{u,1})+b_{u,2}\end{aligned}\end{equation}

In Eq. (24), $W_{u,1}$ and $W_{u,2}$ are projection weight matrices, while $b_{u,1}$ and $b_{u,2}$ are the corresponding biases; GELU is an elementwise activation. For hidden channel $c$, FiLM applies the local control-derived scale ${\gamma}_{k,i}$ and shift ${\beta}_{k,i}$ to the layer-$l$ input $X_{l-1,c,k,i}$, giving the modulated feature $a_{l,c,k,i}$. The modulation parameters may be reused across layers, whereas spectral matrices, pointwise channel matrices, and temporal channel matrices are learned separately for each layer:

\begin{equation}\label{eq:25}\begin{aligned}a_{l,c,k,i}=(1+{\gamma}_{k,i})X_{l-1,c,k,i}+{\beta}_{k,i},c=1,\text{...} ,W\end{aligned}\end{equation}

The main effect of ${\gamma}_{,k,i}$ and ${\beta}_{k,i}$ is to scale and shift the original latent feature. FiLM learns how a given metering rate changes traffic features; it does not impose in advance whether a higher rate must increase or decrease a particular predicted variable. Modulation initially affects only the current future spatiotemporal point. The subsequent spatial and causal temporal operators determine whether that action’s influence spreads to other road cells and later forecast times. This is how control actions enter state prediction.

\noindent\textbf{Spatial Fourier propagation}\par

Fix prediction step $k$, and let $a_{l,k}\in \mathbb{R} ^{W\times N_{x}}$ denote the modulated features arranged along the road at layer $l$. The spatial operator transforms only the road axis, using the same spectral parameters at every forecast time; it does not mix different forecast steps at this stage. The spatial Fourier operation is (\textcolor{myvioletblue}{\textbf{\hyperlink{bib:19}{Li et al., 2021}}}):

\begin{equation}\label{eq:26}\begin{aligned}\hat{B}_{l,k,m}=R_{l,m}[\mathcal{F} _{x}(a_{l,k})]_{m},0\leq m<M_{x}\end{aligned}\end{equation}

\begin{equation}\label{eq:27}\begin{aligned}\widetilde{X}_{l,k}=GELU(\mathcal{F} _{x}^{-1}(\hat{B}_{l,k};N_{x})+W_{l,p}a_{l,k}+b_{l,p})\end{aligned}\end{equation}

In Eqs. (26) and (27), $\mathcal{F} _{x}$ and $\mathcal{F} _{x}^{-1}$ are the real-valued Fourier transform and inverse transform along the road axis. The matrix $R_{l,m}$ contains the learned spectral weights at spatial frequency $m$ in layer $l$. The number of nonnegative modes included in the spectral calculation is $M_{x}$, the smaller of the configured upper limit $K_{x}$ and the number available on the current grid $\lfloor N_{x}/2\rfloor +1$. The resulting spatially propagated feature at that time is $\widetilde{X}_{l,k}\in \mathbb{R} ^{W\times N_{x}}$. Finally, $W_{l,p}\in \mathbb{R} ^{W\times W}$ and $b_{l,p}\in \mathbb{R} ^{W}$ are the weight matrix and bias of the pointwise channel map.

More specifically, Eq. (26) transforms latent features along the road at each prediction time and uses the spectral weights $R_{l,m}$ to mix the $W$ channels of each retained mode, learning cross-cell dependence at different spatial scales. Eq. (27) maps the modified spectrum back to the road grid, yielding features that carry information from other cells. In parallel, the pointwise channel map transforms only the feature at each cell without reading other cells. The two outputs are added and passed through GELU, producing spatially propagated features $\widetilde{X}_{l,k}$ aligned with the original cells.

\noindent\textbf{Causal temporal propagation}\par

After spatial propagation, let $\widetilde{X}_{l,k,i}\in \mathbb{R} ^{W}$ be the feature at cell $i$ and forecast step $k$. Inspired by \textcolor{myvioletblue}{\textbf{\hyperlink{bib:-3}{Bai et al. (2018)}}} and \textcolor{myvioletblue}{\textbf{\hyperlink{bib:2}{J. Chen (2026)}}}, the causal temporal kernel in layer $l$ computes output $X_{l,k,i}$ using only features from the current and preceding steps ${\tau}_{c}\leq k$:

\begin{equation}\label{eq:28}\begin{aligned}{\lambda}_{l}=softplus({\eta}_{l})\end{aligned}\end{equation}

\begin{equation}\label{eq:29}\begin{aligned}{\kappa}_{l,k,{\tau}}=\exp\!\left(-{\lambda}_{l}(t_{k}-t_{{\tau}})^{2}\right)\end{aligned}\end{equation}

\begin{equation}\label{eq:30}\begin{aligned}w_{l,k,{\tau}}=\frac{{\kappa}_{l,k,{\tau}}}{\sum _{s=0}^{k}{\kappa}_{l,k,s}},0\leq {\tau}_{c}\leq k\end{aligned}\end{equation}

\begin{equation}\label{eq:31}\begin{aligned}w_{l,k,{\tau}}=0,{\tau}_{c}>k\end{aligned}\end{equation}

\begin{equation}\label{eq:32}\begin{aligned}X_{l,k,i}=\widetilde{X}_{l,k,i}+\sum _{{\tau}=0}^{k}w_{l,k,{\tau}}W_{l,t}\widetilde{X}_{l,{\tau},i}\end{aligned}\end{equation}

In Eqs. (28)-(32), ${\eta}_{l}$ is an unconstrained scalar learned separately for layer $l$; applying the softplus function gives the positive decay coefficient ${\lambda}_{l}$. The unnormalized weight ${\kappa}_{l,k,{\tau}}$ depends on the physical time difference. The weight $w_{l,k,{\tau}}$ is normalized only over steps satisfying ${\tau}_{c}\leq k$, and is zero for future steps ${\tau}_{c}>k$. The matrix $W_{l,t}\in \mathbb{R} ^{W\times W}$ is the learned temporal channel map in layer $l$. The vector $X_{l,k,i}\in \mathbb{R} ^{W}$ is the output of the causal temporal block; arranging these vectors on the future grid gives the layer output $X_{l}$.

Eq. (28) lets each layer learn how rapidly the effect of earlier actions decays: larger ${\lambda}_{l}$ gives less weight to more distant past times. Eq. (29) scores previous steps by their physical time difference from the current step, and Eq. (30) normalizes these weights. Eq. (31) assigns zero weight to all future steps, so later conditions cannot influence the current prediction. Eq. (32) then averages spatially propagated current and past features and adds a residual connection that retains the current feature. Thus, an action first enters the latent field through FiLM, can propagate along the road, and subsequently influences later forecasts through the temporal kernel. When grid resolution changes over the same physical road and forecast duration, future coordinates and the control field are resampled, FiLM still acts pointwise, spectral matrices still operate on retained modes, and the temporal kernel continues to use physical time differences. The structure can therefore produce outputs with different $H_{f}$ and $N_{x}$.

\subsubsection{Pointwise Probabilistic Output}\label{sec:3.3.5}

The last FNO output $X_{L}$ passes through a shared pointwise head comprising a 1 × 1 channel map and GELU. It converts the $W$-dimensional feature at each future point $(k,i)$ to four raw parameters $o=(o_1,o_2,o_3,o_4)$:

\begin{equation}\label{eq:33}\begin{aligned}o_{k,i}=W_{2}GELU(W_{1}X_{L,k,i}+b_{1})+b_{2}\end{aligned}\end{equation}

In Eq. (33), $W_{1}$ and $W_{2}$ are learnable weight matrices, and $b_{1}$ and $b_{2}$ are their biases. The first two of the four raw parameters $o$ give the density and speed means; the remaining two pass through softplus and a fixed positive floor ${\epsilon}>0$ to produce their standard deviations:

\begin{equation}\label{eq:34}\begin{aligned}{\mu}_{{\rho},k,i}=o_{k,i,1},{\mu}_{v,k,i}=o_{k,i,2}\end{aligned}\end{equation}

\begin{equation}\label{eq:35}\begin{aligned}{\sigma}_{{\rho},k,i}=softplus(o_{k,i,3})+{\epsilon},{\sigma}_{v,k,i}=softplus(o_{k,i,4})+{\epsilon}\end{aligned}\end{equation}

The four output channels at the lower left of \hyperref[fig:2]{\textcolor{myvioletblue}{\textbf{Figure 2}}} represent the means and standard deviations of density and speed. The means describe predicted traffic evolution; the standard deviations quantify uncertainty at each cell and forecast time. A candidate action can therefore affect both the predicted traffic state and its uncertainty.

\subsubsection{Physics-Informed Training}\label{sec:3.3.6}

Training uses future traffic state ground truth and traffic flow priors. A heteroscedastic negative log-likelihood fits available labels. Traffic flow model residual terms check density conservation and speed evolution, and a global conservation term checks changes in the total vehicle inventory on the mainline and at on-ramp queues. These objectives are trained jointly, but the physical priors are soft constraints, they do not require the network to reproduce METANET exactly.

\noindent\textbf{Gaussian likelihood for observed traffic states}\par

Let $\mathcal{D} _{s}$ denote the future spatiotemporal points with ground truth labels. A heteroscedastic Gaussian negative log-likelihood is calculated separately for density and speed:

\begin{equation}\label{eq:36}\begin{aligned}
L_{\mathrm{NLL}}=\sum_{(k,i)\in\mathcal D_s}\sum_{\phi=\rho,\,v}
\left(\frac{(y_{\phi,k,i}-\mu_{\phi,k,i})^2}{2\sigma_{\phi,k,i}^2}
+\log\sigma_{\phi,k,i}\right).
\end{aligned}\end{equation}

In Eq. (36), $\sum_{\phi=\rho,\,v}$ adds the density and speed loss terms at each future point $(k,i)$. The squared-error term draws each predicted mean toward its ground truth, while the standard deviation allows a wider predictive distribution at difficult points. The $\log\sigma_{\phi,k,i}$ term prevents the model from reducing loss simply by making that standard deviation arbitrarily large. The loss is evaluated only at future points with supervisory labels.

\noindent\textbf{Local traffic dynamics residual}\par

The traffic flow model residual uses METANET’s density-conservation and speed-update equations. From forecast step $k$ to the next step, density in cell $i$ changes with the net flow across its upstream and downstream boundaries and with merging or diverging ramp flows. Speed additionally depends on desired speed, the upstream speed difference, and the downstream density gradient.

\begin{equation}\label{eq:37}\begin{aligned}{\rho}_{k+1,i}={\rho}_{k,i}+\frac{{\Delta}t}{n_{i}{\Delta}x}[Q^{\mathrm{m}}_{k,i-1}-Q^{\mathrm{m}}_{k,i}+Q_{k,i}^{e}]\end{aligned}\end{equation}

\begin{equation}\label{eq:38}\begin{aligned}v_{k+1,i}=v_{k,i}+\frac{{\Delta}t}{{\tau}}(V({\rho}_{k,i})-v_{k,i})+\frac{{\Delta}t}{{\Delta}x}v_{k,i}(v_{k,i-1}-v_{k,i})-\frac{{\eta}{\Delta}t}{{\tau}{\Delta}x}\frac{({\rho}_{k,i+1}-{\rho}_{k,i})}{{\rho}_{k,i}+{\kappa}}\end{aligned}\end{equation}

In Eq. (37), ${\Delta}x$ is the length of the equal-sized cells. The mainline flow $Q^{\mathrm{m}}_{k,i}$ can be approximated by ${\rho}_{k,i}v_{k,i}n_{i}$, with per-lane density ${\rho}_{k,i}$ measured in veh/km/lane. The net ramp flow $Q_{k,i}^{e}$ entering the cell is positive for on-ramp merging and negative for off-ramp diverging. At controlled on-ramp $b$, it equals the actual metering rate $r_{b,k}$. In Eq. (38), ${\tau}$ is the speed relaxation time, ${\eta}$ controls anticipation of downstream density, and ${\kappa}>0$ prevents a small denominator at low density. The function $V({\rho}_{k,i})$ gives equilibrium speed.

During training, predicted means are substituted into Eqs. (37) and (38) to assess whether outputs at successive forecast steps follow the traffic-flow model. Eqs. (39) and (40) define the density and speed residuals, respectively:

\begin{equation}\label{eq:39}\begin{aligned}R_{k+1,i}^{{\rho}}={\mu}_{k+1,i}^{{\rho}}-{\mu}_{k,i}^{{\rho}}-\frac{{\Delta}t}{n_{i}{\Delta}x}[{\mu}_{k,i-1}^{{\rho}}{\mu}_{k,i-1}^{v}n_{i-1}-{\mu}_{k,i}^{{\rho}}{\mu}_{k,i}^{v}n_{i}+Q_{k,i}^{e}]\end{aligned}\end{equation}

\begin{equation}\label{eq:40}\begin{aligned}R_{k+1,i}^{v}={\mu}_{k+1,i}^{v}-{\mu}_{k,i}^{v}-\frac{{\Delta}t}{{\tau}}(V({\mu}_{k,i}^{{\rho}})-{\mu}_{k,i}^{v})-\frac{{\Delta}t}{{\Delta}x}{\mu}_{k,i}^{v}({\mu}_{k,i-1}^{v}-{\mu}_{k,i}^{v})+\frac{{\eta}{\Delta}t}{{\tau}{\Delta}x}\frac{({\mu}_{k,i+1}^{{\rho}}-{\mu}_{k,i}^{{\rho}})}{{\mu}_{k,i}^{{\rho}}+{\kappa}}\end{aligned}\end{equation}

At a merging point, the physical residual uses the controlled on-ramp discharge $r_{b,k}$ rather than replacing actual inflow with the Gaussian control field $u_{k,i}$. Cells with unknown off-ramp diverging flows and updates lacking neighboring states or boundary values are excluded from the residual set through a physical mask (\textcolor{myvioletblue}{\textbf{\hyperlink{bib:61}{Yu et al., 2025}}}). Eq. (41) gives the traffic flow model residual based loss:

\begin{equation}\label{eq:41}\begin{aligned}L_{P}=\sum _{(k,i)\in \mathcal{U} }\bigl((R_{k+1,i}^{{\rho}})^{2}+(R_{k+1,i}^{v})^{2}\bigr)\end{aligned}\end{equation}

Here, $\mathcal{U}$ denotes the set of spatiotemporal grid points $(k,i)$ without detector observations.

This loss encourages vehicle conservation and plausible speed dynamics in regions with sparse supervision without forcing the network output to satisfy METANET exactly.

\noindent\textbf{Subsegment vehicle conservation}\par

For global vehicle conservation, off-ramp locations divide the corridor into subsegments that contain no internal off-ramp divergence. Let $e=1,\ldots,E$ index the subsegments, where $E$ is their number. The vehicle inventory of each subsegment is:

\begin{equation}\label{eq:42}\begin{aligned}N_{e,k}=\sum _{i=1}^{N_{e,x}}n_{e,i}{\Delta}x{\rho}_{e,k,i}+\sum _{b=1}^{N_{e,b}}w_{e,b,k}\end{aligned}\end{equation}

In Eq. (42), $N_{e,k}$ is the number of vehicles in a subsegment at time $k$, $N_{e,x}$ is its number of mainline cells, $w_{e,b,k}$ is the number of vehicles waiting at on-ramp $b$, and $N_{e,b}$ is the number of on-ramps in the subsegment.

By vehicle conservation, the change in subsegment inventory over time equals its net boundary inflow:

\begin{equation}\label{eq:43}\begin{aligned}N_{e,k+1}-N_{e,k}={\Delta}t{\Delta}Q_{e,k}\end{aligned}\end{equation}

\begin{equation}\label{eq:44}\begin{aligned}{\Delta}Q_{e,k}=Q_{e,k}^{up}+\sum _{b=1}^{N_{r}}d_{e,b,k}-Q_{e,k}^{down}\end{aligned}\end{equation}

In Eqs. (43) and (44), $Q_{e,k}^{up}$ and $Q_{e,k}^{down}$ are the upstream inflow and downstream outflow of the subsegment (veh/h), and $d_{e,b,k}$ is external demand arriving at on-ramp $b$. Accordingly, the change in inventory between successive times equals upstream mainline inflow plus external on-ramp arrivals, minus downstream mainline outflow. The metering rate does not enter the net boundary flow because it only transfers vehicles from the queue to the mainline within the same subsegment.

Replacing true density with its predicted mean of density and actual ramp queues with queues propagated under the candidate plan yields the predicted subsegment inventory:

\begin{equation}\label{eq:45}\begin{aligned}\hat{N}_{e,k}=\sum _{i=1}^{N_{e,x}}n_{e,i}{\Delta}x{\mu}_{e,k,i}^{{\rho}}+\sum _{b=1}^{N_{e,b}}w_{e,b,k}\end{aligned}\end{equation}

The conservation residual between successive future times is calculated from the predicted inventory and summed over $H_{f}-1$ valid intervals, giving the global vehicle conservation loss:

\begin{equation}\label{eq:46}\begin{aligned}L_{C}=\sum _{e=1}^{E}\sum _{k=0}^{H_{f}-2}(\hat{N}_{e,k+1}-\hat{N}_{e,k}-{\Delta}t{\Delta}Q_{e,k})^{2}\end{aligned}\end{equation}

Eq. (46) compares the predicted inventory change with net boundary flow for each subsegment and pair of adjacent forecast times. A larger discrepancy indicates poorer agreement with system-level vehicle conservation.

The final training objective is  $L_{t}=L_{NLL}+{\lambda}_{P}L_{P}+{\lambda}_{C}L_{C}$. The nonnegative weights ${\lambda}_{P},{\lambda}_{C}\geq 0$ control the traffic flow model residual and global conservation terms.

\subsubsection{Ensemble Prediction and the TTS Surrogate}\label{sec:3.3.7}

A bootstrap deep ensemble (\textcolor{myvioletblue}{\textbf{\hyperlink{bib:17}{Lakshminarayanan et al., 2017}}}) separates dispersion within a member’s conditional prediction from disagreement across independently trained members. Members share the architecture but differ in initialization and bootstrap-resampled training data. Each member $m$ outputs a mean ${\mu}_{m}$ and standard deviation ${\sigma}_{m}$ for every future density and speed point.

At inference, the Gaussian outputs of $N_{e}$ members are mixed with equal weights:

\begin{equation}\label{eq:47}\begin{aligned}p=\frac{1}{N_{e}}\sum _{m=1}^{N_{e}}\mathcal{N}({\mu}_{m},{\sigma}_{m}^{2})\end{aligned}\end{equation}

Eq. (47) defines the equally weighted Gaussian mixture. Its predictive mean is $\overset{-}{{\mu}}=\frac{\sum _{m=1}^{N_{e}}{\mu}_{m}}{N_{e}}$, and its total variance is $\frac{1}{N_{e}}\sum _{m=1}^{N_{e}}({\sigma}_{m}^{2}+{\mu}_{m}^{2})-\overset{-}{{\mu}}^{2}$. The mean within-member variance represents aleatoric uncertainty, while dispersion of the member means represents epistemic uncertainty.

When scoring candidate ramp actions, the observed queue $w_{b,0}$ initializes the queue recurrence. Arrivals during the current forecast window are set to the current measured demand $d_{b,0}$. This constant demand assumption is used only within a short receding-horizon optimization: the next control update reinitializes the forecast from new observations and does not use the simulator's future demand.

\begin{equation}\label{eq:48}\begin{aligned}d_{b,k}=d_{b,0}\end{aligned}\end{equation}

\begin{equation}\label{eq:49}\begin{aligned}w_{b,k+1}=w_{b,k}+{\Delta}t(d_{b,k}-r_{b,k})\end{aligned}\end{equation}

At the beginning of forecast step $k$, releasable vehicles consist of new arrivals plus the existing queue. Thus $0\leq r_{b,k}\leq \min\left(r_{\max,b},d_{b,k}+\frac{w_{b,k}}{\Delta t}\right)$, where $r_{max,b}$ is the maximum discharge rate at ramp $b$. This constraint restricts candidate metering rates to physically available flow.

For each candidate plan, the ensemble average density is converted to mainline inventory as in Eq. (42), and the ramp queues from Eq. (49) are added to give the predicted system inventory $\hat{N}_{k}$ at future step $k$. Integrating this inventory across the forecast window defines the TTS surrogate:

\begin{equation}\label{eq:50}\begin{aligned}E(\hat{TTS}(U_{f}))=\sum _{k=0}^{H_{f}-1}\hat{N}_{k}{\Delta}t\end{aligned}\end{equation}

ATPO thus supplies not only a GTSP method but also comparable TTS estimates over the forecast horizon for alternative ramp metering actions under the same current observations.

\subsection{ATPOUCRM Control}\label{sec:3.4}

This section describes how ATPO is used in ATPOUCRM. It first defines an optimization objective balancing efficiency and risk, then explains the CEM search for multi-ramp plans, and finally describes how the pre-execution safety filter may modify the action sent to the ramps. The predictor estimates traffic consequences, the optimizer chooses a candidate plan, and the safety filter governs execution.

\subsubsection{Optimization Objective}\label{sec:3.4.1}

At control time $t$, a candidate CRM metering rate plan $U_{f}=(u_{0},\text{...} ,u_{H_{f}-1}),u_{k}=(r_{1,k},\text{...} ,r_{N_{r},k})$ is optimized over the next $H_{f}$ prediction steps. After obtaining the best plan, only the rates for the current control interval $H_{c}$, where $H_{c}<H_{f}$, are implemented. Optimization is repeated after $H_{c}$ steps when new observations arrive. Here, $H_{c}$ is the number of forecast steps between control updates.

The objective $J(U_{f})$ uses predicted TTS from Eq. (50) as its main term. To measure relative dispersion in predicted merge area density, the pointwise coefficient of variation (CV) is defined as:

\begin{equation}\label{eq:51}\begin{aligned}CV=\frac{{\sigma}}{{\mu}}\times 100\text{\%}\end{aligned}\end{equation}

In Eq. (51), ${\mu}$ and ${\sigma}$ are the ensemble mean density and total standard deviation at the same future point. CV is dimensionless and is computed only where ${\mu}>0$. The merge area predictive-uncertainty penalty $CV_{m}(U_{f})$ is:

\begin{equation}\label{eq:52}\begin{aligned}CV_{m}(U_{f})=\frac{1}{H_{f}N_{r}}\sum _{b=1}^{N_{r}}\sum _{k=0}^{H_{f}-1}(CV_{k,b}(U_{f}))^{2}\end{aligned}\end{equation}

In Eq. (52), $CV_{k,b}(U_{f})$ is the $CV$ at the merge area of ramp $b$ at forecast step $k$ under candidate plan $U_{f}$. It represents ATPO’s predictive uncertainty near a merge and can indicate less reliable predictions (see Section 5.1.2). To limit abrupt changes in metering rates across control intervals, an action-change penalty is introduced:

\begin{equation}\label{eq:53}\begin{aligned}R_{{\Delta}u}(U_{f})=\frac{1}{H_{f}N_{r}}\sum _{b=1}^{N_{r}}\sum _{k=0}^{H_{f}-1}(\frac{r_{b,k}-r_{b,k-1}}{r_{max,b}-r_{min,b}})^{2}\end{aligned}\end{equation}

Finally, a merge area density-exceedance penalty prevents the optimizer from improving metering rate at the expense of severe mainline congestion:

\begin{equation}\label{eq:54}\begin{aligned}R_{{\rho}}(U_{f})=\frac{1}{H_{f}N_{r}}\sum _{b=1}^{N_{r}}\sum _{k=0}^{H_{f}-1}\max\left(0,\frac{{\mu}_{{\rho},k,i_b}-{\rho}_{thr}}{{\rho}_{thr}}\right)^{2}\end{aligned}\end{equation}

In Eq. (54), $i_b$ is the mainline cell containing the merge area of ramp $b$, and ${\rho}_{thr}$ is the prescribed merge area density threshold. Combining the terms gives the complete objective:

\begin{equation}\label{eq:55}\begin{aligned}J(U_{f})=E(\hat{TTS}(U_{f}))+{\lambda}_{m}CV_{m}(U_{f})+{\lambda}_{r}R_{{\Delta}u}(U_{f})+{\lambda}_{{\rho}}R_{{\rho}}(U_{f})\end{aligned}\end{equation}

In Eq. (55), $CV_{m}(U_{f})$ measures relative uncertainty in predicted merge area density, $R_{{\Delta}u}(U_{f})$ penalizes changes in the metering rate of a given ramp between adjacent control intervals, and $R_{{\rho}}(U_{f})$ penalizes predicted density above ${\rho}_{thr}$ near a merge. The nonnegative coefficients ${\lambda}_{m}$, ${\lambda}_{r}$, and ${\lambda}_{{\rho}}$ weight these penalties relative to predicted TTS. The selected plan is $U_f^{\star}=\operatorname*{arg\,min}_{U_f}J(U_f)$.

\subsubsection{Optimization with the Cross-Entropy Method}\label{sec:3.4.2}

Each candidate plan in Eq. (55) requires ATPO inference, queue propagation, and correction to feasible discharge rates. The objective is therefore inconvenient to express as a simple analytical optimization problem. The cross-entropy method (CEM; \textcolor{myvioletblue}{\textbf{\hyperlink{bib:39}{Rubinstein, 1999}}}) searches for low predicted TTS and low-risk plans through repeated sampling, prediction-based evaluation, and updates of a proposal distribution. It does not require gradients of the predictor with respect to actions.

(1) Decision vector and proposal distribution. Metering rates for $N_{r}$ ramps and $H_{f}$ forecast steps form a single decision vector. CEM samples candidate actions from a truncated diagonal Gaussian with mean ${\mu}_{CEM}$ and standard deviation ${\sigma}_{CEM}$.

(2) CEM iterations. At each iteration, $N_{p}$ candidates are sampled, their rates are corrected using queue propagation, and each plan is mapped to a control field $U_{f}$. The $N_{e}$ ensemble members predict traffic under each candidate, which is scored by Eq. (55). The $N_{c}$ lowest-cost candidates form an elite set $\mathcal{E}$. Their coordinate-wise mean and standard deviation update ${\mu}_{CEM}$ and ${\sigma}_{CEM}$, with smoothing coefficient ${\omega}\in (0,1]$ balancing old and new distribution parameters. Iteration continues until $N_{i}$ iterations or a prescribed convergence criterion is reached.

(3) Receding-horizon execution. After CEM terminates, the feasible candidate with the lowest objective, $U_f^{\star}$, is selected. Only the metering rates for the current control interval $H_{c}$, where $H_{c}<H_{f}$, are implemented. After $H_{c}$ steps and new observations, the optimization is repeated. Here, $H_{c}$ is the number of forecast steps between updates.

\subsubsection{Pre-Execution Safety Filter}\label{sec:3.4.3}

After CEM selects $U_f^{\star}$, the safety filter checks the predicted merge area CV under that plan before execution. If the threshold is reached, the controller switches completely to HERO's coordinated metering rates; otherwise, it implements the current interval of $U_f^{\star}$. The filter changes only the action sent to the ramps, not ATPO’s forecasts or CEM’s candidate scores.

HERO builds on a local ALINEA controller at each on-ramp and recruits upstream helper ramps when the queue at a downstream master ramp reaches a coordination threshold. Here, the local controller uses observed downstream merge area density as feedback to maintain a target density, corresponding to the density-feedback form of ALINEA (\textcolor{myvioletblue}{\textbf{\hyperlink{bib:32}{Papageorgiou et al., 1991}}}; \textcolor{myvioletblue}{\textbf{\hyperlink{bib:34}{Papamichail et al., 2010b}}}; \textcolor{myvioletblue}{\textbf{\hyperlink{bib:60}{Yu et al., 2024}}}). Let $g$ index control updates. The local feedback discharge rate for downstream master ramp $b$ is:

\begin{equation}\label{eq:56}\begin{aligned}r_{b,g}^{A}=r_{b,g-1}^{A}+K_{{\rho},b}({\rho}_{b}^{tar}-{\rho}_{b,g})\end{aligned}\end{equation}

In Eq. (56), $r_{b,g}^{A}$ is the implemented rate in the current control interval; ${\rho}_{b}^{tar}$ and ${\rho}_{b,g}$ are the target critical density and observed density at the merge area of ramp $b$; and $K_{{\rho},b}$ is the feedback gain. A measured density above the target reduces the local metering rate, while a density below it increases the rate.

Let $S_{g}$ be the set of upstream helper ramps coordinated during the current interval. HERO activates this set when the master-ramp queue reaches the coordination threshold $w_{b}^{thr}$ and assigns a minimum queue target to each helper ramp, allowing upstream ramps to share the downstream waiting burden. For an active helper ramp, the queue-coordination rate is:

\begin{equation}\label{eq:57}\begin{aligned}r_{b,g}^{Q}=d_{b,g-1}-K_{w,b}(w_{b,g}^{min}-w_{b,g})\end{aligned}\end{equation}

In Eq. (57), $r_{b,g}^{Q}$ is the helper ramp’s queue-coordination rate for the current interval; $w_{b,g}^{min}$ and $w_{b,g}$ are its minimum queue target and observed queue; and $K_{w,b}$ is the queue-feedback gain. When the observed queue is below target, Eq. (57) reduces discharge so a queue can accumulate upstream. An active helper ramp uses the lower of its local-feedback and queue-coordination rates:

\begin{equation}\label{eq:58}\begin{aligned}r_{b,g}^{H}=\min\left(r_{b,g}^{A},r_{b,g}^{Q}\right),\quad b\in S_{g}\end{aligned}\end{equation}

Once CEM selects $U_f^{\star}$, the safety filter compares the arithmetic mean of predicted merge-area density CV under that plan with a threshold. Reaching $CV_{trig}$ triggers HERO.

\section{Experimental Setup}\label{sec:4}

\subsection{Freeway Corridor and Data Generation}\label{sec:4.1}

The study corridor is an approximately 7-km urban expressway mainline in Shanghai. It has three on-ramps and three off-ramps. All on-ramps are signal controlled, and their discharge rates satisfy the constraints in Eq. (1). The lane configuration, ramp positions, and detector coordinates are listed in \hyperref[tab:1]{\textcolor{myvioletblue}{\textbf{Table 1}}}. The mainline is divided into 28 cells of length 0.25 km each ($N_{x}$ = 28).

\begin{figure}[htbp]\centering
\includegraphics[width=0.95\textwidth,height=0.73\textheight,keepaspectratio]{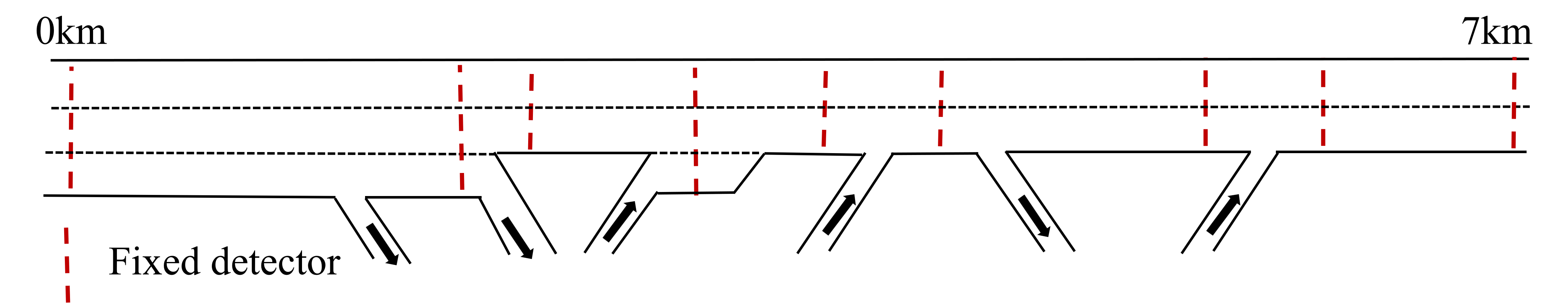}
\caption{\textbf{Corridor topology and locations of on- and off-ramps.}}\label{fig:4}\end{figure}

Base experimental data were generated with a METANET macroscopic simulator. The simulation time step is ${\Delta}t=5$ s, and control is updated every 1 min (12 simulation steps). Key model parameters were calibrated, and the parameter items are given in \hyperref[tab:1]{\textcolor{myvioletblue}{\textbf{Table 1}}}. To represent stochastic traffic fluctuations, Gaussian noise with an amplitude of 2 m/s was added to the speed-dynamics equation, and 5\% noise was applied to model parameters in each scenario.

Varied mainline and ramp demands were used to generate diverse traffic conditions. Each one-hour scenario contains off-peak flow, congestion formation, peak demand, and congestion dissipation. A total of 500 independent scenarios were generated and split into training, validation, and test sets at a 7:1:2 ratio. To cover different control regimes and evaluate out-of-distribution (OOD) actions, each training and validation scenario was simulated under no control, control with a fixed action, PI-ALINEA, and HERO. This yielded $500\times 0.7\times 4=1400$ simulation hours for training and $200$ hours for validation. Each test scenario was simulated under no control, fixed-action control, PI-ALINEA, HERO, and random-action control, where the random actions were used for the OOD evaluation.

\subsection{Sensing and Observation Configuration}\label{sec:4.2}

The observation system combines fixed detectors and probe vehicles.

(1) Fixed detectors. Nine detectors, spaced approximately 0.8 km apart, were installed on the mainline at the locations in \hyperref[fig:4]{\textcolor{myvioletblue}{\textbf{Figure 4}}}. Each provides density and speed measurements at its cross section. Detector locations are identical in training, validation, and testing so that all methods are evaluated under consistent observation conditions.

(2) Probe vehicles. The default penetration rate is 10\%: approximately one in ten vehicles is equipped to report its instantaneous position and speed at each time step. These reports provide speed observations for the mainline cells occupied by the probe vehicles. Vehicle selection is randomized according to the specified penetration rate to emulate sparse and irregular mobile sensing.

At each control or test time $t$, the model uses measurements from the preceding five minutes: sparse density and speed at fixed detectors, sparse probe vehicle speed, mainline and ramp demand flows, and current queues at all on-ramps. It predicts the mean and standard deviation of density and speed in every road cell over the following five minutes, representing both point forecasts and predictive uncertainty.

Selected corridor, simulation, network-training, and control-optimization parameters are summarized in \hyperref[tab:1]{\textcolor{myvioletblue}{\textbf{Table 1}}}.

\begin{table}[H]\centering\fontsize{8.5pt}{10.5pt}\selectfont
\caption{\textbf{Selected experimental parameters.}}\label{tab:1}
\setlength{\tabcolsep}{3.5pt}\renewcommand{\arraystretch}{1.08}
\begin{tabular}{@{}>{\centering\arraybackslash}m{0.17\linewidth}>{\centering\arraybackslash}m{0.14\linewidth}>{\centering\arraybackslash}m{0.45\linewidth}>{\centering\arraybackslash}m{0.17\linewidth}@{}}
\toprule
\textbf{Category} & \textbf{Parameter} & \textbf{Definition} & \textbf{Value / unit} \\
\midrule
 & \(L_{x}\) & Physical corridor length & 7 km \\
\textbf{Corridor configuration} & \(N_{x}\) & Number of cells on the reference spatial grid & 28 \\
 & \({\Delta}x\) & Cell length on the reference grid & 0.25 km \\
 & $r_{min}$, $r_{max}$ & Minimum and maximum ramp metering rates & 300/1800 veh/h \\
\midrule
 & \({\Delta}t\) & Macroscopic simulation and state-update interval & 5 s \\
 & \({\tau}\) & METANET speed relaxation time & 12 s \\
\textbf{Simulation} & \({\eta},{\kappa}\) & METANET anticipation and density-smoothing parameters & 30; 20 \\
 & \(v_{f},{\rho}_{cr},Q_{c}\) & Fundamental-diagram parameters: free-flow speed / critical density / capacity & 105 km/h; 29.4 veh/km; 1918 veh/h \\
\midrule
 & \(N_{s}\) & Number of fixed detectors in the reference configuration & 9 \\
 & \(P_{probe}\) & Reference probe-vehicle penetration rate & 10\% \\
 & \(H_{p}\) & Historical-window steps & 60 \\
 & \(H_{f}\) & Forecast-window steps & 60 \\
 & \(T_{f}\) & Physical forecast duration & 300 s \\
\textbf{Prediction network} & \({\Delta}t_{f}\) & Forecast-grid time interval & 5 s \\
 & \({\Delta}x_{f}\) & Forecast-grid spatial interval & 0.25 km \\
 & \(N_{e}\) & Bootstrap ensemble members & 10 \\
 & ${\lambda}_{P}$, ${\lambda}_{C}$ & Local physical residual / global conservation loss weights & 0.1; 0.1 \\
\midrule
 & $N_{p}$, $N_{c}$ & CEM candidate / elite samples & 50; 12 \\
\textbf{Control optimization} & \(H_{c}\) & Forecast steps between control updates & 12 \\
 & ${\lambda}_{m}$,${\lambda}_{r}$,${\lambda}_{{\rho}}$ & Uncertainty, action-change, and density penalty weights & 10; 0.1; 0.2 \\
 & \({\rho}_{thr}\) & Merge-area density-exceedance threshold & 32 (veh/km/lane) \\
\bottomrule\end{tabular}
\end{table}

\subsection{Evaluation Metrics}\label{sec:4.3}

The proposed method is assessed in terms of prediction accuracy, uncertainty quantification, and control performance.

\subsubsection{Prediction Error Metrics}\label{sec:4.3.1}

Errors are reported separately for density ${\rho}$ and speed $v$ at two spatial scopes:

Detector locations (det): errors are evaluated only at cells containing fixed detectors, measuring fit at observed locations.

Full corridor (global): errors are evaluated at all mainline cells, measuring the ability to extrapolate sparse observations across the corridor.

Two error metrics are used:

Root mean square error (RMSE) measures the typical magnitude of prediction errors and gives greater weight to large deviations:

\begin{equation}\label{eq:59}\begin{aligned}RMSE=\sqrt{\frac{1}{N_{l}}\sum _{i=1}^{N_{l}}(\hat{y}_{i}-y_{i})^{2}}\end{aligned}\end{equation}

where $y_{i}$ and $\hat{y}_{i}$ are the true and predicted values at point $i$, respectively, and $N_{l}$ is the number of evaluated points.

Relative L2 error (L2) is the square root of the ratio of the sum of squared errors to the sum of squared true values. It measures error relative to the scale of the ground truth:

\begin{equation}\label{eq:60}\begin{aligned}L2=\frac{\sqrt{\sum _{i=1}^{N_{l}}(\hat{y}_{i}-y_{i})^{2}}}{\sqrt{\sum _{i=1}^{N_{l}}y_{i}^{2}}}\end{aligned}\end{equation}

Values closer to zero indicate higher prediction accuracy for both metrics.

\subsubsection{Uncertainty Quantification Metrics}\label{sec:4.3.2}

Uncertainty quality is assessed separately at detector locations and globally for density and speed using the following three metrics:

Continuous ranked probability score (CRPS) jointly measures the calibration and sharpness of a predictive distribution. A lower CRPS indicates closer agreement between the predicted distribution and the observed value:

\begin{equation}\label{eq:61}\begin{aligned}\mathrm{CRPS}(F,y)=\int_{\mathbb R}\bigl(F(z)-\mathbf{1}(y\leq z)\bigr)^{2}\,\mathrm{d}z\end{aligned}\end{equation}

where $F$ is the predictive cumulative distribution function, $y$ is the observed value, $\mathbb R$ is the real line, and $\mathbf{1}(\cdot)$ equals one when its condition holds and zero otherwise.

The 95\% empirical coverage probability (PICP95) is the observed fraction of true values lying inside nominal 95\% predictive intervals. A value closer to 0.95 indicates better interval calibration:

\begin{equation}\label{eq:62}\begin{aligned}\mathrm{PICP}_{95}=\frac{1}{n}\sum_{i=1}^{n}\mathbf{1}\bigl(y_i\in[\hat y_i^{\mathrm{low}},\hat y_i^{\mathrm{high}}]\bigr)\end{aligned}\end{equation}

where $[\hat{y}_{i}^{\mathrm{low}},\hat{y}_{i}^{\mathrm{high}}]$ are the bounds of the 95\% predictive interval for sample $i$, and $n$ is the number of evaluated samples.

Quantile interval calibration error (QICE) divides each predictive distribution into equal-probability quantile intervals and compares their empirical coverage frequencies with nominal frequencies. Unlike PICP95, which checks only one nominal coverage level, QICE assesses calibration across the entire quantile partition:

\begin{equation}\label{eq:63}\begin{aligned}\mathrm{QICE}=\frac{1}{N_q}\sum_{i=1}^{N_q}\left|r_i-\frac{1}{N_q}\right|\end{aligned}\end{equation}

where $N_q$ is the number of equal-probability quantile intervals, distinct from the detector count $N_s$, and the empirical frequency of interval $i$ is calculated by Eq. (64).

\begin{equation}\label{eq:64}\begin{aligned}r_i=\frac{1}{M}\sum_{j=1}^{M}\mathbf{1}\bigl(q_{j,i-1}<y_j\leq q_{j,i}\bigr),\qquad 1\leq i\leq N_q\end{aligned}\end{equation}

where $M$ is the number of prediction-truth pairs. For sample $j$, $q_{j,i-1}$ and $q_{j,i}$ are the predictive quantiles at cumulative probabilities $(i-1)/N_q$ and $i/N_q$. The lower endpoint $q_{j,0}$ of the first interval is unbounded below, and the upper endpoint $q_{j,N_q}$ of the last interval is unbounded above. These are interval bounds, not predicted traffic states. The empirical frequency $r_i$ is the proportion of observed values falling in interval $i$ of their respective predictive distributions.

Eq. (64) uses an open left boundary and a closed right boundary to avoid double-counting values on interval boundaries. For a calibrated distribution, each $r_i$ should approach $1/N_q$; a smaller QICE indicates closer agreement with these nominal frequencies.

\subsubsection{Control Performance Metrics}\label{sec:4.3.3}

Two metrics assess practical control effectiveness and computational efficiency:

Total time spent (TTS) is the accumulated time spent by vehicles traveling on the mainline and waiting at on-ramps during the evaluation period. Lower TTS indicates more efficient traffic operations.

Computation time is the mean time (s) needed to perform prediction and solve one control decision, indicating whether the method can respond in real time.

\subsection{Comparison Methods}\label{sec:4.4}

Representative baselines are selected separately for forecasting and control.

\subsubsection{Prediction Baselines}\label{sec:4.4.1}

Physical-model/filtering and learned spatiotemporal methods are compared under the same open-loop forecasting protocol and input interface, using all metrics in Sections 4.3.1 and 4.3.2.

\noindent\textbf{Traffic model and filtering methods}\par

AS (adaptive smoothing) reconstructs traffic states from incomplete detector data using the distinct propagation directions of free-flow and congested disturbances. It does not rely on an explicit traffic-dynamics model (\textcolor{myvioletblue}{\textbf{\hyperlink{bib:44}{Treiber and Helbing, 2003}}}). Under the common forecasting protocol, its reconstructed state is extrapolated over a short horizon.

METANET--EKF propagates traffic states using the METANET macroscopic model and assimilates measurements with an EKF, recursively correcting state estimates (\textcolor{myvioletblue}{\textbf{\hyperlink{bib:49}{Wang and Papageorgiou, 2005}}}; \textcolor{myvioletblue}{\textbf{\hyperlink{bib:64}{Zhao et al., 2022}}}).

\noindent\textbf{Learning-based spatiotemporal methods}\par

MoGERNN combines a mixture of graph experts with a recurrent spatiotemporal encoder to learn dependencies between road nodes and forecast states at unobserved locations (\textcolor{myvioletblue}{\textbf{\hyperlink{bib:65}{Zhou et al., 2025}}}).

IGNNK uses an inductive graph neural network for spatiotemporal kriging to infer states at unobserved nodes (\textcolor{myvioletblue}{\textbf{\hyperlink{bib:56}{Wu et al., 2021}}}). Its original task emphasizes reconstruction, here it is adapted to future-state prediction under the common evaluation protocol.

UIGNN introduces uncertainty-aware mechanisms into an inductive graph network for traffic prediction with missing observations, producing forecasts and uncertainty estimates (\textcolor{myvioletblue}{\textbf{\hyperlink{bib:26}{Mei et al., 2023}}}).

DistPred uses an IGNNK backbone adapted for multistep forecasting and generates predictive samples through DistPred’s distribution-free probabilistic inference mechanism (\textcolor{myvioletblue}{\textbf{\hyperlink{bib:22}{Liang et al., 2025}}}).

The original versions of these learning-based baselines do not jointly condition their forecasts on candidate ramp actions, mainline demand, and ramp queues. For a consistent action-conditioned comparison, MoGERNN, IGNNK, UIGNN, and DistPred are each equipped with a cross-attention adapter. The candidate multi-ramp discharge sequence, mainline-demand history, and on-ramp queue history are encoded as condition tokens. Each model's future latent features serve as queries, while the condition tokens supply keys and values (\textcolor{myvioletblue}{\textbf{\hyperlink{bib:45}{Vaswani et al., 2017}}}). The attended information is fused with the original latent features before prediction.

\subsubsection{Control Baselines}\label{sec:4.4.2}

Four baselines are used for control comparison.

PI-ALINEA adds proportional-integral regulation to local ALINEA feedback. In this implementation, metering rates respond to the difference between downstream merge-area density and its target value (\textcolor{myvioletblue}{\textbf{\hyperlink{bib:51}{Wang et al., 2010}}}).

HERO coordinates local ALINEA controllers through queue management between downstream master ramps and upstream helper ramps. Once a master-ramp queue reaches its activation threshold, the coordinator assigns minimum queue targets upstream and uses the available storage there to relieve the downstream bottleneck (\textcolor{myvioletblue}{\textbf{\hyperlink{bib:34}{Papamichail et al., 2010b}}}). The local controller here is density-feedback ALINEA.

METANET-EKF+CEM replaces the learned forward operator with METANET-EKF for state estimation and candidate-action simulation while retaining the same CEM for action search. Holding the optimizer fixed isolates the effect of the forward model on closed-loop performance.

AMOC assumes that the current global traffic state and future boundary demand are known in the test simulator and optimizes ramp metering sequences with a macroscopic traffic model (Kotsialos et al., 2002). It is an ideal-information performance benchmark against which the TTS of other strategies is compared.

\subsection{Experimental Tasks}\label{sec:4.5}

Five experiments assess forecasting, network components, sensing configuration, closed-loop control, and controller design.

Prediction comparison: ATPO and the baselines in Section 4.4.1 are compared on global traffic states, predicted TTS, and probabilistic predictions in the macroscopic simulation test set.

Predictor ablation: six ATPO components are removed individually to assess their contributions to state and TTS prediction.

Observation sensitivity: detector count and probe-vehicle penetration are varied separately. A model is trained for each configuration and tested on the same scenarios to assess the marginal contribution of each source.

Control comparison: ATPOUCRM and baseline policies are evaluated in multi-ramp closed-loop simulations through TTS, variation across scenarios, and state evolution in representative cases.

Controller ablation and solver comparison: the uncertainty penalty and safety filter are removed separately, while SQP, MPPI, and GRPO replace CEM as methods for generating actions. Mean TTS, maximum scenario TTS, and per-update computation time are compared to examine risk-handling components and solver trade-offs.

\section{Results}\label{sec:5}

\subsection{Prediction Performance}\label{sec:5.1}

\subsubsection{Prediction Accuracy}\label{sec:5.1.1}

GTSP must be evaluated at both detector locations and unobserved cells. \hyperref[tab:2]{\textcolor{myvioletblue}{\textbf{Table 2}}} reports density and speed errors at detector locations and global grid, together with TTS prediction errors derived from predicted density states. ATPO has the lowest values for all these metrics among the listed methods. Its global density, speed, and TTS RMSE values are 2.54 veh/km/lane, 4.56 km/h, and 0.76 veh·h, respectively. Relative to MoGERNN, these errors are lower by 29.1\%, 25.7\%, and 36.1\%. Relative to METANET--EKF, they are lower by 18.3\%, 41.5\%, and 33.3\%. These results show stronger action-conditioned GTSP performance than the baselines.

\begin{figure}[htbp]\centering
\includegraphics[width=0.95\textwidth,height=0.73\textheight,keepaspectratio]{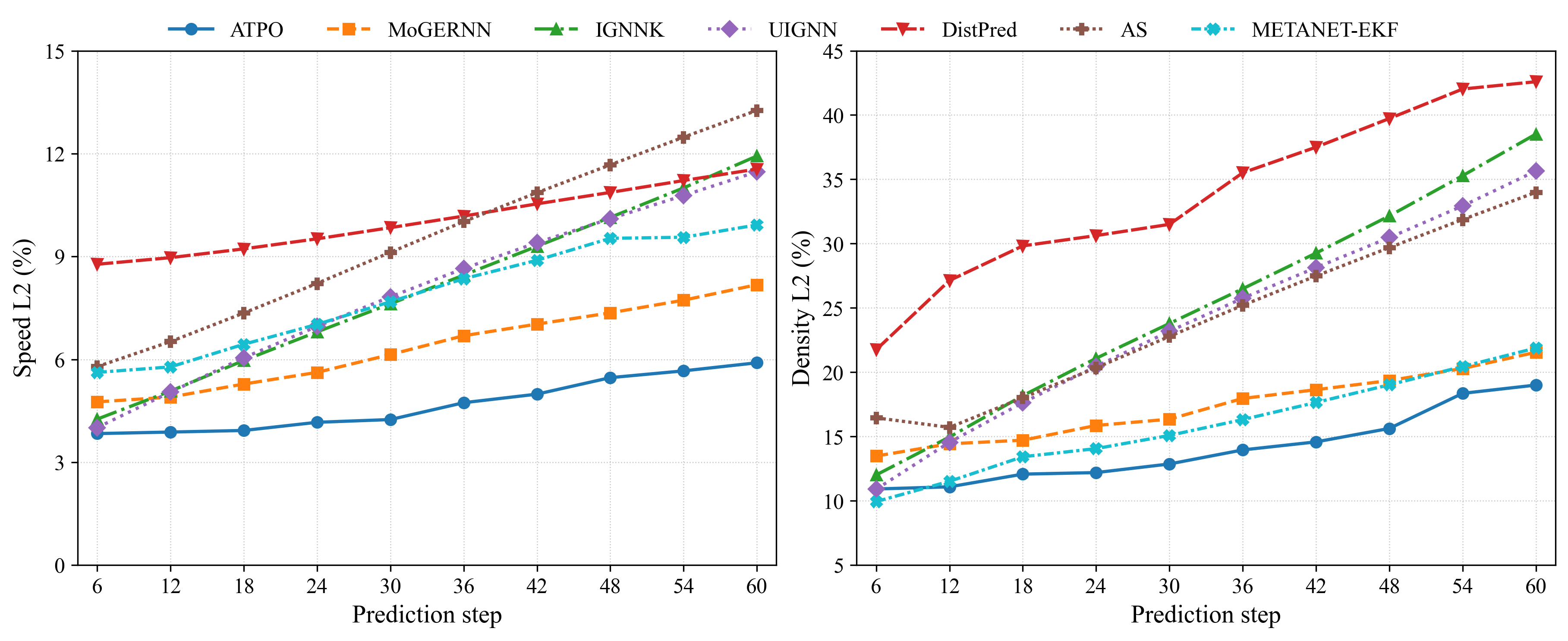}
\caption{\textbf{Global relative L2 error by forecast step for the comparison methods.}}\label{fig:5}\end{figure}

\hyperref[fig:5]{\textcolor{myvioletblue}{\textbf{Figure 5}}} plots global density and speed relative L2 errors by prediction horizon. Errors generally increase with lead time for all methods, but ATPO maintains the lowest overall error across the horizon.

\begin{table}[H]\centering\fontsize{8.5pt}{10.5pt}\selectfont
\caption{\textbf{Traffic state and system-TTS errors over the full forecast window (d: detector cells; g: all cells).}}\label{tab:2}
\setlength{\tabcolsep}{1.8pt}\renewcommand{\arraystretch}{1.12}
\begin{tabular}{@{}c*{10}{c}@{}}
\toprule
\textbf{Method} & \multicolumn{4}{c}{\textbf{Density}} & \multicolumn{4}{c}{\textbf{Speed}} & \multicolumn{2}{c}{\textbf{TTS}} \\
\cmidrule(lr){2-5}\cmidrule(lr){6-9}\cmidrule(l){10-11}
 & RMSE-d & L2-d & RMSE-g & L2-g & RMSE-d & L2-d & RMSE-g & L2-g & RMSE & L2 \\
\midrule
ATPO & \best{2.44} & \best{13.15} & \best{2.54} & \best{12.73} & \best{4.66} & \best{5.16} & \best{4.56} & \best{5.10} & \best{0.76} & \best{3.46} \\
MoGERNN & 3.02 & 14.79 & 3.58 & 17.94 & 5.00 & 5.53 & 6.14 & 6.87 & 1.19 & 5.38 \\
IGNNK & 3.68 & 18.02 & 5.19 & 26.00 & 5.96 & 6.59 & 7.76 & 8.68 & 3.36 & 15.24 \\
UIGNN & 3.91 & 19.14 & 5.22 & 26.17 & 6.23 & 6.89 & 8.05 & 9.00 & 3.68 & 16.69 \\
DistPred & 4.78 & 23.41 & 6.75 & 33.85 & 6.79 & 7.51 & 8.27 & 9.24 & 5.43 & 24.65 \\
AS & 5.43 & 26.62 & 5.84 & 29.30 & 9.32 & 10.32 & 9.01 & 10.08 & 4.56 & 20.72 \\
METANET--EKF & 2.74 & 13.44 & 3.11 & 15.59 & 6.36 & 7.04 & 7.80 & 8.73 & 1.14 & 5.19 \\
\bottomrule\end{tabular}
\par\vspace{2pt}{\scriptsize\emph{Note:} d = detector cells; g = all cells. Density RMSE: veh/km/lane; speed RMSE: km/h; TTS RMSE: veh\,h. L2 values are percentages. Red bold indicates the lowest error in each column.}
\end{table}

\begin{figure}[htbp]\centering
\includegraphics[width=0.95\textwidth,height=0.73\textheight,keepaspectratio]{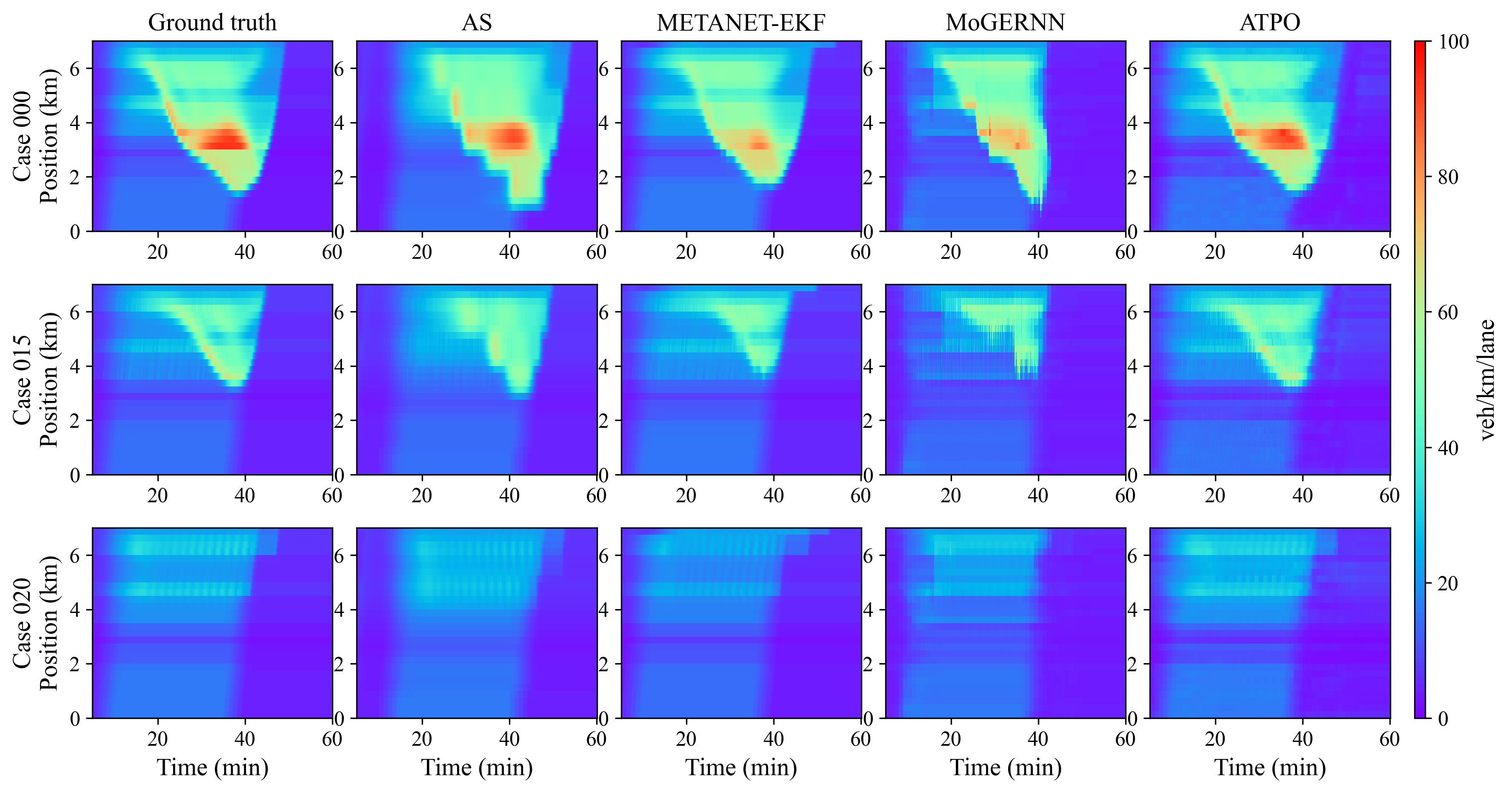}
\caption{\textbf{Ground-truth and predicted mainline density fields in three test scenarios.}}\label{fig:6}\end{figure}

\begin{figure}[htbp]\centering
\includegraphics[width=0.95\textwidth,height=0.73\textheight,keepaspectratio]{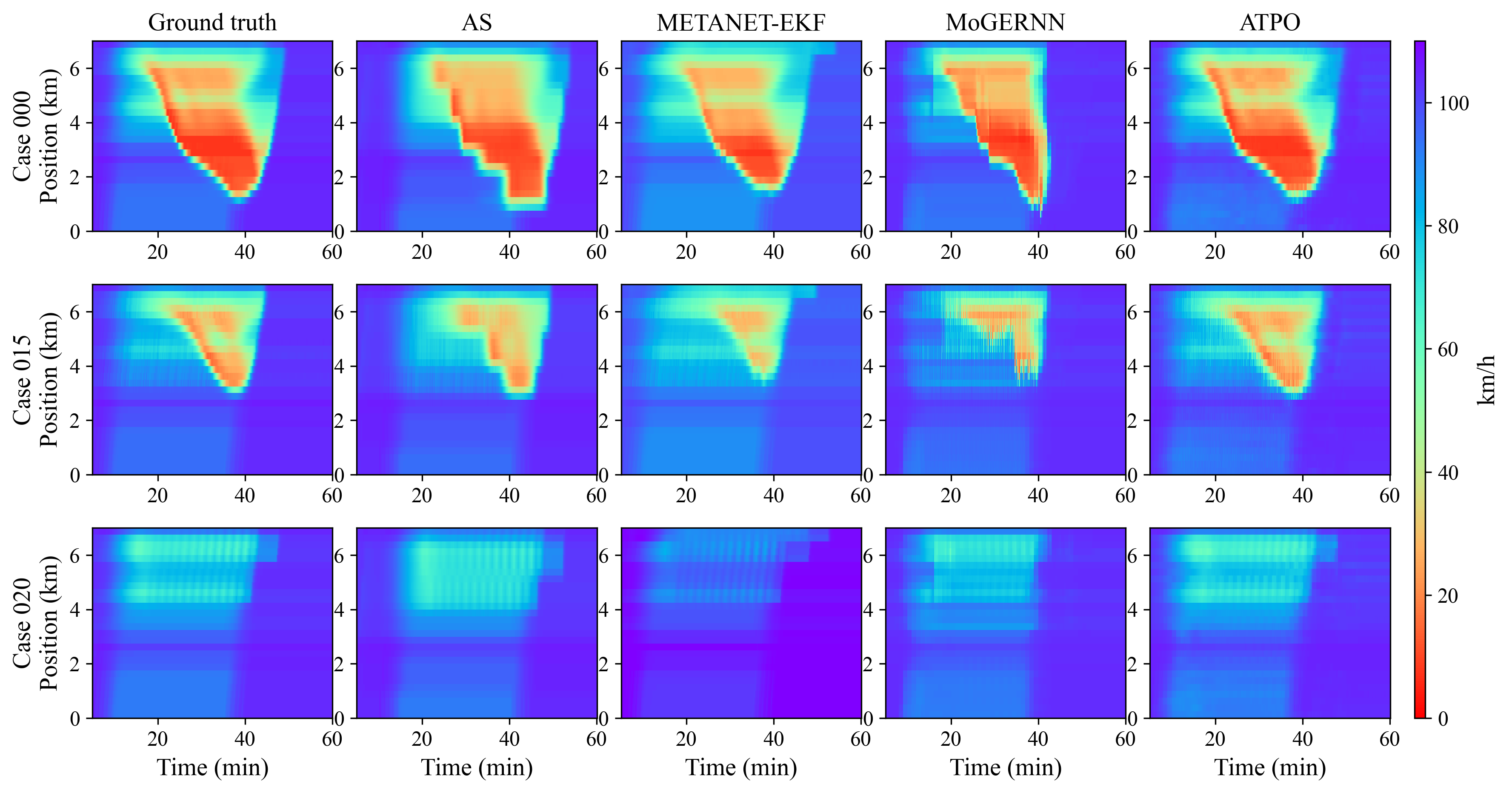}
\caption{\textbf{Ground-truth and predicted mainline speed fields in three test scenarios.}}\label{fig:7}\end{figure}

To examine spatiotemporal dynamics beyond aggregate errors, AS, METANET-EKF, MoGERNN, and ATPO are compared in three scenarios from \hyperref[tab:2]{\textcolor{myvioletblue}{\textbf{Table 2}}}: no control (Case 0), control with a fixed action (Case 15), and HERO (Case 20). \hyperref[fig:6]{\textcolor{myvioletblue}{\textbf{Figures 6}}} and \hyperref[fig:7]{\textcolor{myvioletblue}{\textbf{7}}} show their density and speed fields at the five-minute forecast horizon. ATPO reproduces congestion onset and dissipation close to the reference states, with remaining shifts near local wave fronts and high-density regions. METANET-EKF can represent the formation and dissipation of congestion and recovery waves, but its predicted congestion onset lags the reference. AS and MoGERNN capture the broad congested region less precisely at its boundaries.

\begin{figure}[htbp]\centering
\includegraphics[width=0.95\textwidth,height=0.73\textheight,keepaspectratio]{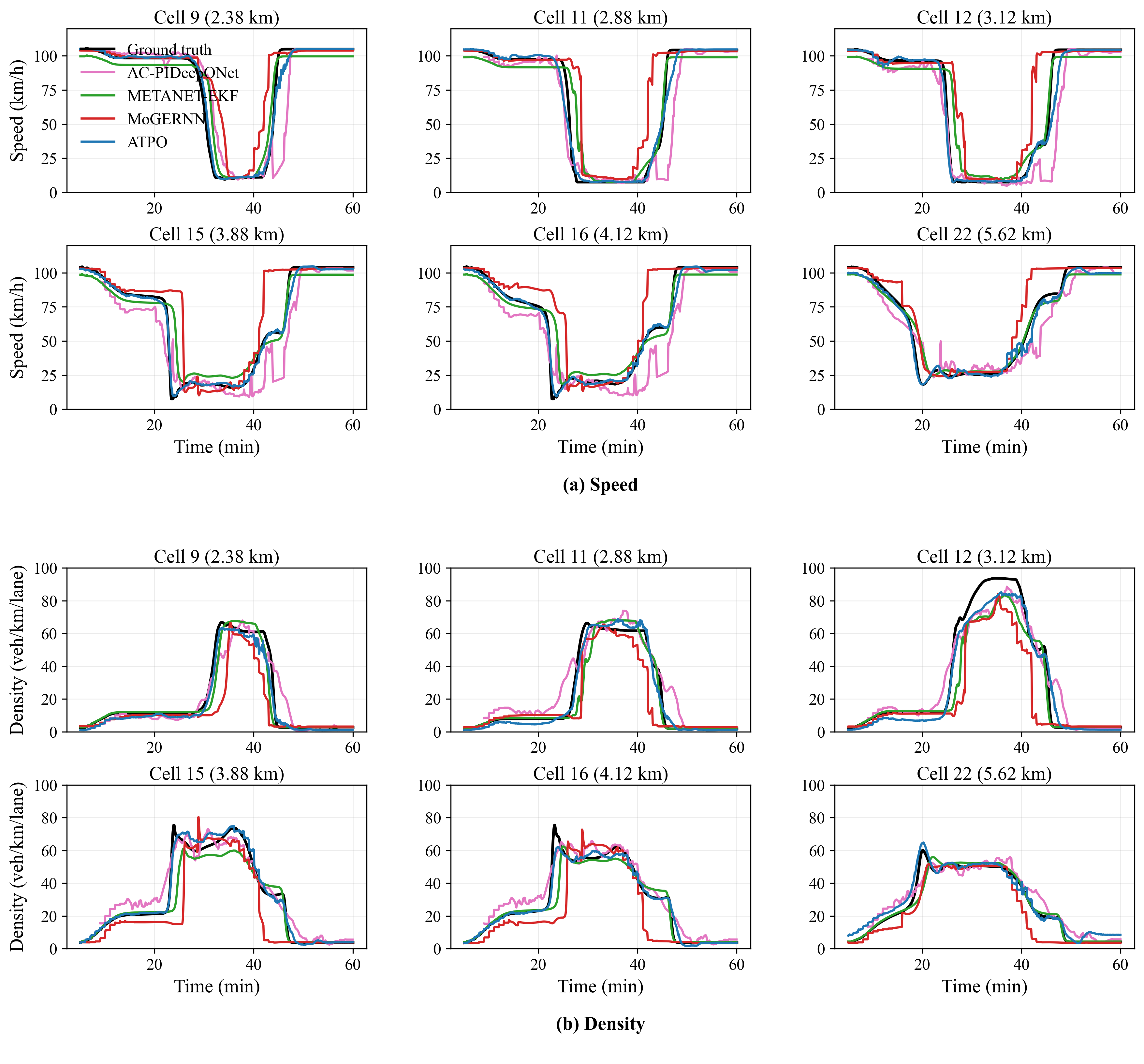}
\caption{\textbf{Speed and density forecasts at cells without detectors: (a) speed; (b) density.}}\label{fig:8}\end{figure}

The differences are more pronounced at cells without detectors. \hyperref[fig:8]{\textcolor{myvioletblue}{\textbf{Figure 8}}} shows forecasts five minutes ahead for six such cells in Case 0. In the ground truth, speed drops rapidly and density rises sharply when congestion reaches cells at 2.38--3.12 km. Changes at 3.88--5.62 km are more gradual. ATPO reproduces both forms of congestion onset and approximates their duration. AC-PIDeepONet delays recovery at several cells, while MoGERNN predicts speed recovery and density reduction too early. These comparisons show how errors in the timing of congestion propagation and dissipation differ across the methods.

\begin{figure}[htbp]\centering
\includegraphics[width=0.95\textwidth,height=0.73\textheight,keepaspectratio]{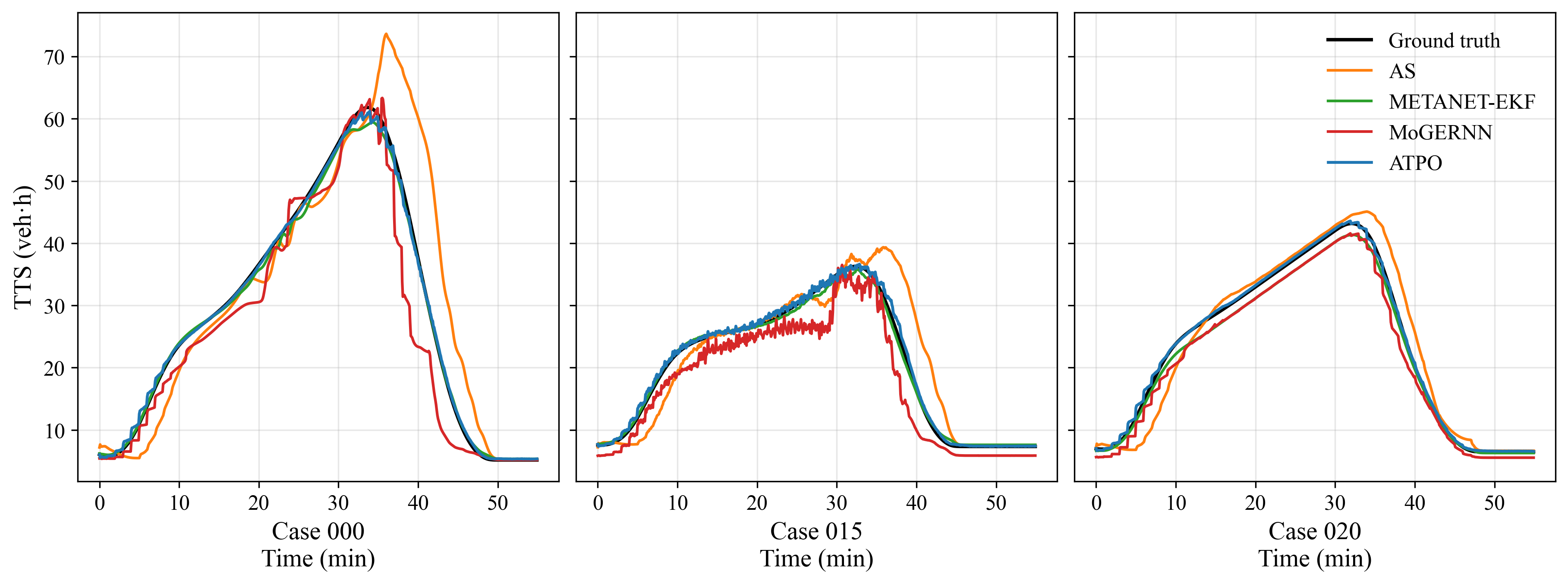}
\caption{\textbf{Ground-truth and predicted TTS trajectories in three test scenarios.}}\label{fig:9}\end{figure}

Prediction errors at unobserved cells also affect predicted TTS. \hyperref[fig:9]{\textcolor{myvioletblue}{\textbf{Figure 9}}} compares TTS over five minutes calculated from the predicted density fields. In Case 0, reference TTS rises with congestion to a peak and then falls rapidly. ATPO follows the rise and peak magnitude, with a small delay during recovery. AS overpredicts the peak and delays the decline, whereas MoGERNN predicts the decrease too early. Case 15 shows similar temporal differences. In Case 20, which remains in free flow, the curves are closer. Differences in TTS prediction are therefore concentrated in congestion growth and dissipation, consistent with the fields and unobserved-cell errors discussed above.

\begin{figure}[htbp]\centering
\includegraphics[width=0.95\textwidth,height=0.73\textheight,keepaspectratio]{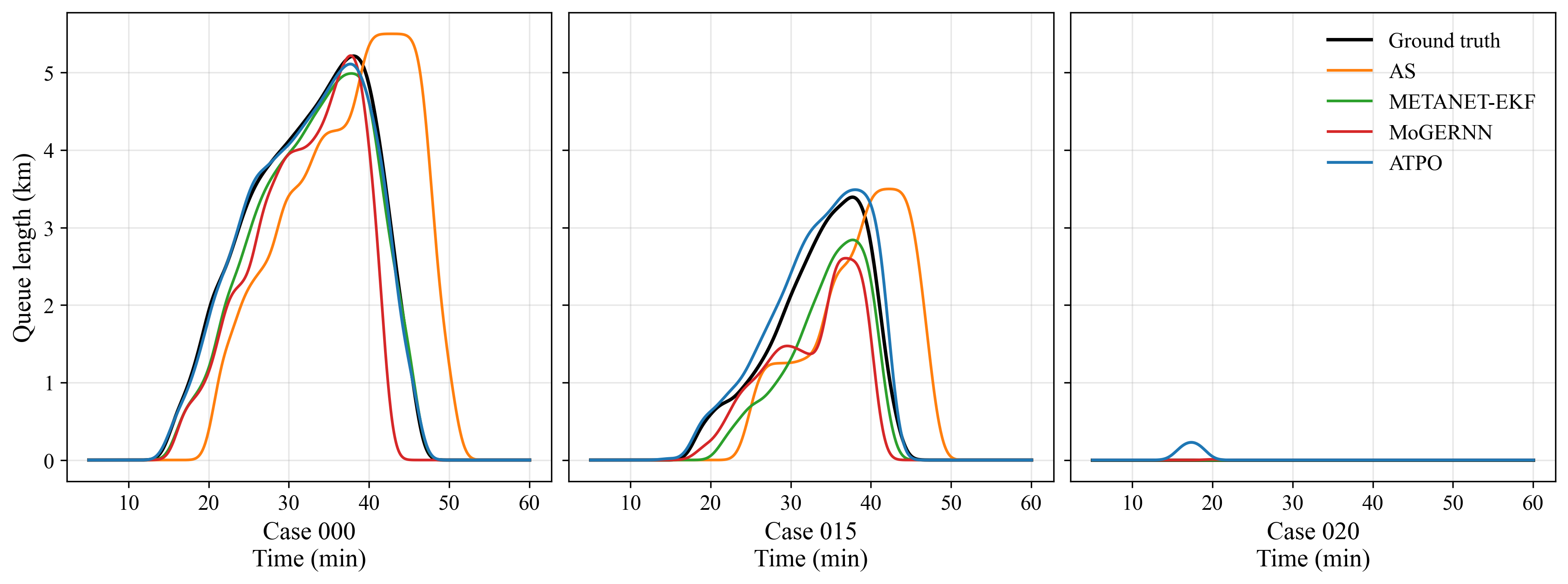}
\caption{\textbf{Ground-truth and predicted mainline queue length trajectories in three test scenarios.}}\label{fig:10}\end{figure}

These spatiotemporal errors also appear in predictions of mainline queue length. A cell with speed below 40 km/h is classified as queued, and \hyperref[fig:10]{\textcolor{myvioletblue}{\textbf{Figure 10}}} compares predicted queue lengths. In Cases 0 and 15, ATPO broadly follows reference queue growth, peak, and clearance time. AS predicts a later peak and clearance, and MoGERNN predicts earlier dissipation. Case 20 has almost no reference queue, so differences are small. The queue length errors corroborate the temporal discrepancies in TTS forecasts.

\begin{figure}[htbp]\centering
\includegraphics[width=0.95\textwidth,height=0.73\textheight,keepaspectratio]{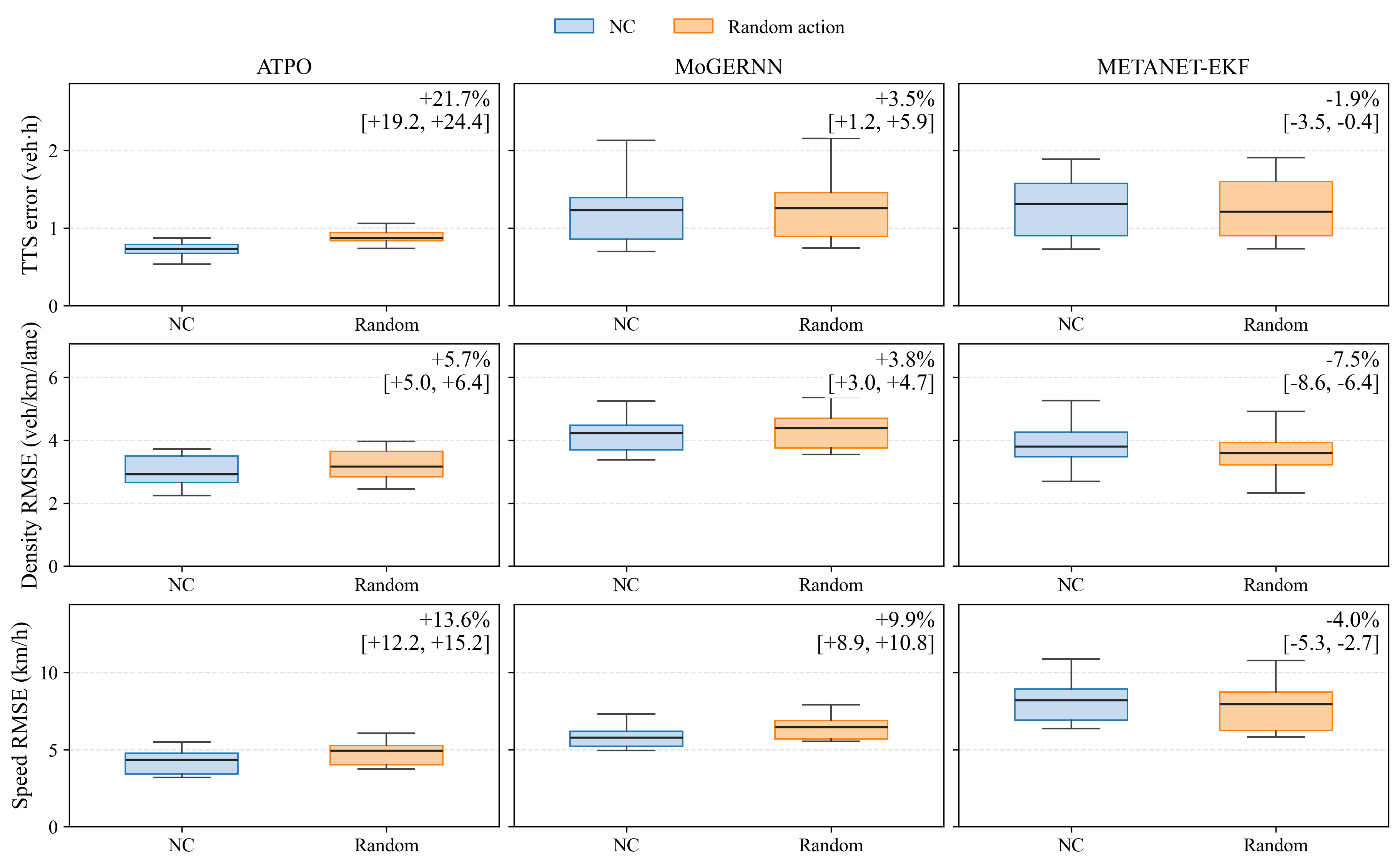}
\caption{\textbf{Prediction error distributions for no control and counterfactual random actions from the same initial states.}}\label{fig:11}\end{figure}

Finally, unseen control actions test the generalization of action-conditioned prediction. In the scenario without control Case 0, the main simulation executes no control throughout. At each decision time, the model receives the same current observations and separately predicts futures under no control and a policy with random actions (OOD policy) that was not used in training. Predictions without control are compared with the main simulation's subsequent states. Ground truth for each random action is generated by a separate simulator branch from the same current state. That action is never executed in the main simulation. \hyperref[fig:11]{\textcolor{myvioletblue}{\textbf{Figure 11}}} shows that ATPO’s mean predicted TTS, density, and speed errors under random actions increase by 21.7\%, 5.7\%, and 17.6\%, respectively, relative to no control. Unseen actions thus make prediction harder. Nevertheless, the corresponding error distributions remain below those of MoGERNN and METANET-EKF, indicating useful predictive accuracy under this unseen policy.

\subsubsection{Uncertainty Quantification}\label{sec:5.1.2}

\hyperref[tab:3]{\textcolor{myvioletblue}{\textbf{Table 3}}} evaluates predictive distributions using probabilistic scores and interval calibration. Lower CRPS means a predictive distribution is closer to observations. PICP95 has a nominal level of 0.95. Proximity to 0.95 indicates better interval calibration. QICE measures calibration across quantile intervals. ATPO’s density and speed CRPS values are 2.001 and 3.356, respectively, 46.3\% and 37.7\% lower than UIGNN and also below DistPred. Its QICE values of 0.0264 and 0.0273 are the lowest among the three methods. Density and speed PICP95 values are 0.870 and 0.901, both below nominal coverage, indicating undercoverage. Density PICP95 is also below DistPred's 0.886. ATPO's advantages in CRPS and QICE therefore do not imply superiority on every calibration measure.

\begin{figure}[htbp]\centering
\includegraphics[width=0.95\textwidth,height=0.73\textheight,keepaspectratio]{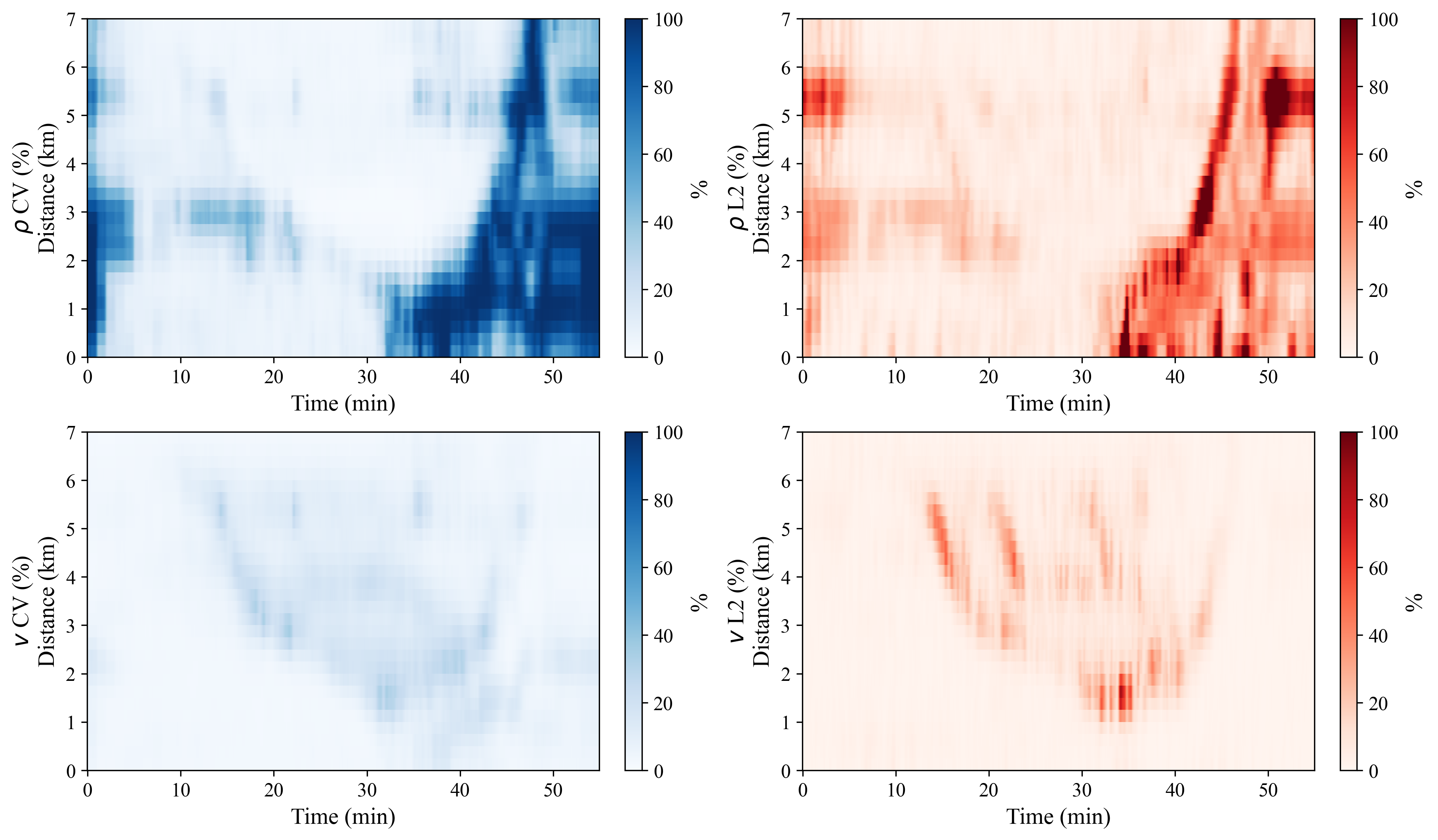}
\caption{\textbf{Spatiotemporal maps of predictive CV and relative error for density and speed in Case 0.}}\label{fig:12}\end{figure}

\begin{figure}[htbp]\centering
\includegraphics[width=0.95\textwidth,height=0.73\textheight,keepaspectratio]{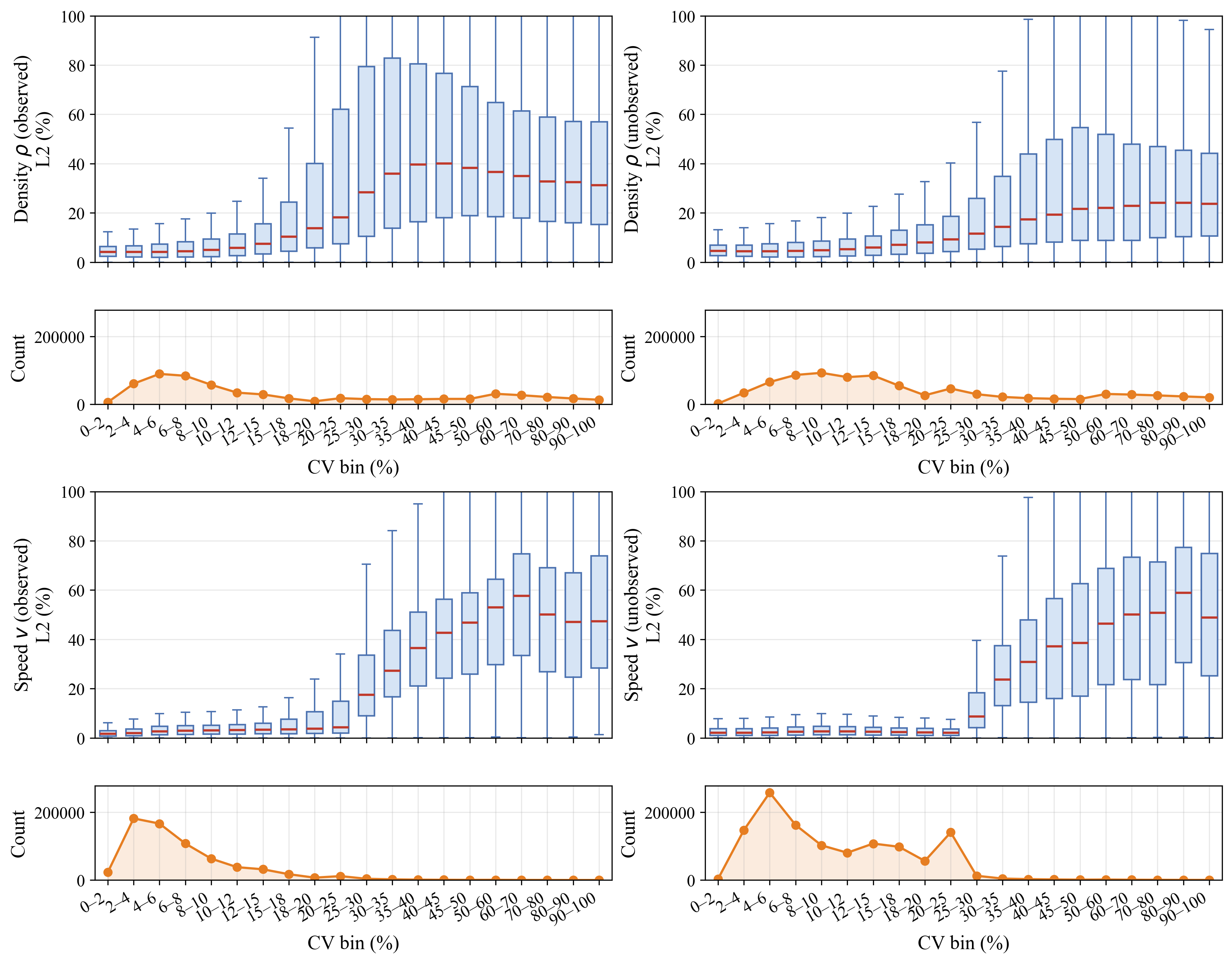}
\caption{\textbf{Relative density and speed error distributions grouped by predictive CV.}}\label{fig:13}\end{figure}

Aggregate probabilistic scores do not show whether predicted uncertainty tracks actual error. \hyperref[fig:12]{\textcolor{myvioletblue}{\textbf{Figure 12}}} presents spatiotemporal maps of the CV and L2 error for density and speed in Case 0. Higher density CV appears mainly where congestion changes or recovers rapidly, and the corresponding error map also shows larger L2 errors. Speed displays a similar association. \hyperref[fig:13]{\textcolor{myvioletblue}{\textbf{Figure 13}}} groups all predictions by CV and shows that median relative density and speed errors generally rise with CV. The model’s uncertainty output therefore identifies locations that are difficult to predict, supporting the uncertainty penalty in the control objective of Section 3.4.1.

\begin{table}[H]\centering\footnotesize
\caption{\textbf{Probabilistic scores and interval calibration for global density and speed forecasts.}}\label{tab:3}
\setlength{\tabcolsep}{3.5pt}\renewcommand{\arraystretch}{1.12}
\begin{tabular}{@{}c*{6}{c}@{}}
\toprule
\textbf{Method} & \multicolumn{3}{c}{\textbf{Density}} & \multicolumn{3}{c}{\textbf{Speed}} \\
\cmidrule(lr){2-4}\cmidrule(l){5-7}
 & CRPS & PICP$_{95}$ & QICE & CRPS & PICP$_{95}$ & QICE \\
\midrule
ATPO & \best{2.001} & 0.870 & \best{0.0264} & \best{3.356} & \best{0.901} & \best{0.0273} \\
UIGNN & 3.729 & 0.764 & 0.0356 & 5.391 & 0.764 & 0.0286 \\
DistPred & 4.406 & \best{0.886} & 0.0267 & 8.295 & 0.898 & 0.0340 \\
\bottomrule\end{tabular}
\par\vspace{2pt}{\scriptsize\emph{Note:} Red bold indicates the lowest CRPS or QICE, or the PICP$_{95}$ closest to its nominal value of 0.95.}
\end{table}

\subsection{Predictor Ablation}\label{sec:5.2}

The proposed predictor combines six components:

Action-conditioned input (AC) introduces candidate actions into the global prediction operator.

Pointwise feature modulation (FiLM) represents control effects within the FNO.

Exogenous cross-attention (CA) incorporates historical upstream mainline demand and on-ramp queues.

Causal FNO blocks (CFNO) enforce the temporal ordering of control effects.

Physics-informed loss (PI) combines traffic-flow priors with data learning.

Uncertainty quantification (UC) uses a heteroscedastic negative log-likelihood and a bootstrap deep ensemble.

Each component is removed individually from the complete ATPO to assess its contribution to prediction: AC, FiLM, CA, CFNO, PI, and UC. In the no-FiLM variant, candidate actions are still supplied through exogenous cross-attention, but no longer modulate the latent features pointwise in FNO layers. In the no-CFNO variant, the causal FNO blocks are replaced with standard FNO blocks without an explicit constraint on temporal causality, allowing the contribution of the causal temporal structure to be assessed.

\begin{figure}[htbp]\centering
\includegraphics[width=0.95\textwidth,height=0.73\textheight,keepaspectratio]{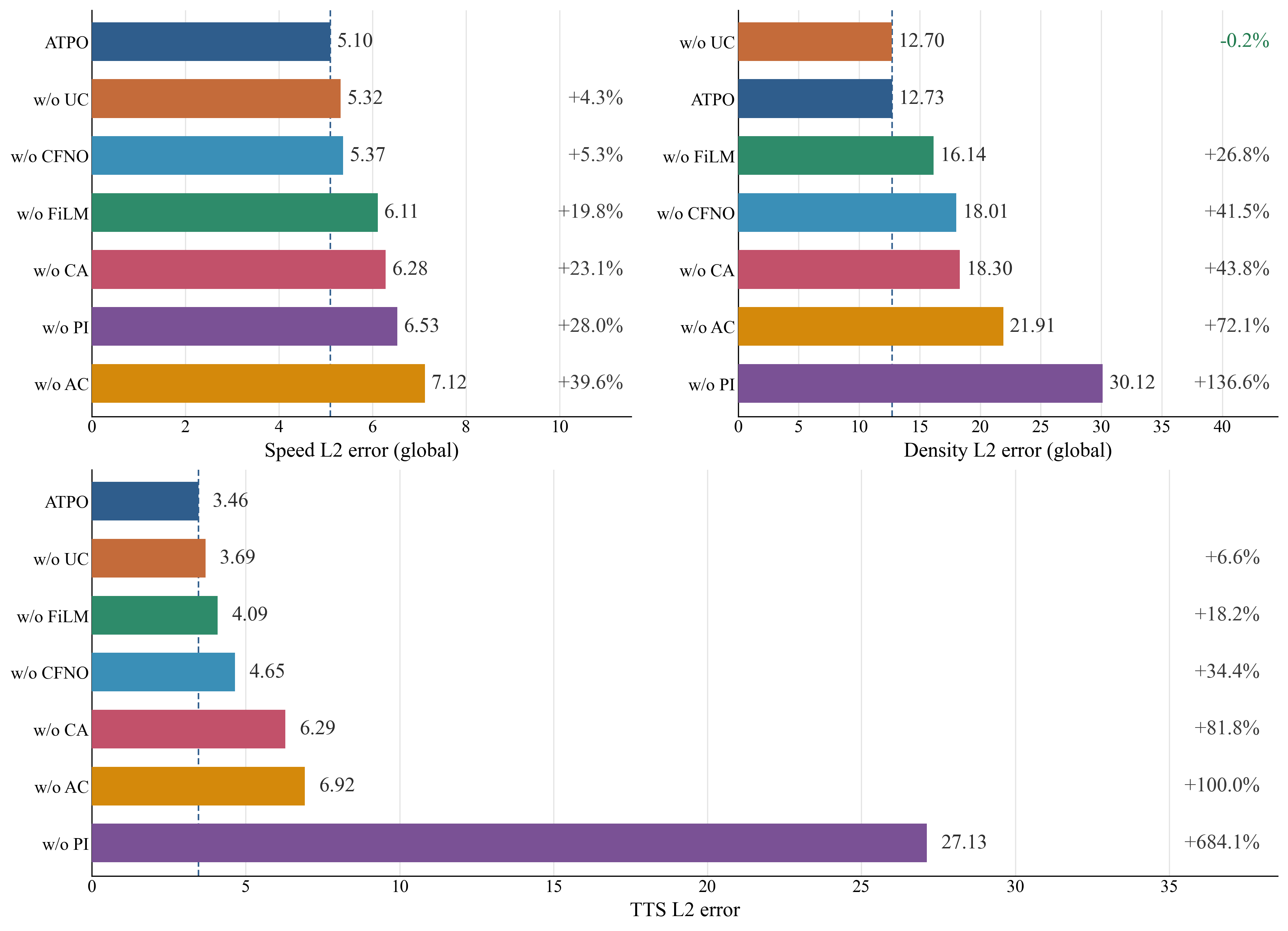}
\caption{\textbf{Contributions of predictor components to accuracy in the ablation study.}}\label{fig:14}\end{figure}

\hyperref[tab:4]{\textcolor{myvioletblue}{\textbf{Table 4}}} reports errors at detector locations, across the global traffic field, and for predicted TTS. \hyperref[fig:14]{\textcolor{myvioletblue}{\textbf{Figure 14}}} compares increases in global density, global speed, and TTS relative L2 errors after individual components are removed. Without AC, the three errors rise by 72.1\%, 39.6\%, and 100.0\%, respectively. Without CA, they rise by 43.8\%, 23.1\%, and 81.8\%. TTS deteriorates more than either state variable in both cases, showing that metering actions, demand, and queue conditions affect not only traffic-state forecasts but also the system cost calculated from them. Removing FiLM increases all three errors by 18.2\%-26.8\%, which indicates retaining an action input alone cannot replace FiLM’s effects.

The influence of PI is most pronounced for global density and TTS: removing it increases their relative L2 errors by 136.6\% and 684.1\%, respectively, while speed error rises by 28.0\%. Detector-density error changes little in \hyperref[tab:4]{\textcolor{myvioletblue}{\textbf{Table 4}}}. Good fit at monitored sites therefore does not ensure accurate forecasts at unobserved cells, global density errors further affect accumulated vehicle inventory and TTS. Removing CFNO increases density and TTS errors by 41.5\% and 34.4\%, compared with 5.3\% for speed, indicating a stronger contribution to density field and cumulative-cost accuracy. Removing UC reduces density error by 0.2\% and increases speed and TTS errors by only 4.3\% and 6.6\%. UC has a limited effect on the point error metrics in \hyperref[fig:14]{\textcolor{myvioletblue}{\textbf{Figure 14}}}, so its contribution should also be judged by predictive-distribution calibration.

\begin{table}[H]\centering\fontsize{8.5pt}{10.5pt}\selectfont
\caption{\textbf{Prediction ablation results for ATPO network components.}}\label{tab:4}
\setlength{\tabcolsep}{1.8pt}\renewcommand{\arraystretch}{1.12}
\begin{tabular}{@{}c*{10}{c}@{}}
\toprule
\textbf{Method} & \multicolumn{4}{c}{\textbf{Density}} & \multicolumn{4}{c}{\textbf{Speed}} & \multicolumn{2}{c}{\textbf{TTS}} \\
\cmidrule(lr){2-5}\cmidrule(lr){6-9}\cmidrule(l){10-11}
 & RMSE-d & L2-d & RMSE-g & L2-g & RMSE-d & L2-d & RMSE-g & L2-g & RMSE & L2 \\
\midrule
ATPO & 2.44 & 13.15 & 2.54 & 12.73 & \best{4.66} & \best{5.16} & \best{4.56} & \best{5.10} & \best{0.76} & \best{3.46} \\
-w/o AC & 3.15 & 16.99 & 4.37 & 21.91 & 5.79 & 6.41 & 6.36 & 7.12 & 1.52 & 6.92 \\
-w/o FiLM & 2.59 & 13.94 & 3.22 & 16.14 & 4.99 & 5.52 & 5.46 & 6.11 & 0.90 & 4.09 \\
-w/o CA & 2.93 & 15.81 & 3.65 & 18.30 & 5.04 & 5.58 & 5.61 & 6.28 & 1.39 & 6.29 \\
-w/o CFNO & 2.49 & 13.44 & 3.59 & 18.01 & 5.45 & 6.03 & 4.80 & 5.37 & 1.02 & 4.65 \\
-w/o PI & \best{2.35} & \best{12.67} & 6.01 & 30.12 & 4.85 & 5.37 & 5.83 & 6.53 & 5.98 & 27.13 \\
-w/o UC & 2.41 & 13.01 & \best{2.53} & \best{12.70} & 5.25 & 5.81 & 4.75 & 5.32 & 0.81 & 3.69 \\
\bottomrule\end{tabular}
\par\vspace{2pt}{\scriptsize\emph{Note:} d = detector cells; g = all cells. Density RMSE: veh/km/lane; speed RMSE: km/h; TTS RMSE: veh\,h. L2 values are percentages. Red bold indicates the lowest error in each column.}
\end{table}

The proposed changes also retain a central property of operator learning: a trained model can predict on grids of different resolutions without retraining. Over the same physical corridor and forecast duration, future queries, observations, and the control field are simply reconstructed on finer grids and evaluated with the same learned parameters.

\begin{figure}[htbp]\centering
\includegraphics[width=0.95\textwidth,height=0.73\textheight,keepaspectratio]{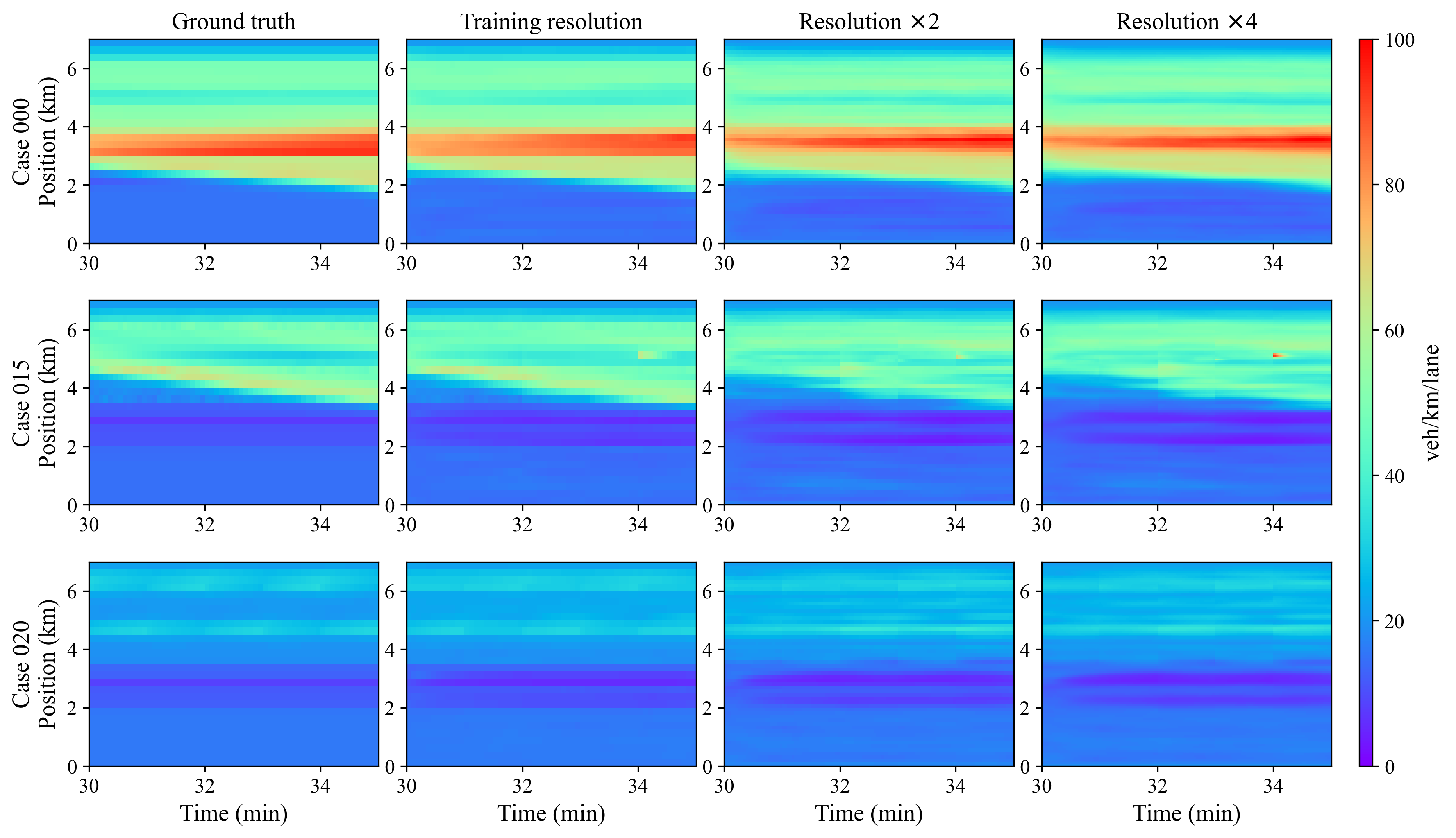}
\caption{\textbf{Predicted and ground-truth density fields at the original and finer grid resolutions.}}\label{fig:15}\end{figure}

\begin{figure}[htbp]\centering
\includegraphics[width=0.95\textwidth,height=0.73\textheight,keepaspectratio]{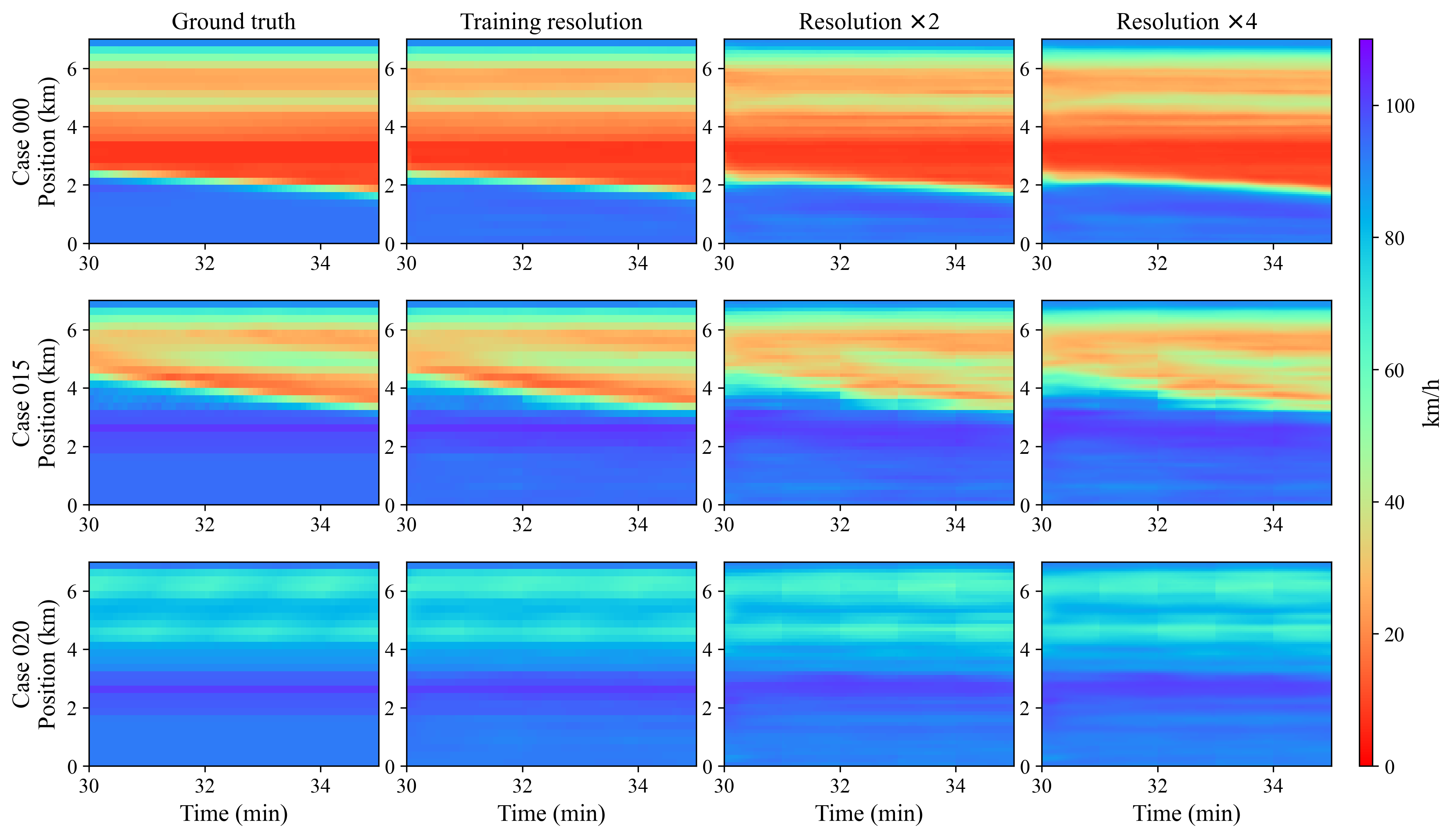}
\caption{\textbf{Predicted and ground-truth speed fields at the original and finer grid resolutions.}}\label{fig:16}\end{figure}

\hyperref[fig:15]{\textcolor{myvioletblue}{\textbf{Figures 15}}} and \hyperref[fig:16]{\textcolor{myvioletblue}{\textbf{16}}} compare density and speed predictions at the original, twofold, and fourfold grid resolutions. Reference fields at each resolution come from an independent simulation of the same traffic scenario on that grid, rather than interpolation of lower-resolution truth. The predicted fields show similar congestion locations, propagation directions, and dissipation stages across all three grids. In \hyperref[tab:5]{\textcolor{myvioletblue}{\textbf{Table 5}}}, TTS RMSE values are 0.76, 0.72, and 0.69 veh·h, and relative L2 errors are 3.46\%, 3.02\%, and 2.71\%. For this fixed physical corridor and forecast horizon, grid refinement requires no retraining and does not increase TTS error. The cross-resolution experiment shows that the ATPO modifications preserve this useful property of FNO-based prediction.

\begin{table}[H]\centering\footnotesize
\caption{\textbf{TTS prediction errors at the original and finer grid resolutions.}}\label{tab:5}
\setlength{\tabcolsep}{3.5pt}\renewcommand{\arraystretch}{1.12}
\begin{tabular}{@{}cccc@{}}
\toprule
\textbf{Resolution} & \textbf{Grid} & \textbf{TTS RMSE (veh·h)} & \textbf{TTS L2} \\
\midrule
Original & 60×28 & 0.76 ± 0.35 & 3.46\% ± 0.76\% \\
×2 & 120×56 & 0.72 ± 0.30 & 3.02\% ± 0.65\% \\
×4 & 240×112 & 0.69 ± 0.33 & 2.71\% ± 0.72\% \\
\bottomrule\end{tabular}
\end{table}

\subsection{Sensitivity to Observation Configuration}\label{sec:5.3}

\begin{figure}[htbp]\centering
\includegraphics[width=0.95\textwidth,height=0.73\textheight,keepaspectratio]{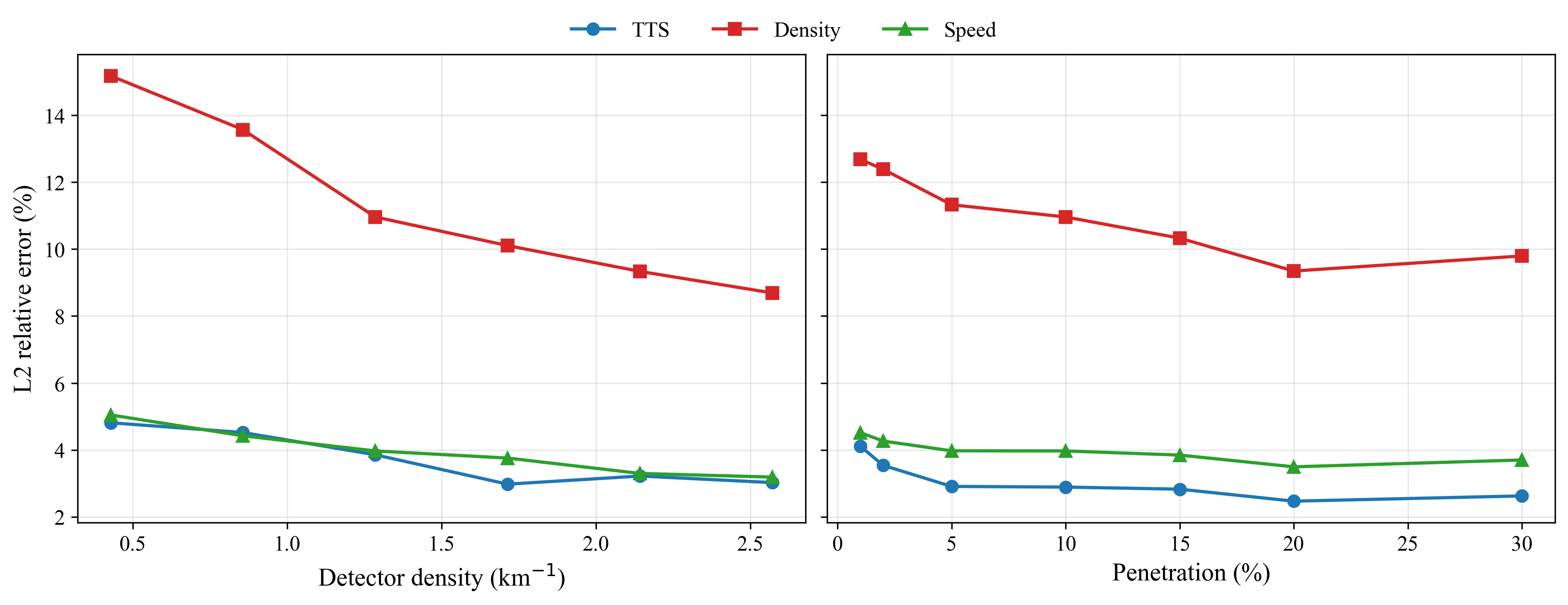}
\caption{\textbf{Effects of detector count and probe-vehicle penetration on relative prediction error.}}\label{fig:17}\end{figure}

\hyperref[fig:17]{\textcolor{myvioletblue}{\textbf{Figure 17}}} examines how fixed detector density and probe vehicle penetration affect ATPO's global prediction accuracy. One source is varied in each experiment. With probe penetration fixed at 10\%, the number of detectors increases from 3 to 18 in increments of 3. With nine detectors fixed, penetration takes values of 1\%, 2\%, 5\%, 10\%, 15\%, 20\%, and 30\%. Data are constructed for each configuration and a separate model is trained. All models are evaluated on the same test scenarios, isolating the effect of observation configuration.

As detector count increases, relative density, speed, and TTS errors all decline, with the largest improvement in density. Fixed detectors repeatedly provide density and speed at the same cross section and therefore constrain the location of congestion over time, directly improving density field reconstruction. Speed and TTS errors fall less, suggesting diminishing returns from additional detectors once speed forecasts stabilize at higher detector coverage.

Higher probe vehicle penetration also generally improves GTSP’s performance, but the response differs by metric. Density error falls rapidly at low penetration and shows little further improvement beyond approximately 20\%. Mobile probes provide scattered spatial samples that fill gaps between fixed detectors rather than replacing their continuous temporal measurements. Additional probes markedly improve spatial coverage when penetration is low. Once coverage approaches saturation, new samples are more redundant and yield smaller marginal gains.

The two experiments indicate complementary information from fixed detectors and probe vehicles. Adding either source can reduce error over part of the tested range. Deployment choices should therefore consider the target metric and the cost of acquiring each data source.

\subsection{Control Performance}\label{sec:5.4}

\begin{figure}[htbp]\centering
\includegraphics[width=0.95\textwidth,height=0.73\textheight,keepaspectratio]{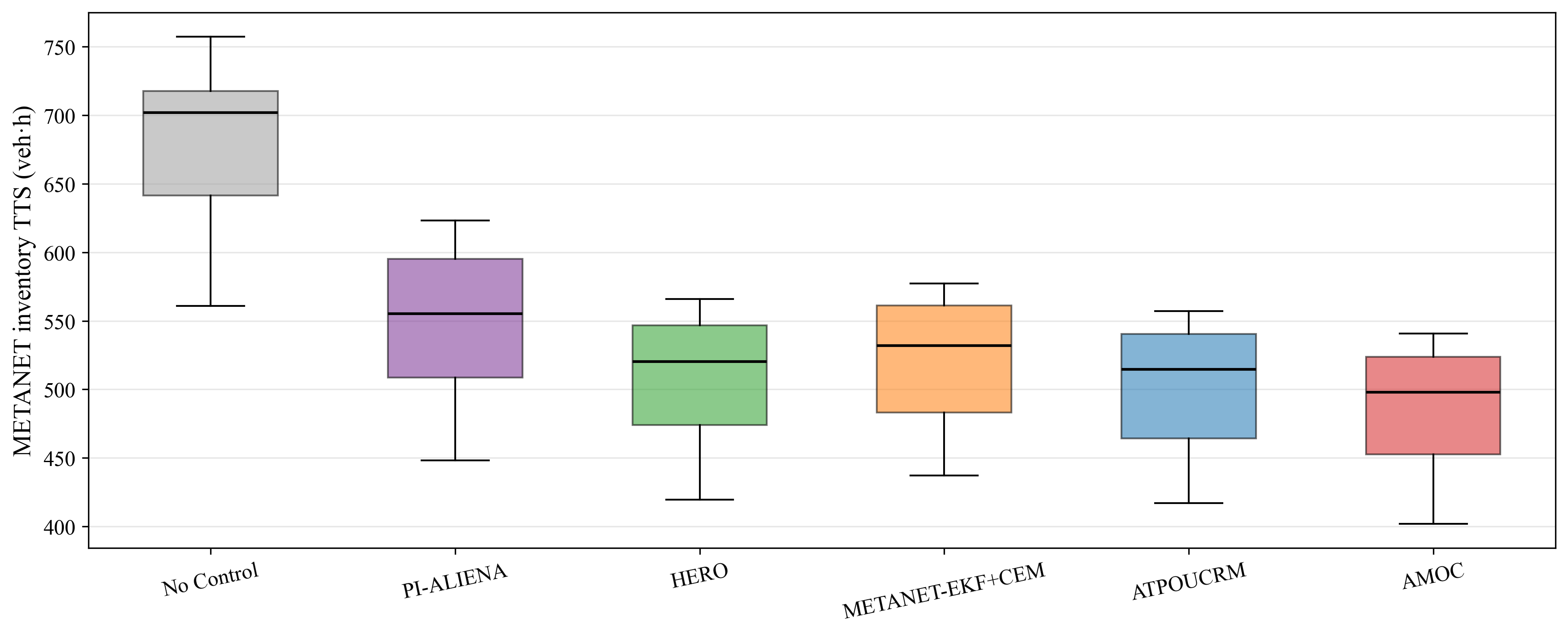}
\caption{\textbf{Control scenario TTS distributions for six control methods under paired test scenarios.}}\label{fig:18}\end{figure}

All strategies are evaluated in the same 100 different demand traffic scenarios to assess control performance. \hyperref[tab:6]{\textcolor{myvioletblue}{\textbf{Table 6}}} and \hyperref[fig:18]{\textcolor{myvioletblue}{\textbf{Figure 18}}} show scenario TTS distributions. ATPOUCRM has mean TTS of 503.74 veh·h, 26.17\% lower than no control and 46.36, 16.66, and 20.56 veh·h lower than PI-ALINEA, HERO, and METANET-EKF+CEM, respectively. AMOC, which has complete state and future-demand information, gives the lowest mean TTS (487.02 veh·h), 16.72 veh·h below ATPOUCRM. Relative to the TTS reduction from no control to AMOC, ATPOUCRM realizes approximately 91.4\% of the improvement. AMOC and ATPOUCRM operate with different information and prediction conditions. Their performance gap therefore reflects combined effects of state observation, future prediction, and action optimization.

ATPOUCRM has a between-scenario TTS standard deviation of 45.55 veh·h, lower than 55.85 veh·h for no control and 46.47 veh·h for METANET-EKF+CEM, and close to HERO and AMOC. Relative to HERO, its mean paired TTS improvement is 16.66 veh·h with a 95\% confidence interval of [12.20, 19.63] veh·h. It achieves lower TTS in 93 of 100 scenarios. The paired results indicate that the average gain is broadly distributed rather than driven primarily by a few unusually favorable scenarios.

\begin{figure}[htbp]\centering
\includegraphics[width=0.95\textwidth,height=0.73\textheight,keepaspectratio]{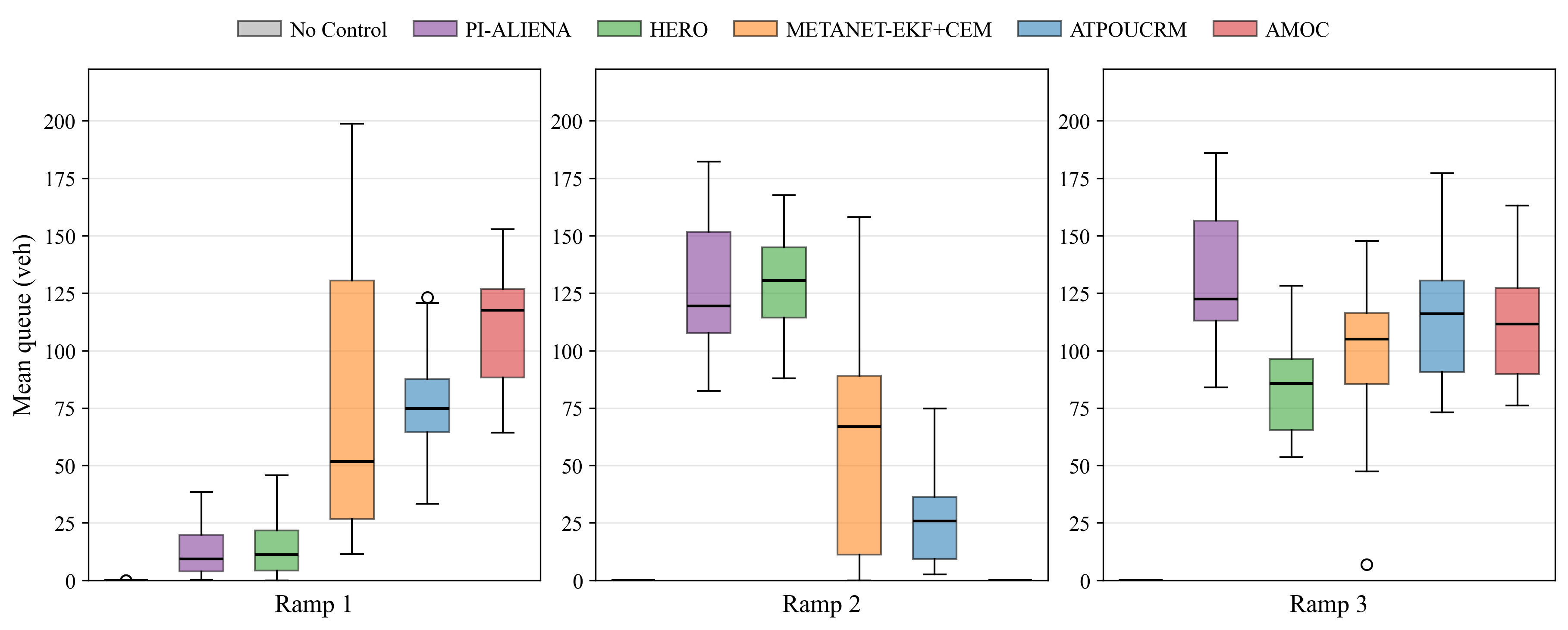}
\caption{\textbf{Scenario-mean queue lengths at the three on-ramps under different control methods.}}\label{fig:19}\end{figure}

\hyperref[fig:19]{\textcolor{myvioletblue}{\textbf{Figure 19}}} compares how the strategies allocate queues across the three on-ramps. No control produces almost no metering queue. Metering policies trade on-ramp waiting against mainline congestion by changing the location and length of queues. METANET-EKF+CEM maintains relatively long queues at ramps 1 and 2, especially ramp 2. Compared with PI-ALINEA and HERO, ATPOUCRM keeps ramp 2’s queue shorter while allocating more waiting to ramps 1 and 3. AMOC also uses queue storage mainly at ramps 1 and 3. Because TTS includes both mainline travel and ramp waiting, queue allocation is necessary for interpreting differences in closed-loop cost.

\begin{table}[H]\centering\footnotesize
\caption{\textbf{TTS of six control methods over all test scenarios.}}\label{tab:6}
\setlength{\tabcolsep}{3.5pt}\renewcommand{\arraystretch}{1.12}
\begin{tabular}{@{}c*{6}{c}@{}}
\toprule
 & \textbf{No Control} & \textbf{PI-ALINEA} & \textbf{HERO} & \textbf{METANET--EKF+CEM} & \textbf{ATPOUCRM} & \textbf{AMOC} \\
\midrule
TTS (veh·h) & 682.26 & 550.10 & 520.40 & 524.30 & \secondbest{503.74} & \best{487.02} \\
SD (veh·h) & 55.85 & 52.65 & 46.41 & 46.47 & \best{45.55} & \secondbest{45.67} \\
TTS Reduction (\%) & / & 19.37 & 23.72 & 23.15 & \secondbest{26.17} & \best{28.62} \\
\bottomrule\end{tabular}
\par\vspace{2pt}{\scriptsize\emph{Note:} Red bold and blue bold indicate the best and second-best values in each row, respectively. Lower TTS and SD and higher TTS reduction are better.}
\end{table}

\begin{figure}[htbp]\centering
\includegraphics[width=0.95\textwidth,height=0.73\textheight,keepaspectratio]{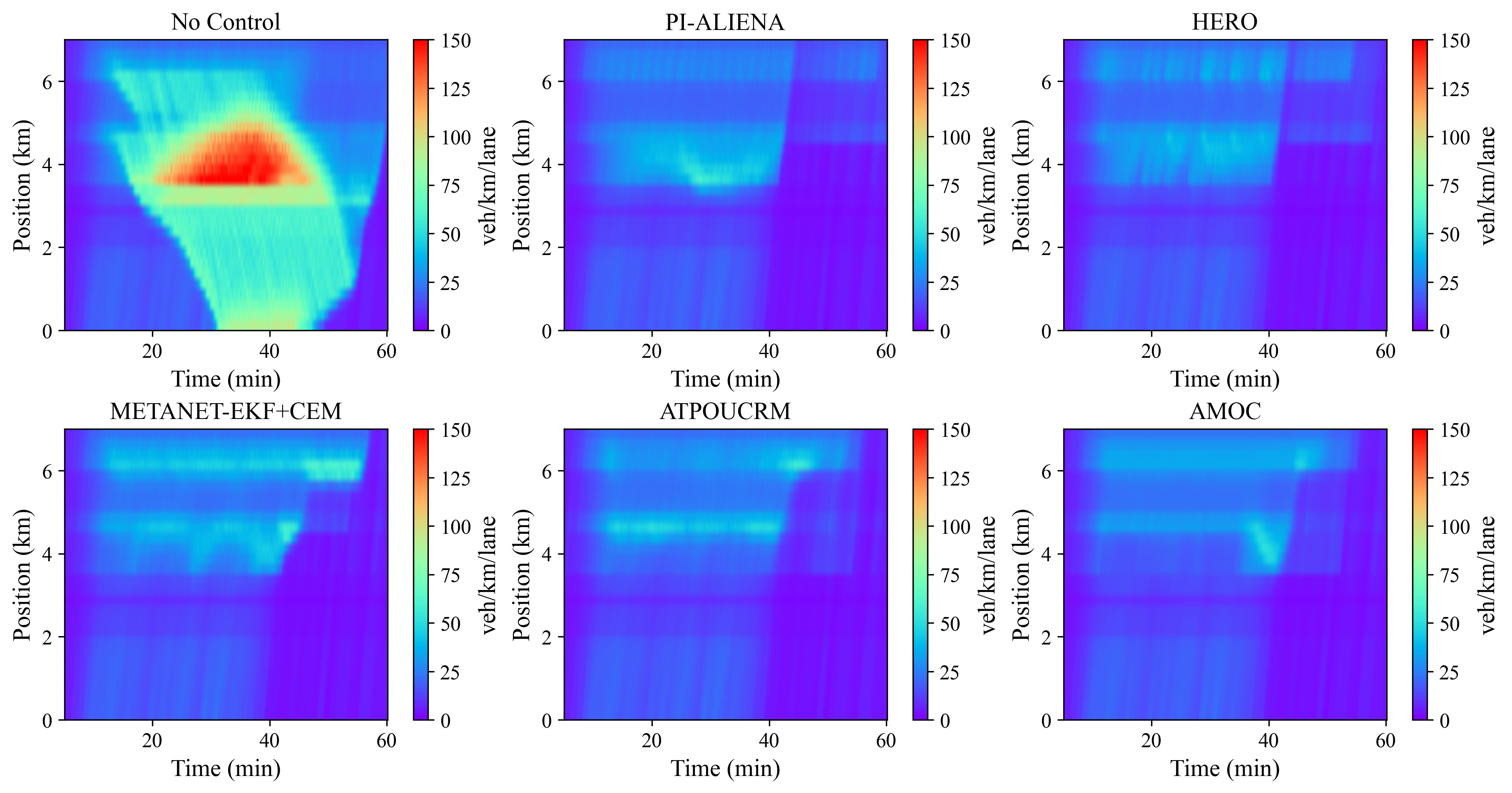}
\caption{\textbf{Mainline density evolution under different control methods in Case 19.}}\label{fig:20}\end{figure}

\begin{figure}[htbp]\centering
\includegraphics[width=0.95\textwidth,height=0.73\textheight,keepaspectratio]{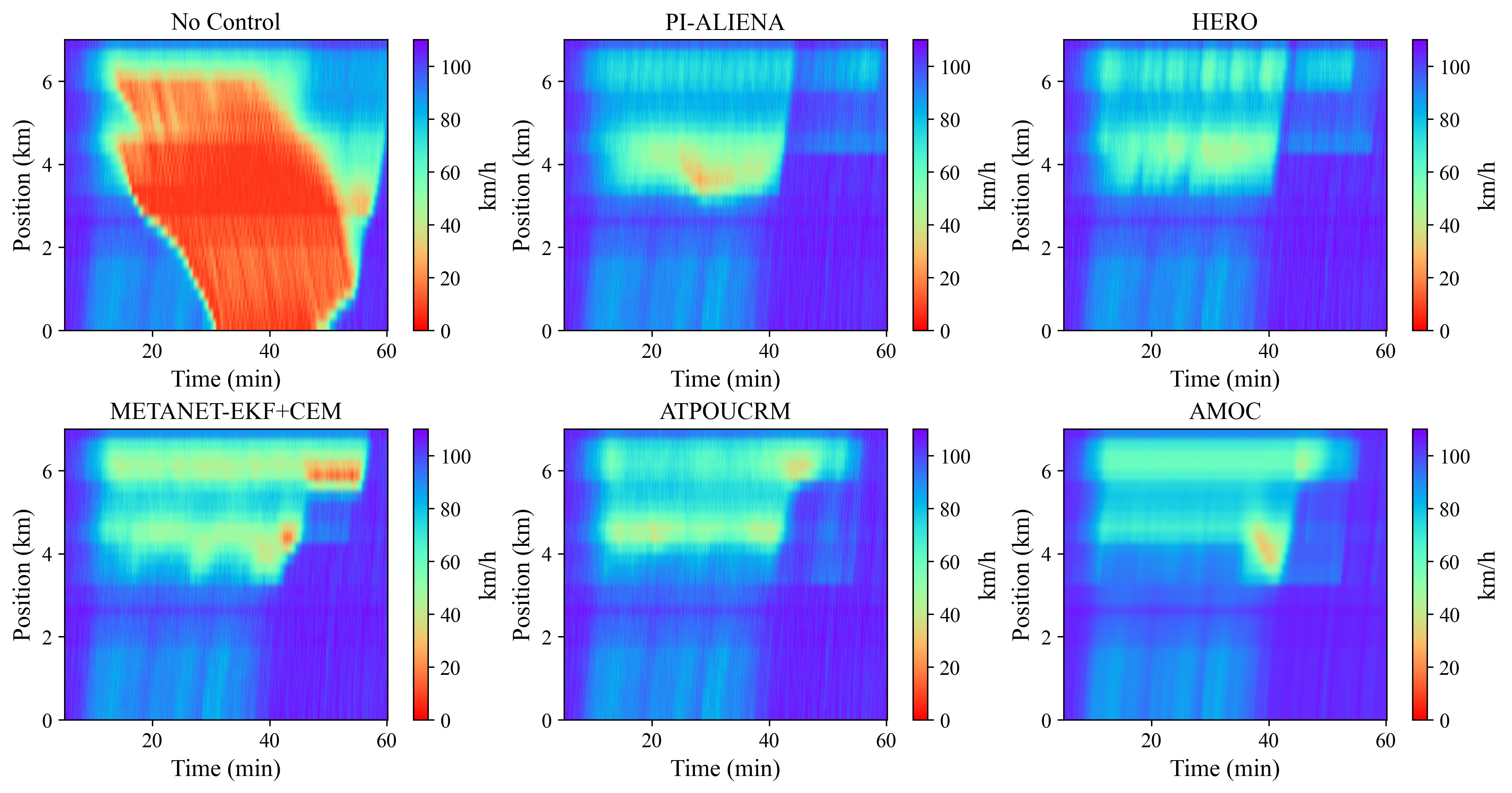}
\caption{\textbf{Mainline speed evolution under different control methods in Case 19.}}\label{fig:21}\end{figure}

Case 19 is a representative test scenario in which extensive, persistent congestion develops without control. \hyperref[fig:20]{\textcolor{myvioletblue}{\textbf{Figures 20}}} and \hyperref[fig:21]{\textcolor{myvioletblue}{\textbf{21}}} explain the Case 19 control results through mainline density and speed fields. Without control, high density and low speed cover a long road section and persist into the latter part of peak demand. All metering policies reduce this region. METANET-EKF+CEM substantially relieves mainline congestion, although local density still rises and speed falls near the end of the peak. ATPOUCRM yields a narrower congestion band and evolution closer to AMOC. PI-ALINEA and HERO also relieve the principal bottleneck, but a region between on-ramps 1 and 2 retains higher density and lower speed. Mainline fields alone cannot establish overall control superiority: ramp waiting must also be counted, as it is in the scenario TTS in \hyperref[tab:6]{\textcolor{myvioletblue}{\textbf{Table 6}}}.

\begin{figure}[htbp]\centering
\includegraphics[width=0.95\textwidth,height=0.73\textheight,keepaspectratio]{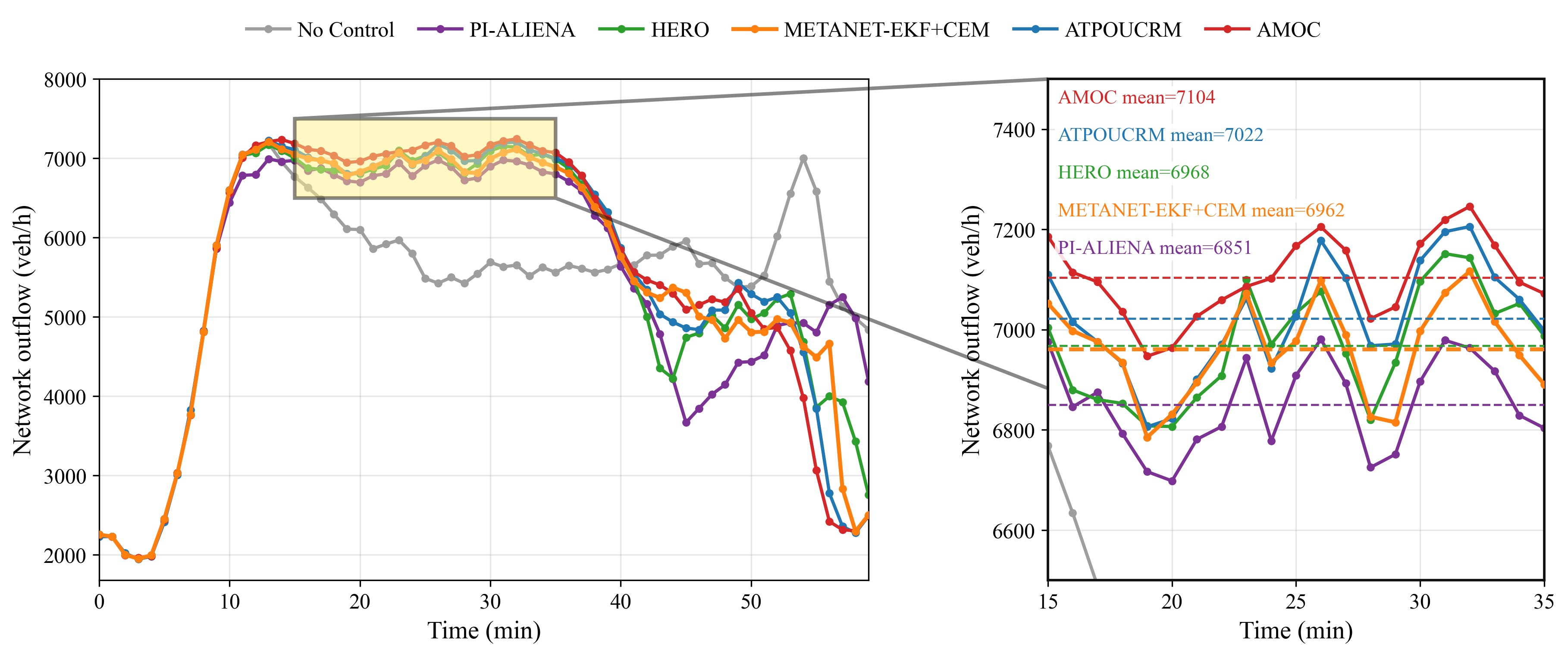}
\caption{\textbf{Network outflow in Case 19, with an enlarged view of the high-demand period.}}\label{fig:22}\end{figure}

\hyperref[fig:22]{\textcolor{myvioletblue}{\textbf{Figure 22}}} adds network outflow to this comparison. The demand peak from minute 15 to 35 is enlarged in the figure. Mean total network outflow under ATPOUCRM during this period is 7022 veh/h, slightly below AMOC but above 6851 veh/h for PI-ALINEA, 6962 veh/h for METANET--EKF+CEM, and 6968 veh/h for HERO. The higher outflow is consistent with the lighter mainline congestion in \hyperref[fig:20]{\textcolor{myvioletblue}{\textbf{Figures 20}}} and \hyperref[fig:21]{\textcolor{myvioletblue}{\textbf{21}}}: more vehicles pass through the network during the peak in this scenario.

\subsection{Controller Ablation and Solver Comparison}\label{sec:5.5}

To distinguish the effects of the uncertainty penalty, pre-execution safety filter, action discretization, and action-generation method on closed-loop performance, six variants are compared with complete ATPOUCRM in the same 100 test demand scenarios.

CEM-noU removes the predictive uncertainty penalty from the candidate-action objective while retaining CEM and the safety filter. CEM-noS retains the original objective and CEM but removes the safety filter applied before execution. These variants respectively test the criteria for selecting a candidate action and the protection applied to the action ultimately executed.

Because practical ramp signals have discrete timing, CEM-d restricts each candidate ramp discharge rate to an integer multiple of 60 veh/h. This variant tests how discrete metering commands affect control performance and computation time. The remaining three variants retain ATPO but use different methods to generate actions. Sequential quadratic programming (SQP; \textcolor{myvioletblue}{\textbf{\hyperlink{bib:31}{Nocedal and Wright, 2006}}}) locally approximates the objective and constraints and repeatedly solves quadratic subproblems to update the metering sequence. Model predictive path integral control (MPPI; \textcolor{myvioletblue}{\textbf{\hyperlink{bib:53}{Williams et al., 2017}}}) samples perturbations of metering sequences and updates the sequence using trajectory cost weighted samples. Group relative policy optimization (GRPO; \textcolor{myvioletblue}{\textbf{\hyperlink{bib:41}{Shao et al., 2024}}}) trains an action policy with ATPO as its world model: actions sampled from the same state are evaluated by ATPO, and their relative returns update the policy. The trained policy outputs actions directly at test time.

\begin{figure}[htbp]\centering
\includegraphics[width=0.95\textwidth,height=0.73\textheight,keepaspectratio]{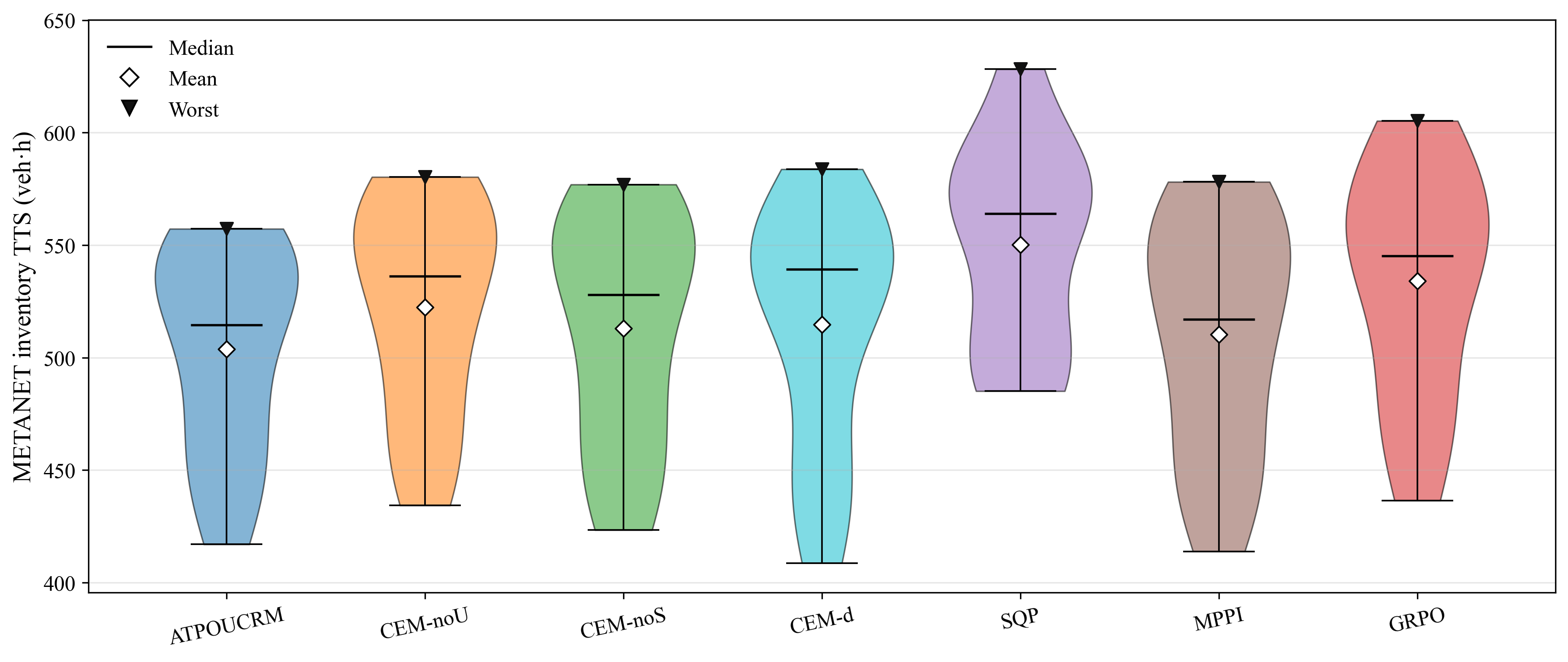}
\caption{\textbf{Scenario TTS distributions for control ablations, discrete metering, and alternative action-generation methods.}}\label{fig:23}\end{figure}

\hyperref[tab:7]{\textcolor{myvioletblue}{\textbf{Table 7}}} summarizes closed-loop TTS and computation time, while \hyperref[fig:23]{\textcolor{myvioletblue}{\textbf{Figure 23}}} shows the scenario TTS distributions. To account for differences in demand among scenarios, each variant's TTS is paired with that of complete ATPOUCRM in the same scenario. A Wilcoxon signed-rank test assesses whether these paired differences systematically depart from zero. Complete ATPOUCRM has mean TTS of 503.74 veh·h, a between-scenario standard deviation of 45.55 veh·h, and worst-scenario TTS of 557.20 veh·h, the lowest values among the compared methods.

Removing the uncertainty penalty (CEM-noU) increases mean TTS by 18.68 veh·h and worst-scenario TTS by 23.05 veh·h. Removing the safety filter (CEM-noS) increases the corresponding values by 9.12 and 19.72 veh·h. Both comparisons with the complete method give paired test values of $p<0.001$, supporting distinct control benefits from the two components in the tested scenarios. The safety filter is triggered at only 10.83\% of all control updates and mainly during congestion formation and dissipation. Its low activation rate, together with the CEM-noS performance difference, indicates that occasional pre-execution corrections can affect outcomes at consequential stages.

With discrete metering rates, CEM-d has mean TTS of 514.67 veh·h, 10.93 veh·h above the complete method but 5.73 veh·h below HERO. Its computation time per update falls from 21.4 to 8.7 s, leaving more computational margin within the one-minute control update interval. The paired test gives $p=0.058$ for CEM-d versus ATPOUCRM; this comparison does not reach the conventional 0.05 threshold.

Among the action-generation methods, MPPI gives mean TTS of 510.23 veh·h, only 1.3\% above CEM, but requires 52.4 s per update, about 2.45 times the CEM cost and close to the one-minute update interval. SQP requires 26.1 s per update; its mean and worst-scenario TTS exceed CEM by 46.50 and 70.97 veh·h, respectively. ATPO allows gradients of the predicted objective with respect to candidate actions, making SQP feasible. Differentiability alone, however, does not ensure that a local solver finds a high-performing sequence under the specified initialization, constraint treatment, and computation budget.

After offline training, GRPO outputs an action in 0.09 s per test update, but its mean and worst-scenario TTS exceed CEM by 30.27 and 47.97 veh·h. The pretrained policy markedly reduces online action-generation time but does not retain CEM's closed-loop performance in the present setting. Because GRPO learns from trajectories generated by ATPO, its performance may depend on world-model error, training-state coverage, and policy optimization. The paired tests give $p=0.044$ for MPPI and $p<0.001$ for SQP and GRPO relative to complete ATPOUCRM.

\begin{table}[H]\centering\footnotesize
\caption{\textbf{Closed-loop control ablations and action-generation comparisons over 100 scenarios.}}\label{tab:7}
\setlength{\tabcolsep}{2.8pt}\renewcommand{\arraystretch}{1.12}
\begin{tabular}{@{}c*{7}{c}@{}}
\toprule
 & \textbf{ATPOUCRM} & \textbf{CEM-noU} & \textbf{CEM-noS} & \textbf{CEM-d} & \textbf{SQP} & \textbf{MPPI} & \textbf{GRPO} \\
\midrule
TTS (veh·h) & \best{503.74} & 522.42 & 512.86 & 514.67 & 550.24 & 510.23 & 534.01 \\
SD (veh·h) & \best{45.55} & 47.16 & 50.71 & 53.20 & 45.83 & 51.80 & 51.17 \\
Worst TTS (veh·h) & \best{557.20} & 580.25 & 576.92 & 583.70 & 628.17 & 578.07 & 605.17 \\
Paired Wilcoxon $p$ & --- & $<0.001$ & $<0.001$ & 0.058 & $<0.001$ & 0.044 & $<0.001$ \\
Computation time (s) & 21.4 & 20.2 & 21.2 & 8.7 & 26.1 & 52.4 & \best{0.09} \\
\bottomrule\end{tabular}
\par\vspace{2pt}{\scriptsize\emph{Note:} Red bold indicates the lowest value for each performance measure; paired Wilcoxon tests compare each variant with ATPOUCRM. Computation time excludes offline training.}
\end{table}

\begin{figure}[htbp]\centering
\includegraphics[width=0.95\textwidth,height=0.73\textheight,keepaspectratio]{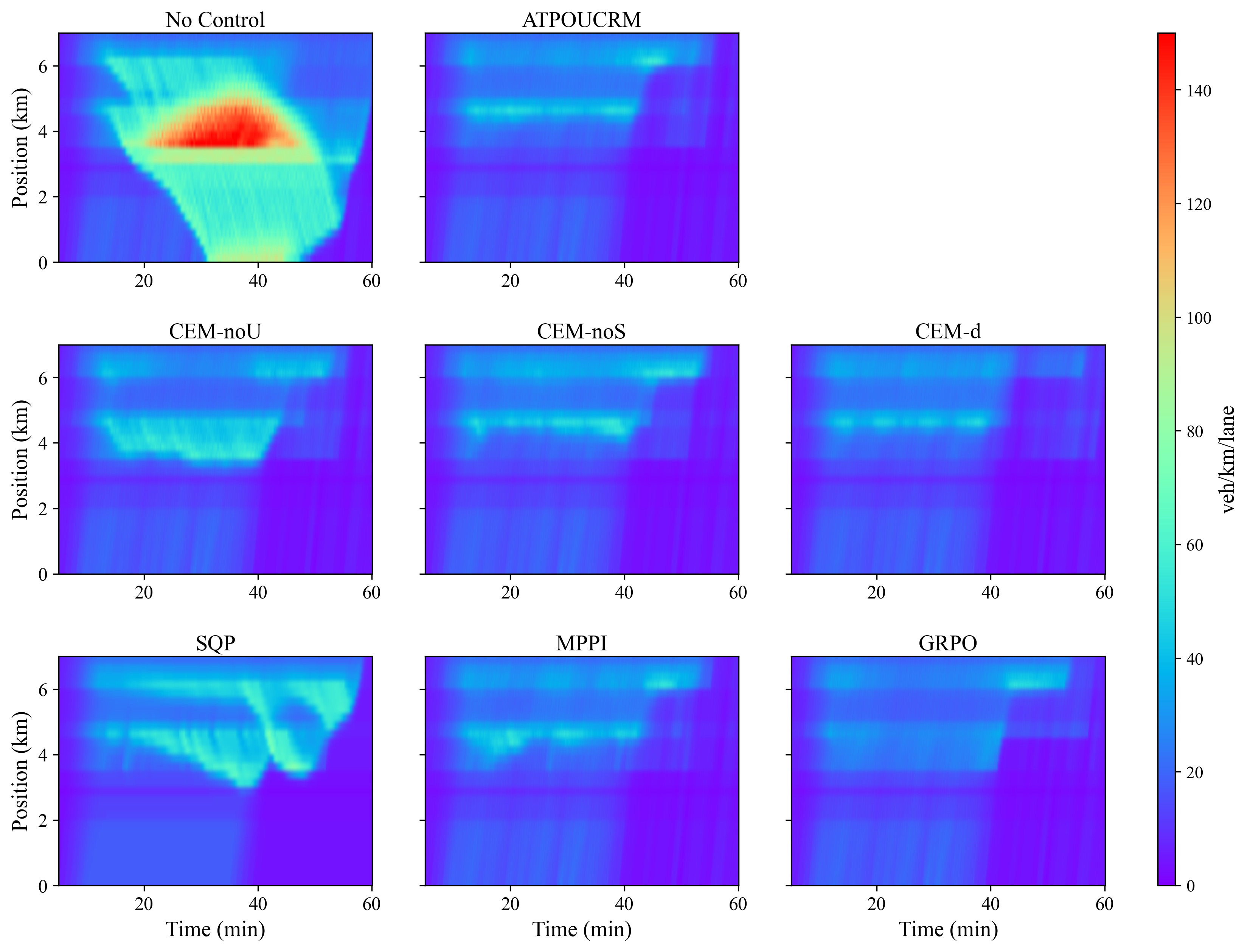}
\caption{\textbf{Mainline density fields in Case 19 under the control ablations and alternative action-generation methods.}}\label{fig:24}\end{figure}

\begin{figure}[htbp]\centering
\includegraphics[width=0.95\textwidth,height=0.73\textheight,keepaspectratio]{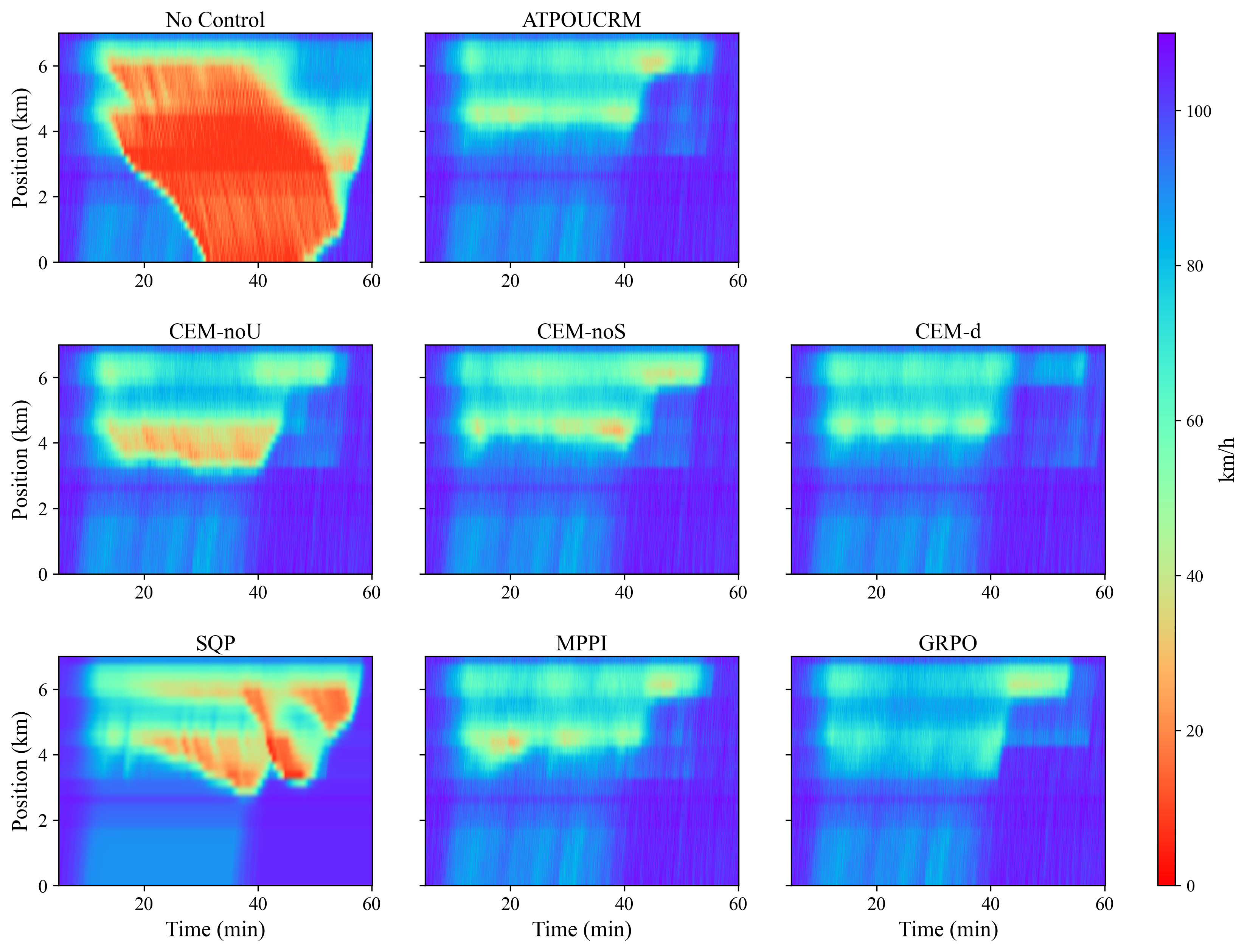}
\caption{\textbf{Mainline speed fields in Case 19 under the control ablations and alternative action-generation methods.}}\label{fig:25}\end{figure}

\hyperref[fig:24]{\textcolor{myvioletblue}{\textbf{Figure 24}}} and \hyperref[fig:25]{\textcolor{myvioletblue}{\textbf{Figure 25}}} show the mainline density and speed fields for the Case 19 control variants. Complete ATPOUCRM confines the high-density and low-speed regions to a relatively narrow area. Removing the uncertainty penalty or safety filter intensifies congestion between on-ramps 1 and 2, while SQP produces a wider low-speed region. GRPO shows a more conservative pattern: restricted ramp admission keeps mainline density low, but the resulting action sequence yields higher network TTS than ATPOUCRM.

\section{Conclusions}\label{sec:6}

This study developed ATPOUCRM, a CRM framework combining the Action-Conditioned Traffic Prediction Operator (ATPO) with an uncertainty-aware controller (UCRM). ATPO forms a sparse four-channel history from density, speed, and their separate masks; future spatiotemporal coordinates query all valid historical observations, and separate branches incorporate upstream mainline demand and on-ramp queues. Candidate actions modulate latent features at each future point. A spatial Fourier operator models dependence along the road, while a causal temporal kernel constrains the direction in which control information propagates. ATPO outputs means and standard deviations for future corridor-wide density and speed, complemented by disagreement among ensemble members. UCRM evaluates candidate actions with a main objective of total time spent (TTS), including mainline travel and on-ramp waiting, and adds penalties for uncertainty, abrupt action changes, and density exceedance. It searches actions with CEM and uses a pre-execution safety filter to reduce the likelihood of executing high-risk plans. ATPO prediction and ATPOUCRM control were evaluated in a 7-km macroscopic freeway simulation. The principal findings are:

(1) ATPO achieves the lowest global density, speed, and TTS prediction errors among the comparison methods. It also tracks the principal congestion dynamics at unobserved locations and under an unseen policy with random actions. Its uncertainty outputs provide useful information about forecast difficulty.

(2) Ablations show that action-conditioned input, exogenous cross-attention, layerwise FiLM, the causal Fourier block, and physics-informed loss each contribute to prediction accuracy. Removing physics-informed loss increases global density and TTS relative L2 errors by 136.6\% and 684.1\%, respectively, while detector-density error changes little. Accuracy at measured locations alone is therefore insufficient to assess global forecasts.

(3) Over the same physical spatiotemporal domain, ATPO applies its trained weights to finer grids without retraining, and TTS prediction error does not increase with grid refinement in the tested cases.

(4) Across 100 test-demand scenarios, ATPOUCRM gives mean closed-loop TTS of 503.74 veh·h, 26.17\% below no control and 16.66 veh·h below HERO. Spatiotemporal results in a representative scenario show that coordinated metering and queue allocation reduce the mainline congestion region.

(5) Controller ablations indicate that the uncertainty objective and pre-execution safety filter both improve closed-loop performance. CEM also offers favorable control accuracy with lower computation time than the compared online solvers in this setting.

Further extensions remain. GTSP is examined here only for a freeway corridor and could be extended to large urban networks. Control decisions still rely on online optimization, whose computation time may limit real-time deployment. Future work may iteratively improve reinforcement-learning policies and GTSP models to obtain low-latency, accurate prediction and control.

\section*{Author Contributions}

The authors confirm the following contributions. Hongxin Yu contributed to study conception and design, results analysis and interpretation, and drafting. Yibing Wang contributed to study conception and design, results analysis and interpretation, and drafting. Fengyue Jin collected data. Tao Jiang conducted the simulation experiments. Klaus Bogenberger contributed to results analysis and interpretation. Jingqiu Guo and Lihui Zhang contributed to drafting. All authors reviewed the results and approved the final manuscript.

\section*{Funding}

This study was supported by the National Natural Science Foundation of China (52272315) and the Zhejiang Provincial Key Research and Development Program (2024C01180).

\section*{References}
\begin{hangparas}{0.5cm}{1}
\hypertarget{bib:1}{}Airaldi, F., De Schutter, B., \& Dabiri, A. (2025). Reinforcement learning with model predictive control for highway ramp metering. IEEE Transactions on Intelligent Transportation Systems, 26(5), 5988--6004. \url{https://doi.org/10.1109/TITS.2025.3549227}\par
\hypertarget{bib:2}{}Bai, S., Kolter, J. Z., \& Koltun, V. (2018). An empirical evaluation of generic convolutional and recurrent networks for sequence modeling. arXiv:1803.01271. \url{https://doi.org/10.48550/arXiv.1803.01271}\par
\hypertarget{bib:3}{}Bhan, L., Shi, Y., \& Krstic, M. (2024). Neural operators for bypassing gain and control computations in PDE backstepping. IEEE Transactions on Automatic Control, 69(8), 5310--5325. \url{https://doi.org/10.1109/TAC.2023.3347499}\par
\hypertarget{bib:4}{}Bhan, L., Qin, P., Krstic, M., \& Shi, Y. (2025). Neural operators for predictor feedback control of nonlinear delay systems. Proceedings of Machine Learning Research, 283, 179--193. \url{https://doi.org/10.48550/arXiv.2411.18964}\par
\hypertarget{bib:5}{}Briani, M., Cristiani, E., \& Onofri, E. (2024). Inverting the fundamental diagram and forecasting boundary conditions: How machine learning can improve macroscopic models for traffic flow. Advances in Computational Mathematics, 50(6), 115. \url{https://doi.org/10.1007/s10444-024-10206-8}\par
\hypertarget{bib:6}{}Chen, S.-M., Ji, X.-M., \& Shao, H.-P. (2026). Neural operator learning methods for traffic state estimation in connected environments. China Journal of Highway and Transport, 39(1), 265--279. \url{https://doi.org/10.19721/j.cnki.1001-7372.2026.01.022}\par
\hypertarget{bib:7}{}Chen, J. (2026). Flux-form spatiotemporal neural operators for coarse-grained dynamics of multiscale PDEs. arXiv:2608.18148. \url{https://doi.org/10.48550/arXiv.2608.18148}\par
\hypertarget{bib:8}{}Chua, K., Calandra, R., McAllister, R., \& Levine, S. (2018). Deep reinforcement learning in a handful of trials using probabilistic dynamics models. Advances in Neural Information Processing Systems, 31. \url{https://doi.org/10.48550/arXiv.1805.12114}\par
\hypertarget{bib:9}{}Daganzo, C. F. (1994). The cell transmission model: A dynamic representation of highway traffic consistent with the hydrodynamic theory. Transportation Research Part B, 28(4), 269--287. \url{https://doi.org/10.1016/0191-2615(94)90002-7}\par
\hypertarget{bib:10}{}de Jong, T. O., Shukla, K., \& Lazar, M. (2025). Deep operator neural network model predictive control. IEEE Open Journal of Control Systems, 4, 501--517. \url{https://doi.org/10.1109/OJCSYS.2025.3614875}\par
\hypertarget{bib:11}{}Gao, Q., \& Di, X. (2024). Comparative studies of neural operators for traffic state estimation. 2024 IEEE 27th International Conference on Intelligent Transportation Systems (ITSC). \url{https://doi.org/10.1109/ITSC58415.2024.10920089}\par
\hypertarget{bib:12}{}Gomes, G., \& Horowitz, R. (2006). Optimal freeway ramp metering using the asymmetric cell transmission model. Transportation Research Part C, 14(4), 244--262. \url{https://doi.org/10.1016/j.trc.2006.08.001}\par
\hypertarget{bib:13}{}Gu, C., Zhou, T., \& Wu, C. (2023). Deep Koopman traffic modeling for freeway ramp metering. IEEE Transactions on Intelligent Transportation Systems, 24(6), 6001--6013. \url{https://doi.org/10.1109/TITS.2023.3248649}\par
\hypertarget{bib:14}{}Han, Y., Ramezani, M., Hegyi, A., Yuan, Y., \& Hoogendoorn, S. P. (2020). Hierarchical ramp metering in freeways: An aggregated modeling and control approach. Transportation Research Part C, 110, 1--19. \url{https://doi.org/10.1016/j.trc.2019.09.023}\par
\hypertarget{bib:15}{}Han, Y., Wang, M., Li, L., Roncoli, C., Gao, J., \& Liu, P. (2022). A physics-informed reinforcement learning-based strategy for local and coordinated ramp metering. Transportation Research Part C, 137, 103584. \url{https://doi.org/10.1016/j.trc.2022.103584}\par
\hypertarget{bib:16}{}Harting, A., Johansson, K. H., \& Barreau, M. (2025). Closed-loop neural operator-based observer of traffic density. 2025 IEEE 64th Conference on Decision and Control. \url{https://doi.org/10.1109/CDC57313.2025.11312700}\par
\hypertarget{bib:17}{}Hegyi, A., De Schutter, B., \& Hellendoorn, J. (2005). Model predictive control for optimal coordination of ramp metering and variable speed limits. Transportation Research Part C, 13(3), 185--209. \url{https://doi.org/10.1016/j.trc.2004.08.001}\par
\hypertarget{bib:18}{}Hu, Z., \& Ma, W. (2024). Demonstration-guided deep reinforcement learning for coordinated ramp metering and perimeter control in large scale networks. Transportation Research Part C, 159, 104461. \url{https://doi.org/10.1016/j.trc.2023.104461}\par
\hypertarget{bib:19}{}Kotsialos, A., \& Papageorgiou, M. (2004). Efficiency and equity properties of freeway network-wide ramp metering with AMOC. Transportation Research Part C, 12(6), 401--420. \url{https://doi.org/10.1016/j.trc.2004.07.016}\par
\hypertarget{bib:20}{}Kotsialos, A., Papageorgiou, M., Mangeas, M., \& Haj-Salem, H. (2002). Coordinated and integrated control of motorway networks via non-linear optimal control. Transportation Research Part C, 10(1), 65--84. \url{https://doi.org/10.1016/S0968-090X(01)00005-5}\par
\hypertarget{bib:21}{}Kovachki, N., Li, Z., Liu, B., Azizzadenesheli, K., Bhattacharya, K., Stuart, A., \& Anandkumar, A. (2023). Neural operator: Learning maps between function spaces with applications to PDEs. Journal of Machine Learning Research, 24(83), 1--97. \url{https://doi.org/10.5555/3648699.3648788}\par
\hypertarget{bib:22}{}Lakshminarayanan, B., Pritzel, A., \& Blundell, C. (2017). Simple and scalable predictive uncertainty estimation using deep ensembles. Advances in Neural Information Processing Systems, 30. \url{https://doi.org/10.48550/arXiv.1612.01474}\par
\hypertarget{bib:23}{}Li, Y., Yu, R., Shahabi, C., \& Liu, Y. (2018). Diffusion convolutional recurrent neural network: Data-driven traffic forecasting. International Conference on Learning Representations. \url{https://doi.org/10.48550/arXiv.1707.01926}\par
\hypertarget{bib:24}{}Li, Z., Kovachki, N. B., Azizzadenesheli, K., Liu, B., Bhattacharya, K., Stuart, A., \& Anandkumar, A. (2021). Fourier neural operator for parametric partial differential equations. International Conference on Learning Representations. \url{https://doi.org/10.48550/arXiv.2010.08895}\par
\hypertarget{bib:25}{}Li, Z., Wang, T., Zou, G., Wang, R., \& Li, Y. (2025). Physics-informed deep operator network for traffic state estimation. Transportmetrica B: Transport Dynamics, 13(1), 2595549. \url{https://doi.org/10.1080/21680566.2025.2595549}\par
\hypertarget{bib:26}{}Li, Z., Wang, T., Zou, G., Wang, R., \& Li, Y. (2026). Across temporal--spectral domains: Graph convolutional network with Fourier neural operator for spatiotemporal traffic flow prediction. Chaos, Solitons \& Fractals, 209, 118559. \url{https://doi.org/10.1016/j.chaos.2026.118559}\par
\hypertarget{bib:27}{}Liang, D., Zhang, H., \& Yuan, D. (2025). DistPred: A distribution-free probabilistic inference method for regression and forecasting. Proceedings of the 31st ACM SIGKDD Conference on Knowledge Discovery and Data Mining, 753--764. \url{https://doi.org/10.1145/3690624.3709286}\par
\hypertarget{bib:28}{}Lu, L., Jin, P., Pang, G., Zhang, Z., \& Karniadakis, G. E. (2021). Learning nonlinear operators via DeepONet based on the universal approximation theorem of operators. Nature Machine Intelligence, 3, 218--229. \url{https://doi.org/10.1038/s42256-021-00302-5}\par
\hypertarget{bib:29}{}Lv, K., Wang, J., Zhang, Y., \& Yu, H. (2025). Neural operators for adaptive control of freeway traffic. Automatica, 182, 112553. \url{https://doi.org/10.1016/j.automatica.2025.112553}\par
\hypertarget{bib:30}{}Mallick, T., Macfarlane, J., \& Balaprakash, P. (2024). Uncertainty quantification for traffic forecasting using deep-ensemble-based spatiotemporal graph neural networks. IEEE Transactions on Intelligent Transportation Systems, 25(8), 9141--9152. \url{https://doi.org/10.1109/TITS.2024.3381099}\par
\hypertarget{bib:31}{}Mei, H., Li, J., Liang, Z., Zheng, G., Shi, B., \& Wei, H. (2023). Uncertainty-aware traffic prediction under missing data. 2023 IEEE International Conference on Data Mining, 1223--1228. \url{https://doi.org/10.1109/ICDM58522.2023.00152}\par
\hypertarget{bib:32}{}Messmer, A., \& Papageorgiou, M. (1990). METANET: A macroscopic simulation program for motorway networks. Traffic Engineering \& Control, 31, 466--470.\par
\hypertarget{bib:33}{}Mo, Z., Fu, Y., Xu, D., \& Di, X. (2022). TrafficFlowGAN: Physics-informed flow based generative adversarial network for uncertainty quantification. Machine Learning and Knowledge Discovery in Databases, Part III, 323--339. \url{https://doi.org/10.1007/978-3-031-26409-2_20}\par
\hypertarget{bib:34}{}Muralidharan, A., \& Horowitz, R. (2015). Computationally efficient model predictive control of freeway networks. Transportation Research Part C, 58, 532--553. \url{https://doi.org/10.1016/j.trc.2015.03.029}\par
\hypertarget{bib:35}{}Nantes, A., Ngoduy, D., Bhaskar, A., Miska, M., \& Chung, E. (2016). Real-time traffic state estimation in urban corridors from heterogeneous data. Transportation Research Part C, 66, 99--118. \url{https://doi.org/10.1016/j.trc.2015.07.005}\par
\hypertarget{bib:36}{}Nocedal, J., \& Wright, S. J. (2006). Numerical optimization (2nd ed.). Springer. \url{https://doi.org/10.1007/978-0-387-40065-5}\par
\hypertarget{bib:37}{}Papageorgiou, M., Hadj-Salem, H., \& Blosseville, J.-M. (1991). ALINEA: A local feedback control law for on-ramp metering. Transportation Research Record, 1320, 58--64.\par
\hypertarget{bib:38}{}Papamichail, I., Kotsialos, A., Margonis, I., \& Papageorgiou, M. (2010a). Coordinated ramp metering for freeway networks: A model-predictive hierarchical control approach. Transportation Research Part C, 18(3), 311--331. \url{https://doi.org/10.1016/j.trc.2008.11.002}\par
\hypertarget{bib:39}{}Papamichail, I., Papageorgiou, M., Vong, V., \& Gaffney, J. (2010b). Heuristic ramp-metering coordination strategy implemented at Monash Freeway, Australia. Transportation Research Record, 2178, 10--20. \url{https://doi.org/10.3141/2178-02}\par
\hypertarget{bib:40}{}Pereira, M., Lang, A., \& Kulcsár, B. (2022). Short-term traffic prediction using physics-aware neural networks. Transportation Research Part C, 142, 103772. \url{https://doi.org/10.1016/j.trc.2022.103772}\par
\hypertarget{bib:41}{}Perez, E., Strub, F., de Vries, H., Dumoulin, V., \& Courville, A. C. (2018). FiLM: Visual reasoning with a general conditioning layer. Proceedings of the AAAI Conference on Artificial Intelligence, 32(1), 3942--3951. \url{https://doi.org/10.1609/aaai.v32i1.11671}\par
\hypertarget{bib:42}{}Roncoli, C., Papamichail, I., \& Papageorgiou, M. (2016). Hierarchical model predictive control for multi-lane motorways in presence of Vehicle Automation and Communication Systems. Transportation Research Part C, 62, 117--132. \url{https://doi.org/10.1016/j.trc.2015.11.008}\par
\hypertarget{bib:43}{}Roth, A., \& Liebig, T. (2022). Forecasting unobserved node states with spatio-temporal graph neural networks. 2022 IEEE International Conference on Data Mining Workshops (ICDMW). \url{https://doi.org/10.1109/ICDMW58026.2022.00101}\par
\hypertarget{bib:44}{}Rubinstein, R. Y. (1999). The cross-entropy method for combinatorial and continuous optimization. Methodology and Computing in Applied Probability, 1, 127--190. \url{https://doi.org/10.1023/A:1010091220143}\par
\hypertarget{bib:45}{}Sengupta, A., Mondal, S., Das, A., \& Guler, S. I. (2024). A Bayesian approach to quantifying uncertainties and improving generalizability in traffic prediction models. Transportation Research Part C, 162, 104585. \url{https://doi.org/10.1016/j.trc.2024.104585}\par
\hypertarget{bib:46}{}Shao, Z., Wang, P., Zhu, Q., Xu, R., Song, J., Bi, X., Zhang, H., Zhang, M., Li, Y. K., Wu, Y., \& Guo, D. (2024). DeepSeekMath: Pushing the limits of mathematical reasoning in open language models. arXiv:2402.03300. \url{https://doi.org/10.48550/arXiv.2402.03300}\par
\hypertarget{bib:47}{}Tancik, M., Srinivasan, P. P., Mildenhall, B., Fridovich-Keil, S., Raghavan, N., Singhal, U., Ramamoorthi, R., Barron, J. T., \& Ng, R. (2020). Fourier features let networks learn high frequency functions in low dimensional domains. Advances in Neural Information Processing Systems, 33, 7537--7547. \url{https://doi.org/10.48550/arXiv.2006.10739}\par
\hypertarget{bib:48}{}Thodi, B. T., Ambadipudi, S. V. R., \& Jabari, S. E. (2024). Fourier neural operator for learning solutions to macroscopic traffic flow models: Application to the forward and inverse problems. Transportation Research Part C, 160, 104500. \url{https://doi.org/10.1016/j.trc.2024.104500}\par
\hypertarget{bib:49}{}Treiber, M., \& Helbing, D. (2003). An adaptive smoothing method for traffic state identification from incomplete information. In H. Emmerich, B. Nestler, \& M. Schreckenberg (Eds.), Interface and transport dynamics (pp. 343--360). Springer. \url{https://doi.org/10.1007/978-3-662-07969-0_33}\par
\hypertarget{bib:50}{}Vaswani, A., Shazeer, N., Parmar, N., Uszkoreit, J., Jones, L., Gomez, A. N., Kaiser, Ł., \& Polosukhin, I. (2017). Attention is all you need. Advances in Neural Information Processing Systems, 30. \url{https://doi.org/10.48550/arXiv.1706.03762}\par
\hypertarget{bib:51}{}Wang, T., Li, Y., Cheng, R., Zou, G., Dantsuji, T., \& Ngoduy, D. (2026a). Knowledge-data fusion oriented traffic state estimation: A stochastic physics-informed deep learning approach. Transportation Research Part C, 182, 105422. \url{https://doi.org/10.1016/j.trc.2025.105422}\par
\hypertarget{bib:52}{}Wang, T., Li, Z., Cheng, R., Guo, L., Dantsuji, T., \& Li, Y. (2026b). Spatiotemporal Fourier neural operator-empowered super-resolution traffic flow field reconstruction from sparse observations. Physica A, 697, 131742. \url{https://doi.org/10.1016/j.physa.2026.131742}\par
\hypertarget{bib:53}{}Wang, Y., Hu, R., Yu, H., Li, J., Zhang, Y., Xu, Z., He, Z., \& Lu, Q. (2025). Global traffic state prediction method for non-sensing locations on freeways. Journal of Traffic and Transportation Engineering, 25(1), 274--294. \url{https://doi.org/10.19818/j.cnki.1671-1637.2025.01.020}\par
\hypertarget{bib:54}{}Wang, Y., \& Papageorgiou, M. (2005). Real-time freeway traffic state estimation based on extended Kalman filter: A general approach. Transportation Research Part B, 39(2), 141--167. \url{https://doi.org/10.1016/j.trb.2004.03.003}\par
\hypertarget{bib:55}{}Wang, Y., Papageorgiou, M., \& Messmer, A. (2008). Real-time freeway traffic state estimation based on extended Kalman filter: Adaptive capabilities and real data testing. Transportation Research Part A, 42(10), 1340--1358. \url{https://doi.org/10.1016/j.tra.2008.06.001}\par
\hypertarget{bib:56}{}Wang, Y., Papageorgiou, M., Gaffney, J., Papamichail, I., Rose, G., \& Young, W. (2010). PI-ALINEA for local ramp metering involved with random downstream bottlenecks. Proceedings of the 24th ARRB Conference, 1--14.\par
\hypertarget{bib:57}{}Wang, Y., Zhao, M., Yu, X., Hu, Y., Zheng, P., Hua, W., Zhang, L., Hu, S., \& Guo, J. (2022). Real-time joint traffic state and model parameter estimation on freeways with fixed sensors and connected vehicles: State-of-the-art overview, methods, and case studies. Transportation Research Part C, 134, 103444. \url{https://doi.org/10.1016/j.trc.2021.103444}\par
\hypertarget{bib:58}{}Williams, G., Aldrich, A., \& Theodorou, E. A. (2017). Model predictive path integral control: From theory to parallel computation. Journal of Guidance, Control, and Dynamics, 40(2), 344--357. \url{https://doi.org/10.2514/1.G001921}\par
\hypertarget{bib:59}{}Wright, M. A., \& Horowitz, R. (2016). Fusing loop and GPS probe measurements to estimate freeway density. IEEE Transactions on Intelligent Transportation Systems, 17(12), 3577--3590. \url{https://doi.org/10.1109/TITS.2016.2565438}\par
\hypertarget{bib:60}{}Wu, F., Cheng, Z., Chen, H., Qiu, Z., \& Sun, L. (2024). Traffic state estimation from vehicle trajectories with anisotropic Gaussian processes. Transportation Research Part C, 163, 104646. \url{https://doi.org/10.1016/j.trc.2024.104646}\par
\hypertarget{bib:61}{}Wu, Y., Zhuang, D., Labbe, A., \& Sun, L. (2021). Inductive graph neural networks for spatiotemporal kriging. Proceedings of the AAAI Conference on Artificial Intelligence, 35(5), 4478--4485. \url{https://doi.org/10.1609/aaai.v35i5.16575}\par
\hypertarget{bib:62}{}Wu, Z., Pan, S., Long, G., Jiang, J., \& Zhang, C. (2019). Graph WaveNet for deep spatial-temporal graph modeling. Proceedings of the 28th International Joint Conference on Artificial Intelligence, 1907--1913. \url{https://doi.org/10.24963/ijcai.2019/264}\par
\hypertarget{bib:63}{}Yang, H., Du, L., Zhang, G., \& Ma, T. (2023). A traffic flow dependency and dynamics based deep learning aided approach for network-wide traffic speed propagation prediction. Transportation Research Part B, 167, 99--117. \url{https://doi.org/10.1016/j.trb.2022.11.009}\par
\hypertarget{bib:64}{}Yang, H., Yu, W., Zhang, G., \& Du, L. (2024). Network-wide traffic flow dynamics prediction leveraging macroscopic traffic flow model and deep neural networks. IEEE Transactions on Intelligent Transportation Systems, 25(5), 4443--4457. \url{https://doi.org/10.1109/TITS.2023.3329489}\par
\hypertarget{bib:65}{}Yu, H., Zhang, L., Zhang, M., Jin, F., \& Wang, Y. (2024). Coordinated ramp metering considering the dynamics of mixed-autonomy traffic. Sustainability, 16(22), 10055. \url{https://doi.org/10.3390/su162210055}\par
\hypertarget{bib:66}{}Yu, H., Wang, Y., Jin, F., Zhang, M., \& Chen, A. (2025). A physics-informed deep operator for real-time freeway traffic state estimation. arXiv preprint, arXiv:2508.08002. \url{https://doi.org/10.48550/arXiv.2508.08002}\par
\hypertarget{bib:67}{}Zhang, J., Mao, S., Yang, L., Ma, W., Li, S., \& Gao, Z. (2024). Physics-informed deep learning for traffic state estimation based on the traffic flow model and computational graph method. Information Fusion, 101, 101971. \url{https://doi.org/10.1016/j.inffus.2023.101971}\par
\hypertarget{bib:68}{}Zhang, Y., Zhong, R., \& Yu, H. (2025). Mitigating stop-and-go traffic congestion with operator learning. Transportation Research Part C, 170, 104928. \url{https://doi.org/10.1016/j.trc.2024.104928}\par
\hypertarget{bib:69}{}Zhao, M., Roncoli, C., Wang, Y., Bekiaris-Liberis, N., Guo, J., \& Cheng, S. (2022). Generic approaches to estimating freeway traffic state and percentage of connected vehicles with fixed and mobile sensing. IEEE Transactions on Intelligent Transportation Systems, 23(8), 13155--13177. \url{https://doi.org/10.1109/TITS.2021.3121181}\par
\hypertarget{bib:70}{}Zhou, Q., Zhang, Y., Makridis, M. A., Kouvelas, A., Wang, Y., \& Hu, S. (2025). MoGERNN: An inductive traffic predictor for unobserved locations. Transportation Research Part C, 174, 105080. \url{https://doi.org/10.1016/j.trc.2025.105080}\par
\hypertarget{bib:71}{}Zhou, T., Gu, C., Lim, C. P., \& Yuan, J. (2026). Koopman-driven linearized model-based offline planning with application to freeway ramp metering. IEEE Transactions on Neural Networks and Learning Systems, 37(1), 284--298. \url{https://doi.org/10.1109/TNNLS.2025.3605015}\par
\end{hangparas}
\end{document}